\documentclass{article}

\usepackage[main,final]{neurips_2026}

\usepackage[utf8]{inputenc} 
\usepackage[T1]{fontenc}    
\usepackage[colorlinks=true,citecolor=teal,linkcolor=purple,urlcolor=purple]{hyperref}       
\usepackage{url}            
\usepackage{booktabs}       
\usepackage{amsfonts}       
\usepackage{nicefrac}       
\usepackage{microtype}      
\usepackage{xcolor}         

\usepackage[caption=false]{subfig}
\usepackage[normalem]{ulem}
\usepackage{
    adjustbox,
    algorithm,
    algpseudocode,
    amscd,
    amsmath,
    amssymb,
    amsthm,
    bm,
    braket,
    comment,
    csquotes,
    enumerate,
    enumitem,
    graphicx,
    IEEEtrantools,
    indentfirst,
    latexsym,
    mathrsfs,
    mathtools,
    multirow,
    placeins,
    relsize,
    soul,
    tabls,
    tikz,
    tikz-cd
}
\usetikzlibrary{quantikz2}

\definecolor{choiceone}{RGB}{60,110,180}
\definecolor{choicetwo}{RGB}{200,120,60}
\definecolor{choicethree}{RGB}{220,110,150}
\definecolor{choicefour}{RGB}{140,100,180}
\definecolor{choicefive}{RGB}{80,160,110}
\usetikzlibrary{quantikz2}

\newtheorem{theorem}{Theorem}
\newtheorem{definition}{Definition}
\newtheorem{lemma}[theorem]{Lemma}

\newtheorem{corollary}[theorem]{Corollary}

\newtheorem{innercustomgeneric}{\customgenericname}
\providecommand{\customgenericname}{}
\newcommand{\newcustomtheorem}[2]{%
  \newenvironment{#1}[1]
  {%
   \renewcommand\customgenericname{#2}%
   \renewcommand\theinnercustomgeneric{##1}%
   \innercustomgeneric
  }
  {\endinnercustomgeneric}
}

\newcustomtheorem{customthm}{Theorem}
\newcustomtheorem{customlemma}{Lemma}
\newcustomtheorem{customproposition}{Proposition}

\newenvironment{remark}[1][Remark]{\begin{trivlist}
\item[\hskip \labelsep {\bfseries #1}]}{\end{trivlist}}

\algnewcommand\algorithmicswitch{\textbf{switch}}
\algnewcommand\algorithmiccase{\textbf{case}}
\algdef{SE}[SWITCH]{Switch}{EndSwitch}[1]{\algorithmicswitch\ #1\ \algorithmicdo}{\algorithmicend\ \algorithmicswitch}%
\algdef{SE}[CASE]{Case}{EndCase}[1]{\algorithmiccase\ #1}{\algorithmicend\ \algorithmiccase}%
\algtext*{EndSwitch}%
\algtext*{EndCase}%

\newif\ifverbose
\verbosetrue

\newcommand{\stkout}[1]{\ifmmode\text{\st{\ensuremath{#1}}}\else\st{#1}\fi}

\newcommand{\naive}{\ddagger}
\newcommand{\optimal}{\star}

\DeclareMathOperator{\sign}{sign}

\DeclareMathOperator{\CNOT}{CNOT}

\renewcommand{\exp}{\ensuremath{\mathrm{exp}}}
\DeclareMathOperator{\Ppoly}{\mathsf{P/poly}}

\newcommand{\pder}[2][]{\frac{\partial#1}{\partial#2}}
\newcommand{\der}[2][]{\frac{\mathrm{d}#1}{\mathrm{d}#2}}

\providecommand{\hF}{\ensuremath{\hat{F}}}

\providecommand{\hX}{\ensuremath{\hat{X}}}
\providecommand{\hY}{\ensuremath{\hat{Y}}}

\providecommand{\ha}{\ensuremath{\hat{a}}}

\providecommand{\hf}{\ensuremath{\hat{f}}}

\providecommand{\hk}{\ensuremath{\hat{k}}}

\providecommand{\tG}{\ensuremath{\Tilde{G}}}

\providecommand{\tX}{\ensuremath{\Tilde{X}}}

\providecommand{\ta}{\ensuremath{\Tilde{a}}}

\providecommand{\tf}{\ensuremath{\Tilde{f}}}

\providecommand{\tk}{\ensuremath{\Tilde{k}}}

\providecommand{\to}{\ensuremath{\Tilde{o}}}

\providecommand{\tphi}{\ensuremath{\Tilde{\phi}}}

\providecommand{\tpsi}{\ensuremath{\Tilde{\psi}}}

\providecommand{\tcalM}{\ensuremath{\Tilde{\mathcal{M}}}}

\providecommand{\tcalO}{\ensuremath{\Tilde{\mathcal{O}}}}

\providecommand{\calH}{\ensuremath{\mathcal{H}}}

\providecommand{\calL}{\ensuremath{\mathcal{L}}}
\providecommand{\calM}{\ensuremath{\mathcal{M}}}

\providecommand{\calO}{\ensuremath{\mathcal{O}}}

\providecommand{\calX}{\ensuremath{\mathcal{X}}}
\providecommand{\calY}{\ensuremath{\mathcal{Y}}}

\providecommand{\bbE}{\ensuremath{\mathbb{E}}}

\providecommand{\bbI}{\ensuremath{\mathbb{I}}}

\providecommand{\bbN}{\ensuremath{\mathbb{N}}}

\providecommand{\bbP}{\ensuremath{\mathbb{P}}}

\providecommand{\bbR}{\ensuremath{\mathbb{R}}}

\graphicspath{ {figures/} }

\title{Optimal algorithmic complexity of inference in quantum kernel methods}

\title{Optimal algorithmic complexity of inference in quantum kernel methods}

\author{%
Elies Gil-Fuster$^{\ast,1,2}$, Seongwook Shin$^{1,3}$, Sofiene Jerbi$^{1,4}$,\\
\textbf{Jens Eisert}$^{1,2,4}$\textbf{, Maximilian J. Kramer}$^{\dagger,1}$\\
\small $^1$ Freie Universität Berlin \\
\small $^2$ Fraunhofer Heinrich Hertz Institute Berlin \\
\small $^3$ Sejong University \\
\small $^4$ Helmholtz-Zentrum Berlin \\
\small $^\ast$ \texttt{emgilfuster@gmail.com} \\
\small $^\dagger$ \texttt{m.kramer@fu-berlin.de}
}

\begin{document}

\maketitle

\begin{abstract}
  Quantum kernel methods are among the leading candidates for achieving quantum advantage in supervised learning.
  A key bottleneck is the cost of inference: evaluating a weighted sum of $N$ kernel values to additive precision $\varepsilon$, where $\alpha$ is the vector of trained coefficients.
    The standard approach estimates each term independently via sampling, yielding a query complexity of $\mathcal{O}(N\lVert\alpha\rVert_2^2/\varepsilon^2)$.
    In this work, we combine two independent improvements: estimating the kernel values via quantum amplitude estimation and collecting the sum under a single observable.
    We show that the improved approach achieves a query complexity of $\mathcal{O}(\lVert\alpha\rVert_1/\varepsilon)$, removing the dependence on $N$ from the query count and yielding a quadratic improvement in both $\lVert\alpha\rVert_1$ and $\varepsilon$.
    We prove a matching lower bound of $\Omega(\lVert\alpha\rVert_1/\varepsilon)$, establishing query-optimality in our oracle access model.
    Beyond query complexity, we also analyze how these improvements translate into gate costs and show that the query-optimal strategy is not always optimal in practice from the perspective of gate complexity.
    We thus provide both a query-optimal algorithm and practical guiding principles for the choice of strategy depending on hardware capabilities, along with a complete landscape of intermediate methods to guide practitioners.
    All algorithms require only amplitude estimation as a subroutine and are thus natural candidates for early-fault-tolerant implementations.
\end{abstract}

\section{Introduction}\label{s:intro}

    \begin{figure}[t]
        \centering
        \includegraphics[width=0.95\linewidth]{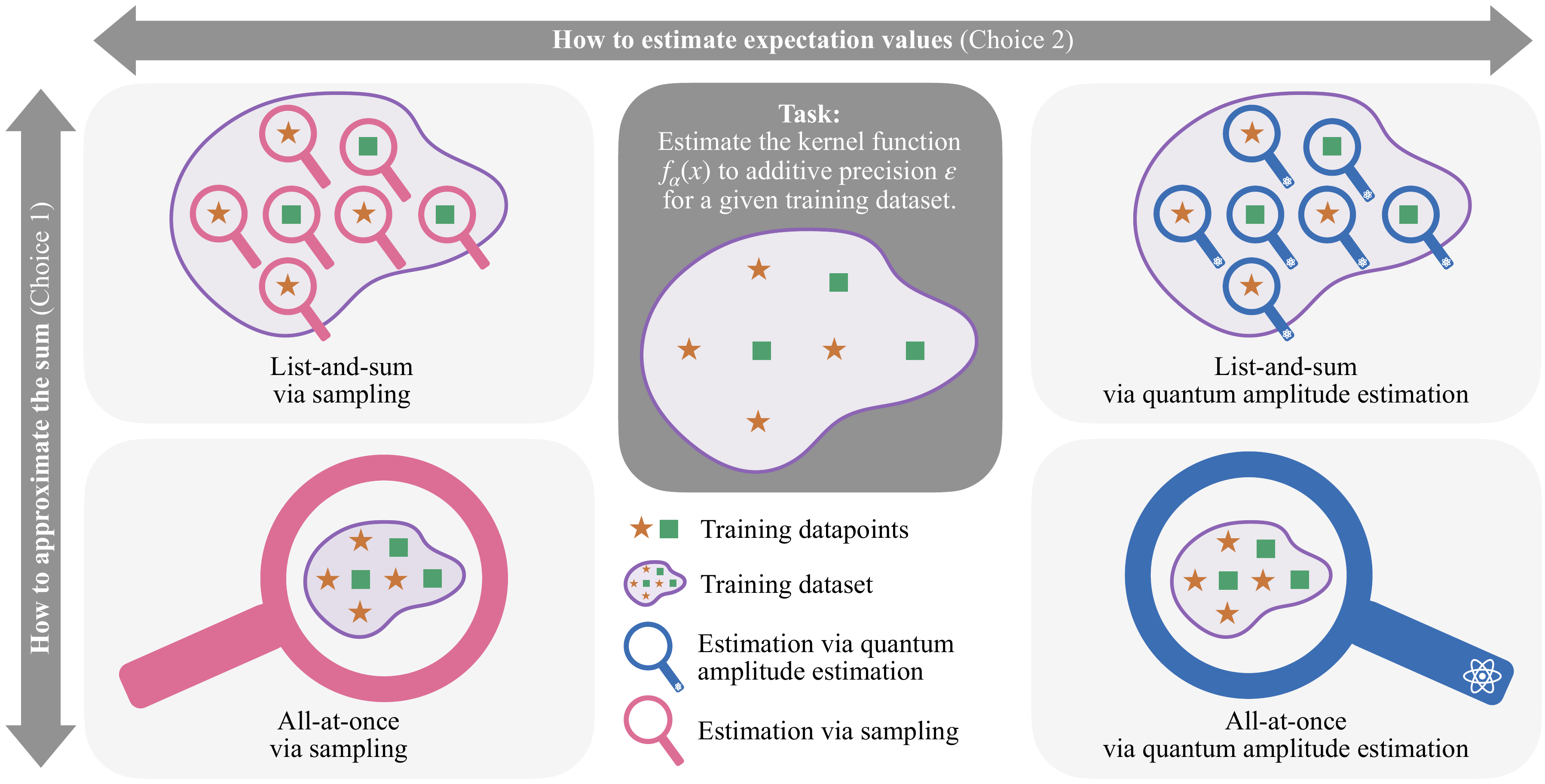}
        \caption{
            Schematic overview of the four algorithmic strategies for inference in quantum kernel methods. The two axes correspond to the two independent design choices: how the sum over training points is approximated (\emph{list-and-sum} versus all-at-once, vertical axis), and how expectation values are estimated (sampling versus quantum amplitude estimation, horizontal axis). In list-and-sum, each kernel value $k(x,x_i)$ is estimated independently and results are combined classically. In the \emph{all-at-once} approach, the full inference sum is encoded as a single observable. Moving from sampling to quantum amplitude estimation yields a quadratic improvement in precision within each row.
        }
        \label{fig:sketch}
    \end{figure}

    Quantum kernel methods~\citep{schuld2021kernels,tanner2026nonvariationalsupervisedquantumkernel} are among the leading candidates for achieving quantum advantage in supervised learning~\citep{biamonte2017quantum,QMLWhitePaper}, as showcased by the provable learning separation of \citet{Liu2021discretelog}.
    The mainstream approach is based on \emph{embedding quantum kernels}~\citep{hubregtsen2021training,gilfuster2024expressivity}, where a kernel function $k$ is specified via an $n$-qubit \emph{quantum feature map} $x\mapsto\rho(x)$ as $k(x,x') = \operatorname{Tr}\{\rho(x)\rho(x')\}$.
    After training -- finding suitable parameters $\alpha\in\bbR^N$ from a training set of size $N$ -- one must evaluate the resulting hypothesis on new data. This \emph{inference} step amounts to computing $f_\alpha(x) = \sum_{i=1}^N \alpha_i k(x,x_i)$, and its cost has been identified as a critical scalability bottleneck~\citep{nakayama2025explicitquantumsurrogatesquantum}.
    Commonly used techniques~\citep{havlicek2019supervised,tanner2026nonvariationalsupervisedquantumkernel} estimate each of the $N$ terms independently via sampling, resulting in a query complexity of $\calO(N\lVert\alpha\rVert_2^2/\varepsilon^2)$.
    This scaling can be improved along two independent axes: In the way individual kernel values are estimated (\emph{sampling} versus \emph{quantum amplitude estimation}), and how the sum over training points is approximated (\emph{term-by-term} versus \emph{via a single observable}). We systematically investigate these choices in Section~\ref{s:results} and Appendix~\ref{a:zoo}.
    Ultimately, we show that in the early fault-tolerant regime~\citep{eisert2025mind}, the optimal combination -- encoding the full sum as a single observable and applying quantum amplitude estimation -- achieves a query complexity of $\calO(\lVert\alpha\rVert_1/\varepsilon)$, where $\lVert\alpha\rVert_1=\sum_{i=1}^N\lvert\alpha_i\rvert$ (Theorem~\ref{thm:meta}).
    We complement this with a matching lower bound of $\Omega(\lVert\alpha\rVert_1/\varepsilon)$, up to constant factors (Theorem~\ref{thm:lower-bound-full}).
    Beyond query complexity, our results provide a practical guide to quantum kernel inference: given hardware capabilities, Algorithm~\ref{alg:unified} and Theorem~\ref{thm:meta} identify the appropriate strategy and its runtime, while Theorem~\ref{thm:meta-gates} characterizes the corresponding gate complexity.
    Among the strategies analyzed here, \emph{list-and-sum} with adaptive budget via quantum amplitude estimation has the smallest leading polynomial gate-count upper bound in the stated regime. Throughout, we use \emph{gate-optimal} in this comparative sense, suppressing polylogarithmic factors.

\section{Preliminaries}\label{s:preliminaries}
\subsection{Problem set-up and notation}\label{ss:problem_setup}
    Suppose we are given a learning task defined on $\calX\times\calY$, where a quantum kernel model has already been trained. We call the parameters of the kernel model $\alpha\in\bbR^N$, where $N$ is the size of the training set.
    We consider quantum feature maps realized via a data-dependent unitary $U(x)$ applied to the all-$0$ state vector as $\lvert\phi(x)\rangle = U(x)\lvert0\rangle$, with $\rho(x) = \lvert\phi(x)\rangle\!\langle\phi(x)\rvert$. In this case, the kernel function becomes $k(x,x') = \lvert\langle\phi(x)\rvert\phi(x')\rangle\rvert^2$
    for $x,x'\in \calX$,
    and the inference task reduces to evaluating
    \begin{align}\label{eq:kernel_function_setup}
        f_\alpha(x) &\coloneqq \sum_{i=1}^N \alpha_i k(x,x_i) 
        = \sum_{i=1}^N \alpha_i 
        \lvert\langle0\rvert U^\dagger(x)U(x_i)\lvert0\rangle\rvert^2.
    \end{align}
    Here, $S=\{(x_i,y_i)\}_{i=1}^N$ is a training set of size $N$, where the co-variates $x_i\in\calX$ are $d$-dimensional real vectors, and the labels $y_i\in\{\pm1\}$ (in the case of binary classification) or real-valued (in the case of regression).
    For simplicity, we restrict our discussion to binary classification and regression; standard approaches beyond these have been discussed by, e.g., \citet{scholkopf2002learning}.
    The vector $\alpha = (\alpha_1, \dots, \alpha_N) \in \mathbb{R}^N$ contains the trained parameters, with each $\alpha_i$ being a weight assigned to a training point $x_i$.

    We refer to $N$ as the \emph{size of the training set} for simplicity. In practice, sparsity is often imposed over the coefficients during training (e.g., via $\ell_1$-regularization), so that $N$ only includes the non-zero coefficients, sometimes dubbed \emph{support vectors}.
    While number of support vectors can be significantly smaller than the size of the training set, we do not make any specific assumptions.
    We summarize the above in the following formal definition for the computational task of inference.
    \begin{definition}[Problem set-up]
    \label{def:problem_setup}
        Let $S=\{(x_i,y_i)\}_{i=1}^N$ be a training set of size $N\in\bbN$, with $(x_i,y_i)\in\calX\times\calY$, let $\alpha=(\alpha_i)_{i=1}^N\in\bbR^N$ be a vector of trained parameters, let $U(x)$ be an input-dependent unitary quantum-feature-map involving $G$-many gates from a fixed gate set, with $U(x)\lvert0\rangle=\lvert\phi(x)\rangle$, and let $\varepsilon>0$ be a precision parameter.
        The goal is to provide a quantum algorithm which, for any new input $x$, evaluates the kernel function $f_\alpha(x)$ in Eq.~\eqref{eq:kernel_function_setup} to additive precision $\varepsilon$ with probability at least $2/3$.
    \end{definition}
    
    \begin{definition}[Norms]\label{def:norms}
        We use the notation 
        $\lVert x\rVert_p\coloneqq \left(\sum_{i=1}^N \lvert x_i\rvert^{p}\right)^{\nicefrac{1}{p}}$
        for $p\in\{\nicefrac{2}{3}, 1, 2\}$, although it is not a norm for $0<p<1$ as it does not fulfill the triangle inequality.
        For a Hermitian operator $\calM$ with eigenvalues $\{\lambda_i\}_i$, we use $\lVert\calM\rVert_\infty=\max_i\lvert\lambda_i\rvert$ and refer to it as the operator norm.
    \end{definition}

    \begin{definition}[Short-hand notation]\label{def:short-hand}
        We separate the coefficient vector into three parts: (1) its norm $\lVert\alpha\rVert_1$, (2) a probability distribution $p=(p_i)_{i=1}^N$ with $p_i = \lvert\alpha_i\rvert/\lVert\alpha\rVert_1$, and (3) a vector of signs $s=(s_i)_{i=1}^N$, with $s_i=\operatorname{sign}(\alpha_i)$.
        With this, we can re-write Eq.~\eqref{eq:kernel_function_setup} as 
        \begin{align}
        f_\alpha(x) = \lVert\alpha\rVert_1\sum_{i=1}^N p_i s_i k(x,x_i)=\lVert\alpha\rVert_1 \bbE_{i\sim p}[s_i k(x,x_i)].
        \end{align}
        Accounting for the positive and negative terms in the sum, we introduce the intermediate variables $f_\pm\in[0,1]$ as $f_\pm\coloneqq\sum_{i=1}^N\delta_{s_i,\pm1}p_ik(x,x_i)$.
        We can re-write Eq.~\eqref{eq:kernel_function_setup} in terms of these normalized quantities as $f_\alpha(x)=\lVert\alpha\rVert_1(f_+-f_-)$.
    \end{definition}

    Below, we define the oracle access model used throughout the work and specify the associated gate costs per query.
    \begin{definition}[Access model with associated gate costs]
    \label{def:access_model}
    To tackle the goal from Definition~\ref{def:problem_setup}, we assume the following quantum oracle access model:
    \begin{itemize}[noitemsep]
        \item \emph{Single-input oracle $U(x)$:} given any input $x$ and an $n$-qubit register, we can implement $U(x)$, $U^\dagger(x)$, and controlled versions thereof, with an associate cost of $G$-many gates.
        \item \emph{Coefficient quantum-state oracle $W(\alpha)$:} Given a $\left(\lceil\log_2(N)\rceil+1\right)$-qubit register, we can implement $W(\alpha)$, which applied onto $\lvert0\rangle$ prepares a quantum state encoding the coefficients as amplitudes $\lvert\alpha\rangle=\sum_{i=1}^N \sqrt{p_i}\lvert i \rangle\otimes\lvert s_i\rangle$, with $p_i$ and $s_i$ as in Definition~\ref{def:short-hand}, and $s_i$ encoding the sign of the $i^\text{th}$ coefficient as an eigenstate of the $Z$ operator as $Z \lvert s_i\rangle=s_i\lvert s_i\rangle$.
        We can also implement the controlled version of $W(\alpha)$, as well as $W^\dagger(\alpha)$.
        The associated cost per query in terms of gates is $N$.
    \end{itemize}

    We consider the following classical oracles to access the coefficient vector $\alpha$: \emph{Reading oracle $O_R(\alpha,i)$}, read the $i^\text{th}$ element $\alpha_i$;
        \emph{Norm oracle $O_N(\alpha)$}, compute the norms $\lVert\alpha\rVert_1$, $\lVert\alpha\rVert_2$, or the pseudo-norm $\lVert\alpha\rVert_{2/3}$; and
        \emph{Sampling oracle $O_C(\alpha)$}, sample an index $i$ with probability $p_i$ (that is, proportional to the magnitude of the coefficients).
    \end{definition}

    In Section~\ref{ss:unified}, we measure algorithmic cost by the total number of \emph{quantum} oracle queries, which include the single-input oracle, the coefficient quantum-state oracle, and the training-set oracle (which we introduce below, in Definition~\ref{def:training-set-oracle}).
    Below, we provide explicit circuit constructions that confirm that the runtime of our algorithms is indeed dominated by the quantum oracle queries.
    We note that the gate cost per query varies across algorithms, yet in a way that is mostly independent of $\varepsilon$ and $\lVert\alpha\rVert_1$, and hence it does not affect the asymptotic comparisons in runtime and query complexity we perform in this work.
    We comment on resource estimates in terms of gates in Section~\ref{ss:resources}.

    The access model we propose in Definition~\ref{def:access_model} not only allows us to perform a formal analysis of query complexity, but it also accurately reflects realistic implementations.
    The oracles we introduce can be directly instantiated on hardware, and they do not require any special computational power, nor do they hide exponential overheads.
    Still, we assume \emph{black-box} access to $U(x)$, $O^\dagger(S)$, and $W(\alpha)$ to remove the possibility of optimized implementations which would exploit the structure of the circuits involved.
    This is a conscious design choice, as it is our goal to focus on the fundamental complexity of the computational task, which in turn provides worst-case guarantees applicable to any specific quantum feature map.
    As is common in query complexity studies, nothing prevents more-efficient procedures from existing under the assumption that $U(x)$ is of a particular known form.

\subsection{Amplitude estimation}\label{s:toolbox}
    The inference task in Definition~\ref{def:problem_setup} can be viewed as an instance of amplitude estimation, which we now define.
    \begin{definition}[Amplitude estimation problem]
    \label{def:amplitude_estimation}
       Let $V \in \mathbb{C}^{d \times d}$ be a unitary such that
       \begin{align}
           V \ket{0\ldots0}\ket{0} = a \ket{\psi_{\mathrm{good}}} \ket{0} + \sqrt{1 - a^2} \ket{\psi_{\mathrm{bad}}} \ket{1},
       \end{align}
       where $\ket{\psi_{\mathrm{good}}}$ and $\ket{\psi_{\mathrm{bad}}}$ are normalized quantum state vectors, and $a \in [0,1]$.
       Given access to $V$, $V^\dagger$, and controlled-$V$, the task is to output an estimate $\hat{a}$ such that $a - \varepsilon < \hat{a} < a + \varepsilon$ with probability at least $2/3$.
    \end{definition}
    Two approaches are relevant: \emph{sampling}, which estimates $a$ from repeated measurements of $V\lvert0\rangle$ with sampling complexity $\calO(\varepsilon^{-2})$, and \emph{quantum amplitude estimation} (QAE), which uses coherent applications of $V$ and $V^\dagger$ to achieve $\calO(\varepsilon^{-1})$~\citep{Brassard_2002,Aaronson_2020}. The latter scaling is optimal~\citep{Nayak1999} and requires inverse access~\citep{tang2025amplitudeamplificationestimationrequire,tang2}. We provide formal statements and further background in Appendix~\ref{a:amplitude_estimation_toolbox}.
    
\subsection{Related work}\label{s:related}
The naive approach to inference, which we call \emph{list-and-sum (fixed budget)} below, was already outlined by, e.g., \citet{havlicek2019supervised} and \citet{schuld2019quantum}, and re-instated countless times \citep{schuld2021kernels, tanner2026nonvariationalsupervisedquantumkernel}.
To the best of our knowledge, a formal runtime analysis of this approach had not been performed before; we provide it in Theorems~\ref{thm:meta} and~\ref{thm:meta-gates}.
\citet{Liu2021discretelog} established a learning separation, proving that quantum kernel methods can learn functions outside of $\Ppoly$ under standard cryptographic assumptions, making the scalability of inference practically relevant.

Beyond the black-box oracle model we consider, existing works have studied reducing inference cost via \emph{surrogates}, either classically~\citep{schreiber2022classical,landman2022classically,Sweke2023Potential,sahebi2025dequantizationsupervisedquantummachine}, exploiting random Fourier features~\citep{rahimi2007random}; or quantumly~\citep{nakayama2025explicitquantumsurrogatesquantum}, also using shadow tomography~\citep{aaronson2018shadow,haug2023quantum,king2025triply}.
Surrogates can reduce inference cost in restricted scenarios, but cannot work in the general setting we consider (Definitions~\ref{def:problem_setup} and~\ref{def:access_model}).

The trade-off between term-by-term and global estimation strategies that we exploit here has parallels in classical signal processing: in compressed sensing, coherent measurements satisfying the restricted isometry property recover $s$-sparse vectors from $\calO(s\log N)$ measurements rather than $\calO(N)$ term-by-term ones~\citep{CompressedSensingGitta}. A related trade-off arises in basis optimization for electronic structure Hamiltonians in quantum chemistry~\citep{rubin2018application,Qubitization}.
 
\section{Results}\label{s:results}

    We highlight two algorithmic choices to make:
    one regarding how the target quantity $f_{\alpha}(x) = \sum_{i=1}^N \alpha_i k(x,x_i)$ (or parts thereof) is estimated, and the other regarding how the sum is approximated.
    In particular, we consider the following possibilities:
    \begin{enumerate}[noitemsep]
        \item[Choice 1.] \textbf{How to estimate expectation values:} either 
        via \emph{sampling} (see Theorem~\ref{thm:sampling_variance_concentration}), yielding precision $\calO(1/\sqrt{M})$ from $M$ shots, or via \emph{quantum amplitude estimation} (see Theorem~\ref{thm:qae-runtime}), yielding precision $\calO(1/M)$ from $M$ coherent queries.
        \item[Choice 2.] \textbf{How to approximate the sum:} the naive option is to \emph{list-and-sum}, i.e., estimate each 
        term independently and add the results classically, either with a \emph{fixed budget} for each term or following an \emph{adaptive budget} scheme.
        Alternatively, all terms can be grouped into a single \emph{all-at-once} observable $\calM$.
    \end{enumerate}
    These algorithmic choices are illustrated schematically in Figure~\ref{fig:sketch} to provide an intuitive overview.
    A priori it could be unclear whether such an \emph{all-at-once} observable exists, or whether it can be efficiently implemented.
    We address this first in Section~\ref{ss:all-at-once}.
    A unified algorithm and a theorem describing the runtime complexities follows in Section~\ref{ss:unified}.
    For a detailed presentation and analysis of each of the ramifications of the unified algorithm, refer to Appendices~\ref{a:all-at-once} and~\ref{a:zoo}.
    We comment on resource estimates in Section~\ref{ss:resources} and end by providing a matching query-complexity lower bound in Section~\ref{ss:lower_bounds}.

\subsection{Existence and implementation of an all-at-once observable}\label{ss:all-at-once}
    Instead of estimating each kernel term individually, we reinterpret the weighted sum as a single expectation value by embedding the coefficients into a quantum state. This allows us to evaluate the entire inference expression in one coherent procedure, rather than aggregating many separate estimates.
    We show that the full inference sum $f_\alpha(x)$ can be cast as the expectation value of a single observable, rather than estimating each kernel value separately.
    Similar versions of the following results were used by, e.g., \citet{gonzalez2022learning,nakayama2025explicitquantumsurrogatesquantum,gilfuster2024understanding,gilfuster2025relation}, though not as formal statements.
    
    We start by introducing another quantum oracle that can be constructed from those in our access model.
    
    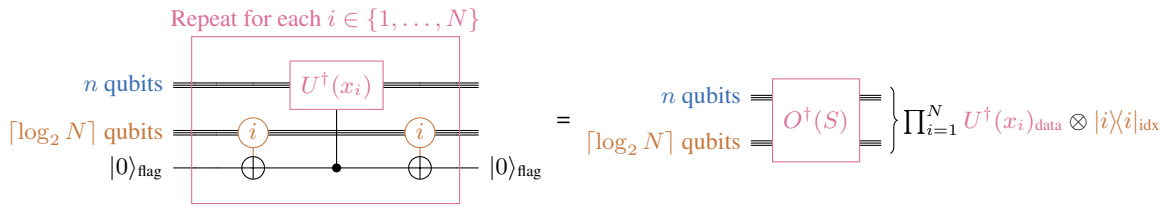
\begin{figure}[t]
        \centering
        \begin{quantikz}[thin lines, column sep = 8, row sep = 3, slice label style={inner sep=1pt,anchor=south west,rotate=40}]
            \setwiretype{q} & \qwbundle{n} & & & \gategroup[3,steps=4,style={draw={choicethree}}]{\color{choicethree}Repeat for each $i\in\{1,\ldots,N\}$} & & \gate[label style={choicethree}, style={draw={choicethree}}]{U^\dagger(x_i)} & & &  \\
        \setwiretype{q} & \qwbundle{\lceil\log_2N\rceil} & & & & \octrl[label style={choicetwo}, style={draw={choicetwo}}]{1}{\vphantom{\big(}\color{choicetwo}i} & & \octrl[label style={choicetwo}, style={draw={choicetwo}}]{1}{\vphantom{\big(}\color{choicetwo}i} && \\
        \setwiretype{q}\lstick{$\lvert0\rangle_\text{flag}$} & & & & & \targ{} & \ctrl{-2} & \targ{} && \rstick{$\lvert0\rangle_\text{flag}$}
        \end{quantikz} = 
        \begin{quantikz}[thin lines, column sep = 8, row sep = 3, slice label style={inner sep=1pt,anchor=south west,rotate=40}]
            \setwiretype{q} & \qwbundle{n} & & & & \gate[2, disable auto height, label style={choicethree}, style={draw={choicethree}}]{O^\dagger(S)} & \rstick[2]{$\sum_{i=1}^N {\color{choicethree}U^\dagger(x_i)} \otimes {\color{choicetwo}\lvert i\rangle\!\langle i\rvert}$}\\
            \setwiretype{q} & \qwbundle{\lceil\log_2N\rceil} & & & & &
        \end{quantikz}
        \caption{
            Realizing the \enquote{training-set oracle} $O^\dagger(S)$ in Definition~\ref{def:training-set-oracle}, as $N$ controlled calls to the inverse of the single-input oracle $U(x)$, one for each input in the training set $x_i\in S$.
            The associated gate cost is $\calO(N(G+\log N))$.
            This construction resembles ideas in the \emph{linear combination of unitaries}  (LCU) framework~\citep{childs2012hamiltonian}.
        }
        \label{fig:training-set-oracle}
    \end{figure}

    \begin{definition}[Training-set oracle $O^\dagger(S)$]\label{def:training-set-oracle}
        Let $S=\{(x_i,y_i)\}_{i=1}^N$ be the training set.
        Given two registers, an $n$-qubit \enquote{data} register and a $\lceil\log_2(N)\rceil$-qubit \enquote{index} register, which indexes each element $i\in\{1,\ldots,N\}$, we call \emph{training-set oracle} $O^\dagger(S)$
        \begin{align}
            O^\dagger(S) &\coloneqq \sum_{i=1}^N U^\dagger(x_i)_\mathrm{data}\otimes\lvert i\rangle\!\langle i\rvert_\mathrm{idx}\equiv \prod_{i=1}^N \left(U^\dagger(x_i)_\mathrm{data}\otimes\lvert i\rangle\!\langle i\rvert_\mathrm{idx} + \bbI_\mathrm{data}\otimes\left(\bbI - \lvert i\rangle\!\langle i\rvert\right)_\mathrm{idx}\right).
        \end{align}
    \end{definition}
    We note that this construction does not require \emph{quantum random-access memory} (QRAM). The training inputs $\{x_i\}$ are classically known and fixed after training, so the controlled unitaries can be compiled into a circuit offline. The $\calO(N)$ gate cost per query is inherent to any procedure -- classical or quantum -- that must touch all $N$ training points.
    \begin{lemma}[Query complexity training-set-oracle]\label{l:training-set-oracle}
        The training-set oracle can be realized with $N$ controlled queries to the single-input oracle $U(x_i)$.
        The associated gate cost is $\calO(N(G+\log N))$, and the procedure uses one extra qubit.
    \end{lemma}
    \begin{proof}
        See Fig.~\ref{fig:training-set-oracle} for a visual proof.
        The associated gate cost is, for each $i\in\{1,\ldots,N\}$: $\calO(\log N)$ gates to perform a multi-controlled bitflip on the flag register conditioned on the index being $i$, $\calO(G)$ to call the controlled version of the $G$ gates in $U(x_i)$, and $\calO(\log N)$ to uncompute the flag bitflip, for a total of $\calO(G+\log N)$.
    Since these must be repeated $N$ times, the total gate cost of the training-set oracle is $\calO(N(G+\log N))$.
    \end{proof}
    We consider the training-set oracle $O^\dagger(S)$ as another quantum oracle in the sense of Definition~\ref{def:access_model}.
    In Appendix~\ref{aa:discrete}, we consider the possibility of a \emph{programmable} embedding unitary $U(x)$, together with a more-efficient implementation of the training-set oracle using the programmable single-input oracle.
    Since such a programmable oracle is an extra assumption, however fair, we do not include it in the main text.

    \begin{theorem}[All-at-once observable]\label{thm:single-observable-existence}
        Consider the problem set-up from Definition~\ref{def:problem_setup}.
        There exists a bounded Hermitian observable $\calM(\alpha,S)$, with $\lVert\calM(\alpha,S)\rVert_\infty\leq\lVert\alpha\rVert_1$, and a quantum state vector $\lvert\phi(x)\rangle$, such that
        \begin{align}
            \langle\phi(x)\rvert\calM(\alpha,S)\lvert\phi(x)\rangle = f_\alpha(x).
        \end{align}
        Moreover, this expectation value can be implemented using $n+\lceil\log_2 N\rceil+2$ qubits and $\calO(N(G+\log N))$ gates.
    \end{theorem}
    \begin{proof}[Proof sketch]
        Set $\calM(\alpha,S)=\sum_{i=1}^N\alpha_i \lvert\phi(x_i)\rangle\!\langle\phi(x_i)\rvert$ and verify directly.
        We give Figure~\ref{fig:circuit_AAO_sampling} as visual proof, and we present full details and a rigorous argument in Appendix~\ref{a:all-at-once}. 
    \end{proof}

    \begin{figure}[t]
        \centering
        \begin{quantikz}[thin lines, column sep = 8, row sep = 3, slice label style={inner sep=1pt,anchor=south west,rotate=40}]
            \setwiretype{q} & \qwbundle{n}  && & & & \gategroup[3,steps=5]{$\tcalM(\alpha,S)/\lVert\alpha\rVert_1$} & &  \gate[2, disable auto height, label style={choicethree}, style={draw=choicethree}]{O^\dagger(S)} &\midstick[3]{$\sum_{i=1}^N {\color{choicetwo}\sqrt{p_i}}{\color{choicethree}U^\dagger(x_i)}{\color{choiceone}\lvert\phi(x)\rangle}{\color{choicetwo}\lvert i\rangle}{\color{choicetwo}\lvert s_i\rangle}$} & \meterD{\lvert0\rangle\!\langle0\rvert\vphantom{0}} \\
            \setwiretype{q} & \qwbundle{\lceil\log_2N\rceil} & & & & & \gate[2,disable auto height, label style={choicetwo}, style={draw=choicetwo}]{W(\alpha)} & \midstick[2]{$\sum_{i=1}^N {\color{choicetwo}\sqrt{p_i}\lvert i\rangle}{\color{choicetwo}\lvert s_i\rangle}$} & & & \\
            \setwiretype{q} & & & & & & & & & & \meterD{Z\vphantom{0}}
        \end{quantikz}
        \caption{
            Quantum circuit encoding the \emph{all-at-once} observable as a single expectation value.
            The \emph{all-at-once} observable $\tcalM(\alpha,S)$ is the collection of gates and fixed observables inside the box, up to a factor of $\lVert\alpha\rVert_1$.
            The intermediate-step formulas serve the purpose of illustrating the action of the gates onto the input state vector $\lvert\phi(x)\rangle$.
        }
        \label{fig:circuit_AAO_sampling}
    \end{figure}
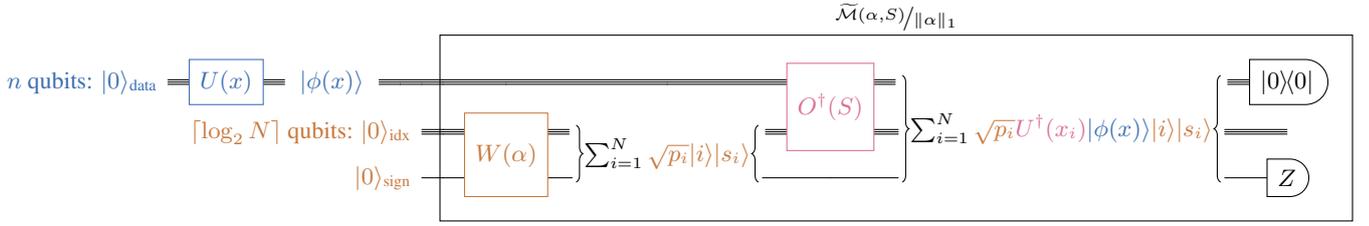

    The implementation follows the \enquote{adjoint} method (using $U(x)$ and $U^\dagger(x)$), as opposed to the SWAP test; the main difference is a trade-off of depth versus width~\citep{schuld2021kernels,hubregtsen2021training}.
    To perform quantum amplitude estimation, recall from Definition~\ref{def:amplitude_estimation} that we must provide a unitary gate $V$ such that $V\lvert0\rangle=a\lvert\psi_\text{good}\rangle\lvert0\rangle + \sqrt{1-a^2}\lvert\psi_\text{bad}\rangle\lvert1\rangle$.
    Here, $a$ is the quantity we would like to estimate, and the precise form of $\lvert\psi_\text{good,bad}\rangle$ is not important, as long as we are able to implement $V$ and $V^\dagger$ multiple times.
    Our general strategy is to relate $a$ to $f_\alpha(x)=\sum_{i=1}^N\alpha_ik(x,x_i)$.
    Following the short-hand notation from Definition~\ref{def:short-hand}, we use quantum amplitude estimation for $f_+$ and $f_-$ independently, and then we reconstruct $f_\alpha(x)$ from their difference.
    We provide a single unitary gate $V$ which allows us to estimate both.
    \begin{definition}[Amplitude-encoding unitary]\label{def:amplitude_encoding_unitary}
        Consider the problem set-up from Definition~\ref{def:problem_setup} and the notation from Definition~\ref{def:short-hand}.
        We define the \emph{amplitude-encoding unitary} $V$ via its action:
        \begin{align}\label{eq:amplitude-encoding-observable}
            V\lvert0\rangle &= \sqrt{f_+}\lvert\psi_+\rangle\lvert0,0\rangle + \sqrt{f_-}\lvert\psi_-\rangle\lvert0,1\rangle +\alpha\lvert\psi_\text{trash}\rangle\lvert1,0\rangle+\beta\lvert\psi'_\text{trash}\rangle\lvert1,1\rangle.
        \end{align}
        We have $\alpha^2+\beta^2 = 1 - f_+-f_-$.
    \end{definition}
    With this, specializing $\lvert\psi_\text{good}\rangle=\lvert\psi_\pm\rangle$ and collecting the rest under $\lvert\psi_\text{bad}\rangle$ for both $f_\pm$ becomes straightforward.
    We give a circuit depiction for how to construct $V$ in Figure~\ref{fig:circuit_AAO_QAE}, further details can be found in Appendix~\ref{a:all-at-once}.
    \begin{lemma}[Single-observable encoded in amplitude]\label{l:single-observable-amplitude}
        Under the conditions of Theorem~\ref{thm:single-observable-existence}, there exists a quantum circuit using $n+\lceil\log_2(N)\rceil+3$ qubits which implements the amplitude-encoding unitary in Definition~\ref{def:amplitude_encoding_unitary}.
    \end{lemma}
    \begin{proof}[Proof sketch]
        Refer to Figure~\ref{fig:circuit_AAO_QAE} as visual proof, full details are in Appendix~\ref{aaa:QAE}.
    \end{proof}
    
    \begin{figure}[t]
        \centering
        \begin{quantikz}[thin lines, column sep = 8, row sep = 3, slice label style={inner sep=1pt,anchor=south west,rotate=40}]
            \setwiretype{q} & \qwbundle{n} & & & & \gate[label style={choiceone}, style={draw={choiceone}}]{U(x)} & \gate[2,disable auto height, label style={choicethree}, style={draw=choicethree}]{O^\dagger(S)}\slice{$\sum_{i=1}^N {\color{choicetwo}\sqrt{p_i}}{\color{choicethree}U^\dagger(x_i)}{\color{choiceone}\lvert\phi(x)\rangle}{\color{choicetwo}\lvert i\rangle}{\color{choicetwo}\lvert s_i\rangle}\lvert{\color{choicefive}0},{\color{choicefour}0}\rangle$} & \octrl{3} & & & \rstick[5]{$\splitfrac{\sqrt{f_+}\lvert\psi_+\rangle\lvert{\color{choicefive}0},{\color{choicefour}0}\rangle + \sqrt{f_-}\lvert\psi_-\rangle\lvert{\color{choicefive}0},{\color{choicefour}1}\rangle}{ +\alpha\lvert\psi_\mathrm{trash}\rangle\lvert{\color{choicefive}1},{\color{choicefour}0}\rangle + \beta\lvert\psi_\mathrm{trash}'\rangle\lvert{\color{choicefive}1},{\color{choicefour}1}\rangle}$} \\
            \setwiretype{q} & \qwbundle{\lceil\log_2N\rceil} & & & & \gate[2,disable auto height, label style={choicetwo}, style={draw=choicetwo}]{W(\alpha)} & & & & & \\
            & & & & & & & & & \ctrl{2}  & \\
            & & & & & & & \targ{} & \gate[label style={choicefive}, style={draw=choicefive}]{X} & &  \\
            & & & & & & & & & \targ{} &
        \end{quantikz}
        \caption{
            Quantum circuit encoding the \emph{all-at-once} observable as amplitudes.
            Explicit implementation of the unitary $V$ described in Lemma~\ref{l:single-observable-amplitude}.
        }
        \label{fig:circuit_AAO_QAE}
    \end{figure}
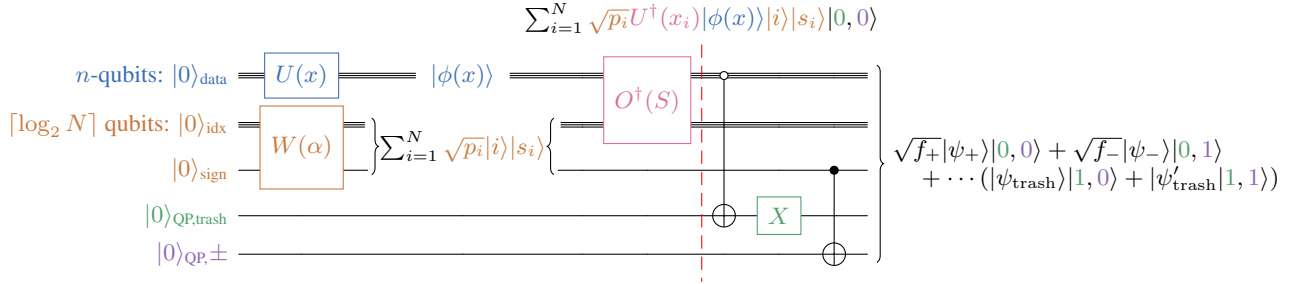

\subsection{Unified algorithm and runtime complexity}\label{ss:unified}

    With the \emph{all-at-once} observable in hand, the two algorithmic choices can be combined freely, giving rise to a landscape of algorithms (further details can be found in Appendix~\ref{a:zoo}, in particular in the unified algorithm summarized in Algorithm~\ref{alg:unified}). For each combination, we prove the resulting runtime complexity; the results are collected in terms of query complexity in Theorem~\ref{thm:meta} and in terms of gates in Theorem~\ref{thm:meta-gates}.
    
    Moreover, for completeness, we also analyze a \emph{sample-and-average} strategy in Appendix~\ref{a:sample-average}, which conceptually interpolates between \emph{list-and-sum} and \emph{all-at-once}.
    Our analysis yields the same query-complexity upper bound for sampling and quantum amplitude estimation within this template.

    As a technical comment, we note that analyzing the \emph{list-and-sum} strategy involves manipulating individual estimators for each of the terms in the sum.
    Specifically for estimators based on quantum amplitude estimation, each of the individual estimators may be biased.
    In our results, we use the nondestructive estimator of \citet{Cornelissen2023} (see also \citet{Rall_2023}), which is nearly unbiased, with a bias that can be suppressed at multiplicative-logarithmic cost, see Theorem~\ref{thm:list-sum-template} and Lemma~\ref{l:unbiased-kernel-est} for full details.
    The estimator used in the \emph{all-at-once} strategy is unaffected by this issue, since it combines two amplitude estimates through their tail bounds rather than by averaging.

    The displayed list-and-sum bounds apply in the regime where the continuous total query budget is at least $N$; per-term allocations are rounded up to at least one query. For the gate-complexity bounds, we assume $n\in\calO(G)$, so the $\calO(n)$ reflection overhead is absorbed into $\calO(G)$.

\begin{theorem}[Query complexity of inference in quantum kernel methods]\label{thm:meta}
    There exist algorithms for estimating the quantum kernel model prediction in Eq.~\eqref{eq:kernel_function_setup} to additive precision $\varepsilon$ with constant success probability (see Definition~\ref{def:problem_setup}).
    The algorithmic runtime depends on the algorithmic choices as follows:
    \begin{center}
    \renewcommand{\arraystretch}{1.0}
    \begin{tabular}{lcc}
        \toprule
        \textbf{Strategy} & \textbf{Sampling} & \textbf{Quantum amplitude estimation} \\
        \midrule
        1(a): list-and-sum, fixed & $\calO\!\left(\tfrac{N\lVert\alpha\rVert_2^2}{\varepsilon^2}\right)^\naive$ {\small[Cor.~\ref{cor:list-sum-fixed-sampling}]} & $\tcalO\!\left(\tfrac{N\lVert\alpha\rVert_2}{\varepsilon}\right)$ {\small[Cor.~\ref{cor:list-sum-fixed-qae}]} \\[4pt]
        1(b): list-and-sum, adaptive & $\calO\!\left(\tfrac{\lVert\alpha\rVert_1^2}{\varepsilon^2}\right)$ {\small[Cor.~\ref{cor:list-sum-adapt-sampling}]} & $\tcalO\!\left(\tfrac{\lVert\alpha\rVert_{\nicefrac{2}{3}}}{\varepsilon}\right)^{\diamond}$ {\small[Cor.~\ref{cor:list-sum-adapt-qae}]} \\[4pt]
        1(c): all-at-once & $\calO\!\left(\tfrac{\lVert\alpha\rVert_1^2}{\varepsilon^2}\right)$ {\small[Thm.~\ref{thm:all-at-once-sampling}]} & $\calO\!\left(\tfrac{\lVert\alpha\rVert_1}{\varepsilon}\right)^{\optimal}$ {\small[Thm.~\ref{thm:AAOQ}]} \\
        \bottomrule
    \end{tabular}
    \end{center}
    We highlight $\naive$ as the naive approach (1(a), sampling), $\optimal$ as query-optimal (1(c), quantum amplitude estimation), and $\diamond$ as gate-optimal (1(b), quantum amplitude estimation; see Theorem~\ref{thm:meta-gates}).
\end{theorem}
\begin{proof}
    Each case follows from the corresponding statement in square brackets; the full proofs are given in Appendix~\ref{a:zoo}.
\end{proof}

\subsection{Resource estimates}\label{ss:resources}

    In terms of runtime complexity, Theorem~\ref{thm:meta} dictates that the best algorithmic choice depends on the available hardware.
    On near-term devices without coherent multi-query access, using an adaptive budget for \emph{list-and-sum} already yields the same runtime scaling as measuring the more-costly \emph{all-at-once} observable.
    Conversely, the \emph{all-at-once} approach is more beneficial in the early-fault-tolerant regime, where the runtime scaling is directly manifested by the $1$-norm of the vector of coefficients, versus the $\nicefrac{2}{3}$-norm achieved by \emph{list-and-sum} with an adaptive budget.
    Explicitly, within any case for Choice 1, quantum amplitude estimation yields a quadratic improvement over sampling.
    We expand beyond query-complexity analysis in Theorem~\ref{thm:meta-gates}, where we show the asymptotic scaling of the number of gates associated to each choice.

    \begin{theorem}[Number of gates for inference in quantum kernel methods]\label{thm:meta-gates}
        There exist algorithms for estimating the quantum kernel model prediction in Eq.~\eqref{eq:kernel_function_setup} to additive precision $\varepsilon$ with constant success probability (see Definition~\ref{def:problem_setup}).
        The number of gates depends on the algorithmic choices as follows:
        \begin{center}
        \renewcommand{\arraystretch}{1.0}
        \begin{tabular}{lcc}
            \toprule
            \textbf{Strategy} & \textbf{Sampling} & \textbf{Quantum amplitude estimation} \\
            \midrule
            1(a): list-and-sum, fixed & $\calO\!\left(\tfrac{NG\lVert\alpha\rVert_2^2}{\varepsilon^2}\right)^\naive$ & $\tcalO\!\left(\tfrac{NG\lVert\alpha\rVert_2}{\varepsilon}\right)$ \\[4pt]
            1(b): list-and-sum, adaptive & $\calO\!\left(\tfrac{G\lVert\alpha\rVert_1^2}{\varepsilon^2}\right)$ & $\tcalO\!\left(\tfrac{G\lVert\alpha\rVert_{\nicefrac{2}{3}}}{\varepsilon}\right)^{\diamond}$ \\[4pt]
            1(c): all-at-once & $\tcalO\!\left(\tfrac{NG\lVert\alpha\rVert_1^2}{\varepsilon^2}\right)$ & $\tcalO\!\left(\tfrac{(NG+n)\lVert\alpha\rVert_1}{\varepsilon}\right)^{\optimal}$ \\
            \bottomrule
        \end{tabular}
        \end{center}
        We highlight $\naive$ as the naive approach (1(a), sampling), $\optimal$ as query-optimal (1(c), quantum amplitude estimation; see Theorem~\ref{thm:meta}), and $\diamond$ as gate-optimal (1(b), quantum amplitude estimation).
    \end{theorem}
    
    \begin{proof}
        We import the query complexity runtime results from Theorem~\ref{thm:meta} and then we account for the differing gate costs of each oracle query.
        Our access model in Definition~\ref{def:access_model} specifies that the single-input oracle $U(x)$ requires $G$ gates, and the coefficient quantum-state oracle $W(\alpha)$ requires $N$ gates.
        Further, we have shown how to construct the training-set oracle $O^\dagger(S)$ using a number of gates in $\calO(N(G+\log N))$.
        Since the \emph{list-and-sum} algorithms only use the single-input oracle, once for $x$ and once for $x_i$, their gate cost is $2G$ times the number of queries.
        Via sampling, this results in $\calO(\varepsilon^{-2}NG\lVert\alpha\rVert_2^2)$ gates with a fixed budget, and $\calO(\varepsilon^{-2}G\lVert\alpha\rVert_1^2)$ with an adaptive budget.
        Via quantum amplitude estimation, the scaling in the number of gates is, in turn, $\tcalO(\varepsilon^{-1}NG\lVert\alpha\rVert_2)$ with a fixed budget and $\tcalO(\varepsilon^{-1}G\lVert\alpha\rVert_{\nicefrac{2}{3}})$ with an adaptive budget.
    
        In the \emph{all-at-once} approaches, the circuits used for sampling and quantum amplitude estimation are not identical, as seen when comparing Figures~\ref{fig:circuit_AAO_sampling} and~\ref{fig:circuit_AAO_QAE}.
        Nevertheless, both circuits make the same number of oracle queries, visually: they contain the same number and types of boxes.
        Namely, each measurement of $\tcalM(\alpha,S)$ and each query to the unitary $V$ both require one call to the single-input oracle $U(x)$, one call to the quantum-state oracle $W(\alpha)$, and one call to the training-set oracle $O^\dagger(S)$.
        Implementing the unitary $V$ requires one multi-controlled $X$ gate, which has an associated cost of $\calO(n)$ gates.
        For sampling, the gate costs per measurement add up to $G+N+\calO(N(G+\log N))$, which scales as $\calO(N(G+\log N))$.
        For quantum amplitude estimation, the extra multi-controlled gate results in a scaling of $\calO(N(G+\log N) + n)$.
        The total number of gates used in the \emph{all-at-once} algorithms is, then, $\tcalO(\varepsilon^{-2}NG\lVert\alpha\rVert_1^2)$ via sampling and $\tcalO(\varepsilon^{-1}(NG+n)\lVert\alpha\rVert_1)$ via quantum amplitude estimation, where we have removed the logarithmic factors for simplicity.
    \end{proof}
    
    \begin{remark}[Query-optimality versus gate-optimality.]
        When comparing the \emph{naive} and \emph{query-optimal} approaches in Theorem~\ref{thm:meta}, there is an $N$-factor speed-up in the query complexity, on top of the quadratic improvement over the norm of $\alpha$ and the precision $\varepsilon$.
        In terms of gates, as stated in Theorem~\ref{thm:meta-gates}, the same is no longer true: both \emph{list-and-sum} with fixed budget and \emph{all-at-once} have an $N$ factor in their gate cost scaling.
        From the perspective of number of gates, it is not immediately clear anymore that \emph{all-at-once} is better than \emph{list-and-sum} with adaptive budget.
        In the case of sampling, \emph{list-and-sum} with adaptive budget now has an $N$-factor \emph{improvement} over \emph{all-at-once}.
        In the case of quantum amplitude estimation, \emph{all-at-once} has both an $N$ and an $n$ factor. Below, we make a principled comparison between the runtime scalings.
        Let us define $\tG_L\coloneqq\varepsilon^{-1}G\lVert\alpha\rVert_{\nicefrac{2}{3}}$ and $G_A\coloneqq\varepsilon^{-1}(NG+n)\lVert\alpha\rVert_1$ the respective scaling of the numbers of gates.
        The subscript $L$ stands for \enquote{list-and-sum}, and the subscript $A$ stands for \enquote{all-at-once}, and the notation $\tG_L$ indicates that we drop logarithmic factors.
        \begin{align}
            \begin{rcases}
                \frac{G_L}{G_A} = \frac{\varepsilon^{-1}G\lVert\alpha\rVert_{\nicefrac{2}{3}}}{\varepsilon^{-1}(NG+n)\lVert\alpha\rVert_1} = \frac{\lVert\alpha\rVert_{\nicefrac{2}{3}}}{(N+\frac{n}{G})\lVert\alpha\rVert_1}\\
                \lVert\alpha\rVert_1\leq\lVert\alpha\rVert_{\nicefrac{2}{3}}\leq\sqrt{N}\lVert\alpha\rVert_1
            \end{rcases} & \Rightarrow \left(\sqrt{N}+\frac{n}{G\sqrt{N}}\right) G_L \leq G_A\leq \left(N+\frac{n}{G}\right) G_L.
        \end{align}
        Thus, after suppressing logarithmic factors, the displayed gate-count upper bound for \emph{all-at-once} is no smaller than that for \emph{list-and-sum} with adaptive budget.
        See Appendix~\ref{aa:discrete} for a refined discussion using a programmable quantum embedding $U(x)$.
    \end{remark}
    \begin{remark}[Practical guidelines.]
        As a guide to the eye for the practitioner, our results suggest the following guidelines for choosing an inference strategy:
        \begin{itemize}[noitemsep]
            \item Use \emph{all-at-once} combined with quantum amplitude estimation when coherent multi-query access is available, as this achieves query-optimal scaling.
            \item Use \emph{list-and-sum with adaptive budget} combined with quantum amplitude estimation when gate cost dominates, as it has the best leading gate-count bound among the strategies analyzed here.
            \item Sampling-based approaches are primarily competitive in near-term regimes without coherent access, but are asymptotically suboptimal in precision.
        \end{itemize}
    \end{remark}

\subsection{Matching lower bound}\label{ss:lower_bounds}
    The query complexity of the \emph{all-at-once} via quantum amplitude estimation algorithm (cf.\  Algorithm~\ref{alg:AAOQ} and Theorem~\ref{thm:AAOQ}) cannot be improved in its dependence on $\lVert\alpha\rVert_1$ and $\varepsilon$.
    
    In Appendix~\ref{app:lower_bounds} we construct a family of oracle instances, with a single training point, on which any quantum algorithm that estimates $f_\alpha(x)$ to additive precision $\varepsilon$ with constant success probability must make
    \begin{align}
        \Omega\!\left(\frac{\lVert\alpha\rVert_1}{\varepsilon}\right)
    \end{align}
    queries to the single-input oracle, counting inverses, controlled versions, and the training-set oracle alike (Theorem~\ref{thm:lower-bound-full}).
    The bound constrains every algorithm equipped with the oracles of Definition~\ref{def:access_model}, not only those routing their computation through the amplitude-encoding unitary $V$: the hardness sits inside the feature-map oracle, leaving all other oracles instance-independent and hence free of charge, and the proof reduces to the two-point approximate-counting bound of \citet{Nayak1999}, with each oracle call simulated at unit cost by bit-oracle queries via phase kickback.
    Since each application of $V$ or $V^\dagger$ makes two such calls, any routine accessing the instance only through $V$, $V^\dagger$, and controlled versions thereof -- in particular any quantum-amplitude-estimation routine run on $V$ -- must apply them $\Omega(\lVert\alpha\rVert_1/\varepsilon)$ times (Corollary~\ref{cor:AAOQ-optimal}).
    Together with the matching upper bound of Theorem~\ref{thm:AAOQ}, Algorithm~\ref{alg:AAOQ} is therefore query-optimal in these two parameters, up to constant factors.

    Some caveats delimit the claim.
    The bound holds for $64\leq\lVert\alpha\rVert_1/\varepsilon\leq2^{n-1}$, and neither end restricts the conclusion: below it the task is solved with $\calO(1)$ queries, while above it the number of feature qubits $n$ is a free parameter of the construction, so hard instances exist at every ratio.
    Since the hard family has $N=1$, where all norms of $\alpha$ coincide, we do not claim optimality with respect to a finer parameterization of $\alpha$.
    Quantum amplitude estimation is likewise already optimal for each individual kernel term -- estimating $k(x,x_i)$ to additive error $\varepsilon_i$ requires $\Omega(\varepsilon_i^{-1})$ queries whenever $2^{1-n}\leq\varepsilon_i\leq1/64$ -- so any improvement over \emph{list-and-sum} must change how the sum is approximated, not how its terms are estimated.

\section{Discussion}\label{s:discussion}

    In this work, we have scrutinized and revisited evaluation methods for inference in quantum kernel methods, which are among the top candidates to achieve quantum advantage in machine learning.
    Our results have practical implications for the long-term viability of quantum kernels versus other approaches in \emph{quantum machine learning} (QML).
    The cost of inference in kernel methods has been cited as a critical flaw that could prevent scalability~\citep{nakayama2025explicitquantumsurrogatesquantum}.
    Our results show that the dependence on the training set size can be confined to the per-query gate cost, so that the number of oracle queries -- and hence the scaling with precision -- becomes independent of $N$.
    We show that the query-complexity scaling of the naive approach $\calO\!\left(N\lVert\alpha\rVert_2^2/\varepsilon^2\right)$ can be improved down to $\calO\!\left(\lVert\alpha\rVert_1/ \varepsilon\right)$.
    By providing matching lower bounds, we showed that this improved scaling is optimal within our access model.
    Importantly, our lower bound concerns query complexity only; no matching bound accompanies the gate-count statements, and an inference algorithm with gate cost sublinear in $N$ is not ruled out by our bound.

    In the regime of exponential kernel concentration~\citep{thanasilp2024exponential}, where $k(x,x')\approx\delta_{x,x'}$, our upper bounds remain valid but the model itself loses predictive power. Our results are therefore most relevant for kernels whose values do not vanish asymptotically~\citep{Liu2021discretelog, henderson2025quantumadvantageexponentialconcentration}, which are precisely the settings where quantum kernel methods yield \emph{quantum advantage}. Extending our construction to non-unitary feature maps (mixed states $\rho(x)$) remains an open question.

    Our results confirm that using quantum amplitude estimation for the \emph{all-at-once} observable is optimal in terms of query complexity.
    Taking into account hardware requirements for both noisy and early-fault-tolerant devices, it may well be that this approach is beyond budget.
    Accordingly, we provide a zoo of algorithms that nicely interpolate between the \emph{naive} and \emph{optimal} ones, and formally analyze their runtimes.
    As such, this work aims to constitute a \emph{useful handbook for practitioners}.
    Indeed, we hope that our analysis clarifies some of the critical design choices involved in realistic implementations for the inference step in quantum kernel methods.

    Finally, an exciting open question is whether our improvements for inference can be extended to speed-up the \emph{training} phase of quantum kernels.
    The usual pipeline in training kernel methods involves inverting an $N\times N$ matrix, and the end-to-end complexity is usually characterized as $\calO(N^3)$.
    Anticipating potential paths to reduce the cost, one could take Eq.~\eqref{eq:kernel_function_setup} without assuming that the parameters $\alpha$ have already been trained, and instead define a loss function with respect to those parameters.
    Then, one option would be to perform (stochastic) gradient descent~\citep{shalev2014understanding} on those parameters.
    Gradient-based optimization of kernel methods has been considered also in classical machine learning, mostly as a proof technique, and it comes with performance guarantees owing to the fact that usual regularized loss functions are convex with respect to $\alpha$~\citep{scholkopf2002learning}.
    The representer theorem~\citep{schuld2021kernels,scholkopf2002learning} further states that the optimal choice of $\alpha$ actually corresponds to the best possible linear function in terms of fitting the training set.
    Beyond straightforward gradient-based optimization, one could consider more advanced forms of quantum optimization~\citep{leng2023quantumhamiltoniandescent,catli2025exponentiallybetterboundsquantum}, which in turn enjoy favorable performance guarantees for convex loss functions.

\subsubsection*{Note added}
    During the final stages of publication, new results by \citet{ambainis2026tight} provide unbiased estimators for quantum amplitude estimation.
    With these, our analysis of \emph{list-and-sum} strategies can be simplified.

\begin{ack}
    This work has been inspired by conversations with Tom O'Brien of Google Quantum AI.
    The authors thank Armando Bellante, Carlos Bravo-Prieto, Vedran Dunjko, Tom O'Brien, Louis Schatzki and Franz Schreiber for useful references and comments in an earlier version of this manuscript.
    E.~G.-F. is a recipient of the 2023 Google PhD Fellowship.
    Moreover, the authors acknowledge support by the BMFTR (PraktiQOM, QuSol, HYBRID++), the Quantum Flagship (Millenion, PasQuans2), 
    the DFG (SPP 2514, CRC 183), the Munich Quantum Valley, Berlin Quantum, the Clusters of Excellence (MATH+, ML4Q), and the European Research Council (DebuQC).

    The authors declare no other competing interests.
\end{ack}

\bibliographystyle{plainnat}
\bibliography{sources}

\clearpage
\appendix

\section*{Supplementary Material}
\addcontentsline{toc}{section}{Supplementary Material}

\numberwithin{theorem}{section}
\numberwithin{algorithm}{section}
\numberwithin{figure}{section}
\numberwithin{table}{section}
\numberwithin{definition}{section}
\numberwithin{equation}{section}

\section{Amplitude estimation toolbox}\label{a:amplitude_estimation_toolbox}
    In this appendix, we collect standard results on amplitude estimation that are used throughout the paper. We present two approaches to the problem defined in Definition~\ref{def:amplitude_estimation}: a \emph{sampling-based method} achieving a scaling of $\calO(\varepsilon^{-2})$, and \emph{quantum amplitude estimation} achieving $\calO(\varepsilon^{-1})$. We also state the matching lower bound and the necessity of inverse access. Our presentation closely follows \citet{tang2025amplitudeamplificationestimationrequire}.

\subsection{Sampling-based amplitude estimation}
    
    In the classical setting, the amplitude $a$ is not directly accessible; one can only sample from the distribution induced by measuring $V\ket{0}$. Each measurement of the auxiliary register yields outcome $\ket{0}$ with probability $p = a^2$, so estimating $a$ reduces to estimating the Bernoulli parameter $p$ and taking a square root.
    
    \begin{theorem}[Sampling estimator: variance and concentration]
    \label{thm:sampling_variance_concentration}
        Let $X \in [0,1]$ and let $Y_1,\dots,Y_M$ be i.i.d.\ Bernoulli random variables with $\mathbb{E}[Y_k] = X$.
        Define the estimator $\hX := \sum_{k=1}^M Y_k/M$.
        Then the mean of the estimator fulfills $\mathbb{E}[\hat X] = X$ and its variance $\mathrm{Var}(\hat X) \leq 1/4M$.
        Further, for all $\varepsilon > 0$, the following tail bound holds $\bbP(\lvert\hX - X\rvert \geq \varepsilon) \leq 2 \ \exp(-2 M \varepsilon^2)$.
        In particular, to guarantee $\bbP(\lvert\hX - X\rvert \leq \varepsilon) \geq 1-\delta$, it suffices to choose $M \in \calO(\varepsilon^{-2}\log(\delta^{-1}))$.
    \end{theorem}

    The proof is via standard Chernoff-based arguments, together with the facts that bounded random variables are sub-Gaussian, and that the observations $Y_1,\ldots,Y_M$ are indeed independent.

    \begin{corollary}[Estimation of amplitudes]
    Let $a \in [0,1]$ and define $X = a^2$. Let $\hat X$ be as in Theorem~\ref{thm:sampling_variance_concentration}, and define $\ha := \sqrt{\hX}$.
    Then for $a \in \Omega(\varepsilon)$, the estimator $\hat a$ satisfies $\lvert\ha - a\rvert \leq \varepsilon$
    with probability at least $1-\delta$ using, $M \in \calO(\varepsilon^{-2}\log(\delta^{-1}))$ samples.
    \end{corollary}

\subsection{Quantum amplitude estimation}
    
    Quantum amplitude amplification and quantum amplitude estimation, introduced by \citet{Brassard_2002}, generalize the quadratic speedup achieved by Grover's search algorithm~\citep{Grover1996}. These techniques serve as fundamental building blocks in quantum algorithm design to speed-up certain subroutines. 

    \begin{theorem}[Quantum amplitude estimation~\citep{Brassard_2002,Aaronson_2020}]\label{thm:qae-runtime}
    Let $a\in[0,1]$ and $V$ an $n_V$-qubit unitary as in Definition~\ref{def:amplitude_estimation}.
    Given access to $V$ and $V^\dagger$, there exists an estimator $\ha$ such that for all $\varepsilon>0$ and $\delta\in(0,1)$, $\bbP(\lvert\ha-a\rvert\geq\varepsilon)\leq\delta$, using $M\in\calO(\varepsilon^{-1}\log(\delta^{-1}))$ applications of $V$ and $V^\dagger$.
    More precisely, there exists a constant $C>0$ such that $\bbP(\lvert\ha-a\rvert\geq\varepsilon)\leq 2\ \exp(-CM\varepsilon)$ for all $a\in\Omega(\varepsilon)$.
    The procedure from \citet{Aaronson_2020} uses $n_V+1$ qubits, and the total number of gates is $\calO(M(G_V+n_V))$, where $G_V$ is the number of gates necessary to implement $V$ or $V^\dagger$.
    \end{theorem}

    \begin{remark}
        The concentration bound in Theorem~\ref{thm:qae-runtime} implies that the estimator behaves as if it had variance of order $1/M^2$, in contrast to the $1/M$ variance of sampling-based amplitude estimation from Theorem~\ref{thm:sampling_variance_concentration}.
    \end{remark}
    
    \citet{Aaronson_2020} provide constructions for quantum amplitude estimation without using quantum phase estimation or controlled applications of $V$ (which was required by \citet{Brassard_2002}). 
    Their analysis assumes $a = \Omega(\varepsilon)$, ensuring the estimation problem is non-trivial: in this regime, additive error $\varepsilon$ corresponds to constant relative error.
    \citet{tang2025amplitudeamplificationestimationrequire} argue that related techniques can be extended to regimes where $a$ is comparable to or smaller than $\varepsilon$.
    Finally, the query complexity of Theorem~\ref{thm:qae-runtime} is optimal up 
    to constant factors.
    \begin{theorem}[Optimality of quantum amplitude estimation~{\citep{Nayak1999}}]
    \label{thm:qae-lower-bound}
        Any quantum algorithm that estimates $a$ to additive error $\varepsilon$ with constant success probability requires $\Omega(\varepsilon^{-1})$ applications of $V$ and $V^\dagger$.
    \end{theorem}
    Theorem~\ref{thm:qae-lower-bound} follows from lower bounds for quantum counting~\citep{Nayak1999}, which reduce to phase estimation lower bounds.
    \citet{lin2023notequantumphaseestimation} and \citet{mande2023} provide refinements and extensions to arbitrary success probabilities. In particular, the dependence on the failure probability $\delta$ is also tight~\citep{mande2023}.
    \citet{tang2025amplitudeamplificationestimationrequire} and \citet{tang2} recently proved that the quadratic speedup for quantum amplitude estimation requires access to the inverses.
    
    \begin{theorem}[Amplitude estimation requires inverse access~\citep{tang2025amplitudeamplificationestimationrequire}]
        Suppose we have only access to the unitary $V  \in \mathbb{C}^{d\times d}$ from above and no access to the inverse, then any algorithm for amplitude estimation must use $\Omega(\min(d,\varepsilon^{-2}))$ applications of $V$.
    \end{theorem}

Finally, we record a variant of amplitude estimation that controls not only the tails but also the bias of the estimate, suppressing the latter at logarithmic cost.
We use this variant in our analysis of the list-and-sample algorithms, in Section~\ref{ss:unified} and Appendix~\ref{aa:list-and-sum}.
\begin{theorem}[Nearly unbiased amplitude estimation~{\citep{Cornelissen2023}}]\label{thm:ndub-qae}
    Let $\lvert\psi\rangle$ be a quantum state and $\Pi$ a projector, and write $p=\lVert\Pi\lvert\psi\rangle\rVert^2$.
    For every $t\geq4$ and $b\in(0,1)$ there is a quantum algorithm that outputs an estimate $\hat{p}\in[-2\pi,2\pi]$ satisfying
    \begin{align}
        \bigl\lvert\bbE[\hat{p}]-p\bigr\rvert \leq b
        \qquad\text{and}\qquad
        \mathrm{Var}[\hat{p}] \leq \frac{91\,p}{t^{2}}+b,
    \end{align}
    using $\calO\!\left(t\log\log(t)\log(\tfrac{t}{b})\right)$ applications, in expectation, of the reflections $\bbI-2\lvert\psi\rangle\!\langle\psi\rvert$ and $\bbI-2\Pi$ and controlled versions thereof; it consumes a single copy of $\lvert\psi\rangle$, which is restored at the end with probability at least $1-b$.
\end{theorem}

\section{The all-at-once observable}
\label{a:all-at-once}

This appendix provides the formal results behind the \emph{all-at-once} approach.
We first establish the proofs that the inference sum $f_\alpha(x) = \sum_{i=1}^N \alpha_i k(x,x_i)$ admits an exact representation as a single expectation value, and give an explicit circuit implementation with gate cost $\calO(N(G+\log N))$.
We then show that, if the feature-map unitary $U(x)$ admits a programmable implementation controlled by an $s$-bit description of the input, this cost can be reduced to $\tcalO((N+G)s)$.

\subsection{Existence and implementation} \label{aa:existence}

    \begin{customthm}{\ref{thm:single-observable-existence}}[Single-observable representation]
        Consider the problem set-up from Definition~\ref{def:problem_setup}, in particular the task of estimating the kernel function in Eq.~\eqref{eq:kernel_function_setup}.
        Then, there exists a bounded Hermitian observable $\calM(\alpha,S)$, with $\lVert\calM(\alpha,S)\rVert_\infty\leq\lVert\alpha\rVert_1$, and 
        a quantum state $\lvert\phi(x)\rangle$, such that
        \begin{align}           \bbE_{\lvert\phi(x)\rangle}\left[\calM(\alpha,S)\right] &= \langle\phi(x)\rvert\calM(\alpha,S)\lvert\phi(x)\rangle = \sum_{i=1}^N\alpha_ik(x,x_i) \eqcolon f_\alpha(x).
        \end{align}
    \end{customthm}
    \begin{proof}
        A simple calculation confirms that it is sufficient to set $\calM(\alpha,S)=\sum_{i=1}^N\alpha_i \lvert\phi(x_i)\rangle\!\langle\phi(x_i)\rvert$, since
        \begin{align}
            \langle\phi(x)\rvert\calM(\alpha,S)\rvert\phi(x)\rangle
            &= \langle\phi(x)\rvert\left(\sum_{i=1}^N\alpha_i\lvert\phi(x_i)\rangle\!\langle\phi(x_i)\rvert\right)\lvert\phi(x)\rangle \\
            &= \sum_{i=1}^N \alpha_i \langle\phi(x)\vert\phi(x_i)\rangle\!\langle\phi(x_i)\vert\phi(x)\rangle \nonumber \\
            &= \sum_{i=1}^N\alpha_i k(x,x_i)
            = f_\alpha(x).
            \nonumber
        \end{align}
        The operator norm of the observable 
        \begin{align}
            \lVert\calM(\alpha,S)\rVert_\infty &= \left\lVert\sum_{i=1}^N\alpha_i\lvert\phi(x_i)\rangle\!\langle\phi(x_i)\rvert\right\rVert_\infty
            \leq\sum_{i=1}^N \left\lVert\alpha_i\lvert\phi(x_i)\rangle\!\langle\phi(x_i)\rvert\right\rVert_\infty
            = \sum_{i=1}^N \lvert\alpha_i\rvert 
            = \lVert\alpha\rVert_1
        \end{align}
        results from the triangle inequality.
    \end{proof}

    Theorem~\ref{thm:single-observable-existence} confirms that, in principle, the inference step in kernel methods, i.e., estimating $f_{\alpha}(x)$ for a data input $x$, can be achieved via a single expectation value, instead of taking a different expectation value for each element in the training set and then summing the results, each weighted by the corresponding $\alpha_i$.
    As presented, it may be unclear how to estimate the expectation value of the given observable with respect to the quantum state on a quantum device.
    A direct approach may attempt to construct a unitary gate that maps the $i^\text{th}$ computational basis element $\lvert i\rangle$ into the quantum feature vector $\lvert\phi(x_i)\rangle$, for $i\in\{1,\ldots,N\}$.
    Such a transformation may be costly to obtain just from having access to the state-preparation unitary gate $U(x)$.
    We next present Algorithm~\ref{alg:single-observable-sampling}, which can be used to implement the \emph{all-at-once} observable using a few auxiliary qubits.
    We spell out each oracle call separately to reflect the sequential circuit structure.
    Note that Algorithm~\ref{alg:single-observable-sampling} follows the \enquote{adjoint} method to estimate the kernel function (using $U(x)$ and its inverse $U^\dagger(x)$), as opposed to the SWAP test.
    This is an arbitrary choice, and the main difference between the two approaches is a trade-off of depth versus width~\citep{hubregtsen2021training, schuld2021kernels}.
    In the adjoint method, the depth of the circuit is at least twice the depth of $U(x)$.
    In the SWAP method, the number of qubits in the circuit is at least twice the number of qubits $U(x)$ acts on.

    \begin{algorithm}[t]
        \caption{Single-observable sampling.}
        \label{alg:single-observable-sampling}
        \begin{algorithmic}[1]
            \Require $x \in \calX$ \Comment New input.
            \Ensure $f_\alpha(x)=\sum_{i=1}^N\alpha_ik(x,x_i)$ \Comment Inference, Eq.~\eqref{eq:kernel_function_setup}.
            \State Initialize $\lvert0\rangle_\mathrm{data}\lvert0\rangle_\mathrm{idx}\lvert0\rangle_\mathrm{sign}$. \Comment $n+\lceil\log_2(N)\rceil+1$ qubits.
            \State Single-input oracle $U(x)_\mathrm{data}$: $\lvert\tphi(x)\rangle=\lvert\phi(x)\rangle_\mathrm{data}\lvert0\rangle_\mathrm{idx}\lvert0\rangle_\mathrm{sign}$ \Comment $G$ gates.
            \State Quantum-state coefficient oracle $W(\alpha)_\mathrm{idx,sign}$:  \Comment $N$ gates. 
            \begin{align} \lvert\phi(x)\rangle_\mathrm{data}\lvert0\rangle_\mathrm{idx}\lvert0\rangle_\mathrm{sign}\mapsto \sum_{i=1}^N \sqrt{p_i}\lvert\phi(x)\rangle_\mathrm{data}\lvert i\rangle_\mathrm{idx}\lvert s_i\rangle_\mathrm{sign} \nonumber \end{align}
            \State Training-set oracle $O^\dagger(S)$: \Comment $\calO(N (G +\log N))$ gates.
            \begin{align} \sum_{i=1}^N \sqrt{p_i}\lvert\phi(x)\rangle_\mathrm{data}\lvert i\rangle_\mathrm{idx}\lvert s_i\rangle_\mathrm{sign}\mapsto \sum_{i=1}^N \sqrt{p_i} U^\dagger(x_i)\lvert\phi(x)\rangle_\mathrm{data}\lvert i\rangle_\mathrm{idx}\lvert s_i\rangle_\mathrm{sign} \nonumber \end{align} 
            \State Norm coefficient oracle $O_N(\alpha)$: $\lVert\alpha\rVert_1$.
            \State\Return Expectation value of observable: $\lVert\alpha\rVert_1\lvert0\rangle\!\langle0\rvert_\mathrm{data}\otimes\bbI_\mathrm{idx}\otimes Z_\mathrm{sign}$  \Comment Total number of gates $\calO(N ( G + \log N))$
        \end{algorithmic}
    \end{algorithm}

    \begin{lemma}[Single-observable implementation]\label{l:single-observable}
        Under the conditions of Theorem~\ref{thm:single-observable-existence}, there exists a quantum circuit with $n+\lceil\log_2(N)\rceil+2$ qubits and $\calO(N(G + \log N))$ gates which implements $\calM(\alpha,S)$, as specified in Algorithm~\ref{alg:single-observable-sampling}.
        The algorithm makes one call to each of the following oracles: single-input oracle, training-set oracle, norm coefficient oracle, and quantum-state coefficient oracle.
        This algorithm involves measuring the expectation value of an auxiliary observable $\tcalM(\alpha,S)$ fulfilling $\lVert\tcalM(\alpha,S)\rVert_\infty = \lVert\alpha\rVert_1$.
    \end{lemma}
    \begin{proof}
        We prove the statement directly by providing a construction.
        Consider three registers: an $n$-qubit \enquote{data} register $\calH_\mathrm{data}$ and an $N$-dimensional \enquote{index} register $\calH_\mathrm{idx}$ (equivalently, a $\lceil\log_2(N)\rceil$-qubit register), and a $1$-qubit \enquote{sign} register.
        Recall the notation $\alpha_i=\lVert\alpha\rVert_1s_ip_i$, where $s_i=\sign{\alpha_i}$ and $p_i = \lvert\alpha_i\rvert/\lVert\alpha\rVert_1$. 
        Calling the single-input oracle on the data register, we obtain the quantum state vector
        \begin{align} U(x)\lvert0\rangle_\mathrm{data}\lvert0\rangle_\mathrm{idx,sign}\mapsto\lvert\phi(x)\rangle_\mathrm{data}\lvert0\rangle_\mathrm{idx,sign}.
        \end{align}
        We refer to this state vector as $\lvert\tphi(x)\rangle_\mathrm{data,idx,sign}$.
        We next introduce an observable $\tcalM(\alpha,S)$ which fulfills $f_\alpha(x)=\langle\tphi(x)\rvert\tcalM(\alpha,S)\lvert\tphi(x)\rangle$.
        It then follows that 
        \begin{align}     \langle\tphi(x)\rvert\tcalM(\alpha,S)\lvert\tphi(x)\rangle = \langle\phi(x)\rvert\calM(\alpha,S)\lvert\phi(x)\rangle, 
        \end{align}
        where $\calM(\alpha,S)$ is the observable defined in Theorem~\ref{thm:single-observable-existence}, and the 
        r.h.s.\ expectation value is defined solely on the data register.
        To specify $\calM(\alpha,S)$, we make use of three oracle calls, namely to the quantum-state coefficient oracle $W(\alpha)$, the training-set oracle $O^\dagger(S)$, and the norm coefficient oracle $O_N(\alpha)$.
        With these, it is sufficient to set
        \begin{align}
            &\tcalM(\alpha,S) \\
            =& \lVert\alpha\rVert_1\left(O^\dagger(S)_\mathrm{data,idx} \cdot W(\alpha)_\mathrm{idx,sign}\right)^\dagger\left(\lvert0\rangle\!\langle0\rvert_\mathrm{data}\otimes\bbI_\mathrm{idx}\otimes Z_\mathrm{sign}\right)\left(O^\dagger(S)_\mathrm{data,idx} \cdot W(\alpha)_\mathrm{idx,sign}\right). \nonumber
        \end{align}
        Intuitively, we recognized that the weighted sum over $\alpha$ can be instantiated using a construction reminiscent of the \emph{linear combination of unitaries} (LCU) framework~\citep{childs2012hamiltonian}.
        The extra sign register allows us to flexibly combine both positive and negative coefficients.
        We show that $\langle\tphi(x)\rvert\tcalM(\alpha,S)\lvert\tphi(x)\rangle=f_\alpha(x)$.
        For clarity, as depicted in Fig.~\ref{fig:circuit_AAO_sampling}, instead of keeping the Heisenberg picture of $\tcalM(\alpha,S)$, we apply the unitary gates $W(\alpha)$ and $O^\dagger(S)$ on the quantum state, one at a time,
        \begin{align}
            &\left(O^\dagger(S)_\mathrm{data,idx} \cdot W(\alpha)_\mathrm{idx,sign}\right)\lvert\tphi(x)\rangle \\ \nonumber
            =& O^\dagger(S)_\mathrm{data,idx}\left(\lvert\phi(x)\rangle_\mathrm{data}\otimes\left(\sum_{i=1}^N\sqrt{p_i}\lvert i\rangle_\mathrm{idx}\lvert s_i\rangle_\mathrm{sign}\right)\right) \\
            \nonumber
            =& \left(\sum_{j=1}^N U^\dagger(x_j)_\mathrm{data}\otimes\lvert j\rangle\!\langle j\rvert_\mathrm{idx}\right)\left(\lvert\phi(x)\rangle_\mathrm{data}\otimes\left(\sum_{i=1}^N\sqrt{p_i}\lvert i\rangle_\mathrm{idx}\lvert s_i\rangle_\mathrm{sign}\right)\right) \\
            \nonumber
            = &\sum_{i,j=1}^N \sqrt{p_i}\left(U^\dagger(x_j)\lvert\phi(x)\rangle_\mathrm{data}\right)\otimes\left(\lvert j\rangle_\mathrm{idx}\underbrace{\langle j\vert i\rangle}_{\delta_{i,j}}\right)\otimes \lvert s_i\rangle_\mathrm{sign} \\
            \nonumber
            =& \sum_{i=1}^N\sqrt{p_i} \left(U^\dagger(x_i)\lvert\phi(x)\rangle_\mathrm{data}\right)\lvert i\rangle_\mathrm{idx}\lvert s_i\rangle_\mathrm{sign}.
            \nonumber
        \end{align}
        We consider the expectation value of the fixed observable $\lVert\alpha\rVert_1\left(\lvert0\rangle\!\langle0\rvert_\mathrm{data}\otimes\bbI_\mathrm{idx}\otimes Z_\mathrm{sign}\right)$ with respect to this quantum state, 
        to get
        \begin{align}
            &\left(\sum_{i=1}^N \sqrt{p_i} \left(U^\dagger(x_i)\lvert\phi(x)\rangle_\mathrm{data}\right)\lvert i\rangle_\mathrm{idx}\lvert s_i\rangle_\mathrm{sign}\right)^\dagger \lVert\alpha\rVert_1\left(\lvert0\rangle\!\langle0\rvert_\mathrm{data}\otimes\bbI_\mathrm{idx}\otimes Z_\mathrm{sign}\right) \\
            & \quad \quad \times \left(\sum_{j=1}^N\sqrt{p_j} \left(U^\dagger(x_j)\lvert\phi(x)\rangle_\mathrm{data}\right)\lvert j\rangle_\mathrm{idx}\lvert s_j\rangle_\mathrm{sign}\right) \nonumber \\
             \nonumber
            &= \lVert\alpha\rVert_1\sum_{i,j=1}^N \sqrt{p_ip_j}\langle\phi(x)\vert \underbrace{U(x_i)\lvert0\rangle}_{\lvert\phi(x_i)\rangle}\!\underbrace{\langle0\rvert U^\dagger(x_j)}_{\langle\phi(x_j)\rvert}\lvert\phi(x)\rangle \underbrace{\langle i\rvert\bbI\lvert j\rangle}_{\delta_{i,j}} \langle s_i\rvert \underbrace{Z\lvert s_j\rangle}_{s_j \lvert s_j \rangle}\\
            & = \sum_{i=1}^N \underbrace{s_i\underbrace{\lVert\alpha\rVert_1p_i}_{\lvert\alpha_i\rvert}}_{\alpha_i} \lvert\langle\phi(x)\vert\phi(x_i)\rangle\rvert^2 \nonumber\\
            &= \sum_{i=1}^N \alpha_i k(x,x_i)
            = f_\alpha(x). \nonumber
        \end{align}
        We next bound the operator norm of $\tcalM(\alpha,S)$.
        Making use of the invariance of norms under unitary transformations, the norm of $\tcalM(\alpha,S)$ is equal to the norm 
        \begin{align}
            \lVert\tcalM(\alpha,S)\rVert_\infty &= \Big\lVert\lVert\alpha\rVert_1\left(\lvert0\rangle\!\langle0\rvert_\mathrm{data}\otimes\bbI_\mathrm{idx}\otimes Z_\mathrm{sign}\right)\Big\rVert_\infty 
            = \lVert\alpha\rVert_1 \lVert \lvert 0\rangle\!\langle0\rvert\rVert_\infty\lVert\bbI\rVert_\infty\lVert Z\rVert_\infty
            = \lVert\alpha\rVert_1
        \end{align}
        of the fixed observable.
        We exploit the factoring of norms under the tensor product in the second line.
        Indeed, the Hermitian observable $\calM(\alpha,S)$ fulfills $\lVert\calM(\alpha,S)\rVert_\infty\leq \lVert\alpha\rVert_1$.
    \end{proof}
These statements together prove Theorem~\ref{thm:single-observable-existence}. 

\subsection{Reduced gate cost via programmable feature maps}\label{aa:discrete}

    \begin{figure}[t]
        \centering
        \begin{quantikz}[thin lines, column sep = 8, row sep = 3, slice label style={inner sep=1pt,anchor=south west,rotate=40}]
        \setwiretype{q} & \qwbundle{\lceil\log_2N\rceil} & & & & & \gategroup[3,steps=4,style={draw={choicetwo}}]{\color{choicetwo}Repeat for each $i\in\{1,\ldots,N\}$} & \octrl[label style={choicetwo}, style={draw={choicetwo}}]{1}{\vphantom{\big(}\color{choicetwo}i} & & \octrl[label style={choicetwo}, style={draw={choicetwo}}]{1}{\vphantom{\big(}\color{choicetwo}i} & & & & \gategroup[3,steps=4,style={draw={choicetwo}}]{\color{choicetwo}Repeat for each $i\in\{1,\ldots,N\}$} & \octrl[label style={choicetwo}, style={draw={choicetwo}}]{1}{\vphantom{\big(}\color{choicetwo}i} & & \octrl[label style={choicetwo}, style={draw={choicetwo}}]{1}{\vphantom{\big(}\color{choicetwo}i} & & \\
        \setwiretype{q}\lstick{$\lvert0\rangle_\text{flag}$} & & & & & & & \targ{} & \ctrl{1} & \targ{} & & & & & \targ{} & \ctrl{1} & \targ{} & & \rstick{$\lvert0\rangle_\text{flag}$} \\
        \setwiretype{q}\lstick{$\lvert0\rangle_\mathrm{prog}$} & \qwbundle{s} & & & & & & & \gate[label style={choicetwo}, style={draw={choicetwo}}]{i\mapsto b(x_i)} & & & \gate[2,disable auto height, label style={choicethree}, style={draw=choicethree}]{\mathrm{PU}} & & & & \gate[label style={choicetwo}, style={draw={choicetwo}}]{i\mapsto b(x_i)} & & & \rstick{$\lvert0\rangle_\mathrm{prog}$} \\
        \setwiretype{q} & \qwbundle{n} & & & & & & & & & & & & & & & & &
        \end{quantikz} 
        \caption{
            Realizing the \enquote{training-set oracle} $O^\dagger(S)$ in Definition~\ref{def:training-set-oracle}, using the programmable single-input oracle in Definition~\ref{def:programmable}.
            The input-data information is encoded onto the $s$-qubit program register as a sequence of $N$ controlled multi-target bitflips, and the (programmable) single-input oracle $\mathrm{PU}_\mathrm{prog,data}$ is called only once, coherently in superposition over programs.
            The associated gate cost is $\calO((N + G)s)$.
            This construction improves on that presented in Fig.~\ref{fig:training-set-oracle}, where no access to the programmable oracle is assumed.
        }
        \label{fig:training-set-oracle-cheaper}
    \end{figure}

    The number of gates required to implement the training-set oracle $O^\dagger(S)$ could be further reduced with an extra assumption, namely that the encoding unitary $U(x)$ can be programmed with a length-$s$ bitstring $b(x)\in\{0,1\}^s$ that depends only on $x$.
    In this case, we consider an auxiliary $s$-qubit \enquote{program} register, which controls the single-input oracle $U(x)$ acting on the $n$-qubit data register.
    To be precise, we assume we have access to the following programmable oracle:

    \begin{definition}[Programmable single-input oracle]\label{def:programmable}
        Let $b:\calX\to\{0,1\}^s$ be an efficiently computable map.
        The \emph{programmable single-input oracle} $\mathrm{PU}_\mathrm{prog,data}$ is the unitary
        \begin{align}
            \mathrm{PU}_\mathrm{prog,data} &= \sum_{p\in\{0,1\}^s} \lvert p\rangle\!\langle p\rvert_\mathrm{prog}\otimes U_p,
        \end{align}
        where $\{U_p\}_{p\in\{0,1\}^s}$ is a family of $n$-qubit unitaries satisfying $U_{b(x)} = U(x)$ for all $x\in\calX$.
        In particular, on inputs of the form $\lvert b(x)\rangle_\mathrm{prog}\otimes\lvert\psi\rangle_\mathrm{data}$, the oracle acts as
        \begin{align}
            \mathrm{PU}_\mathrm{prog,data}\left(\lvert b(x)\rangle_\mathrm{prog}\otimes\lvert\psi\rangle_\mathrm{data}\right) &= \lvert b(x)\rangle_\mathrm{prog}\otimes\left(U(x)\lvert\psi\rangle_\mathrm{data}\right).
        \end{align}
        The associated gate cost is $\calO(Gs)$.
    \end{definition}
    We emphasize that the $\calO(Gs)$ gate cost is not automatic: a black-box family $\{U_p\}$ of $2^s$ arbitrary unitaries would require $\calO(2^s G)$ gates to implement coherently via multi-controlled gates.
    The improved scaling reflects a structural assumption on $U(x)$ -- namely, that it admits an implementation as a sequence of $G$ elementary gates, each parameterized by $\calO(1)$ bits of the program register $b(x)$.
    Under this assumption, each elementary gate is realized with $\calO(s)$ controlled primitives, yielding a total cost of $\calO(Gs)$ rather than the naive $\calO(2^s G)$.
    This is a meaningful assumption: it is weaker than quantum random-access memory but stronger than the black-box access of Definition~\ref{def:access_model}.

    \begin{lemma}[Training-set-oracle from programmable single-input oracle]\label{l:training-set-oracle-cheaper}
        The training-set oracle can be realized with a single query to the programmable single-input oracle $\mathrm{PU}_\mathrm{prog,data}$.
        The associated gate cost is $\tcalO((N + G)s)$, and the procedure uses $s+1$ extra qubits.
    \end{lemma}
    \begin{proof}
        See Figure~\ref{fig:training-set-oracle-cheaper} for a visual proof.
        The associated gate cost to load the programs in the auxiliary register is, for each $i\in\{1,\ldots,N\}$: $\calO(\log N)$ gates to perform a multi-controlled bitflip on the flag register conditioned on the index being $i$ plus $\calO(s)$ controlled bitflips on the program register conditioned on the flag register.
        Then, calling the programmable single-input oracle in superposition uses $\calO(Gs)$ gates.
        Uncomputing the program register has the same cost as loading it.
        In total, the number of gates scale as $\calO(N \log N + (N+G)s) = \tcalO((N + G) s)$.
    \end{proof}

    \begin{remark}[Gate complexity from programmable single-input oracle.]
        Assuming access to the programmable single-input oracle from Definition~\ref{def:programmable}, the number of gates for the \emph{all-at-once} algorithms in Theorem~\ref{thm:meta-gates} reduces to $\tcalO((N+G)s\lVert\alpha\rVert_1^2/\varepsilon^2)$ via sampling and $\tcalO(((N+G)s+n)\lVert\alpha\rVert_1/\varepsilon)$ via quantum amplitude estimation. Compared to the bounds in Theorem~\ref{thm:meta-gates}, the multiplicative factor of $N$ on $G$ is replaced by a multiplicative factor of $s$, at the cost of an additive $\calO(Ns)$ term per query. In the regime $s\ll N$ -- the natural setting in which the programmable oracle assumption is meaningful -- this represents a substantial improvement. Relative to \emph{list-and-sum adaptive}, \emph{all-at-once} now incurs only an additive $\calO(Ns)$ overhead and an $s$ multiplicative factor on $G$, rather than the $N$ multiplicative factor of Theorem~\ref{thm:meta-gates}.
    \end{remark}

\section{A zoo of algorithms}\label{a:zoo}

\begin{algorithm}[t]
    \caption{Unified inference for quantum kernel methods. 
    }
    \label{alg:unified}
    \begin{algorithmic}[1]
        \Require $x \in \calX$ (new input), $\varepsilon > 0$ (additive precision), trained coefficients $\alpha \in \bbR^N$.
        \Ensure Estimate of $f_\alpha(x) = \sum_{i=1}^N \alpha_i k(x,x_i)$ to $\varepsilon$ additive precision.
        \Statex
        \Statex \textcolor{choiceone}{\textbf{Choice 1: How to approximate the sum.}}
        \Statex \hspace{\algorithmicindent} \textcolor{choiceone}{(a) \textsc{list-and-sum (fixed budget)}}: estimate every $k(x,x_i)$ with the same budget $M$.
            \hfill [Algorithms~\ref{alg:list-sum-fixed-sampling} and~\ref{alg:list-sum-fixed-qae}]
        \Statex \hspace{\algorithmicindent} \textcolor{choiceone}{(b) \textsc{list-and-sum (adaptive budget)}}: estimate $k(x,x_i)$ with a separate budget $M_i$.
            \hfill [Algorithms~\ref{alg:list-sum-adaptive-sampling} and~\ref{alg:list-sum-adaptive-qae}]
        \Statex \hspace{\algorithmicindent} \textcolor{choiceone}{(c) \textsc{All-at-once}}: encode $f_\alpha(x)$ as a single expectation value $\langle\tphi(x)\rvert\tcalM(\alpha,S)\lvert\tphi(x)\rangle$.
            \hfill [Algorithms~\ref{alg:single-observable-sampling} and~\ref{alg:AAOQ}]
        \Statex
        \Statex \textcolor{choicetwo}{\textbf{Choice 2: How to estimate expectation values.}}
        \Statex \hspace{\algorithmicindent} \textcolor{choicetwo}{(I) \textsc{Sampling}}: measure $M$ times; precision $\calO(1/\sqrt{M})$ per estimate.
            \hfill [Theorem~\ref{thm:sampling_variance_concentration}]
        \Statex \hspace{\algorithmicindent} \textcolor{choicetwo}{(II) \textsc{Quantum amplitude estimation}}: use $V^{(\dagger)}$ coherently $M$ times; precision $\calO(1/M)$.
            \hfill [Theorem~\ref{thm:qae-runtime}]
        \Statex

    \Switch{\textcolor{choiceone}{Choice 1:}}
        \Case{\textcolor{choiceone}{(a) or (b)}} \Comment{\textcolor{choiceone}{List-and-sum}}
            \For{$i \in \{1, \dots, N\}$}
            \State Reading oracle $O_R(\alpha, i)$: obtain $\alpha_i$.
            \Switch{\textcolor{choicetwo}{Choice 2:}}
                \Case{\textcolor{choicetwo}{(I)}} \Comment{\textcolor{choicetwo}{Sampling}}
                    \State Estimate $k(x,x_i)$ using $M_i$ shots: $\hk^{(M_i)}(x,x_i)$.
                \EndCase
                \Case{\textcolor{choicetwo}{(II)}} \Comment{\textcolor{choicetwo}{Quantum amplitude estimation}}
                    \State Estimate $k(x,x_i)$ using $M_i$ coherent queries: $\hk^{(M_i)}(x,x_i)$. 
                \EndCase
            \EndSwitch
            \State Compute $\alpha_i \hat{k}^{(M_i)}(x, x_i)$.
            \EndFor
            \State \Return $\hat{f}_\alpha(x) = \sum_{i=1}^N \alpha_i \hat{k}(x, x_i)$.
        \EndCase
        \Case{\textcolor{choiceone}{(c)}} \Comment{\textcolor{choiceone}{All-at-once}}
            \Switch{\textcolor{choicetwo}{Choice 2:}}
                \Case{\textcolor{choicetwo}{(I)}} \Comment{\textcolor{choicetwo}{Sampling}}
                    \State Measure $M$ times the \emph{all-at-once} observable (Figure~\ref{fig:circuit_AAO_sampling}, Algorithm~\ref{alg:single-observable-sampling}).
                    \State \Return $\hf_\alpha^{(M)}(x)$.
                \EndCase
                \Case{\textcolor{choicetwo}{(II)}} \Comment{\textcolor{choicetwo}{Quantum amplitude estimation}}
                    \State Consider the \emph{amplitude-encoding} unitary $V$ from Figure~\ref{fig:circuit_AAO_QAE} (Algorithm~\ref{alg:amplitude-encoding-observable}).
                    \State Run quantum amplitude estimation on $V$ to estimate $\sqrt{f_\pm}$ to precision $\calO(1/M)$.
                    \State \Return $\hf_\alpha^{(2M)}(x) = \lVert\alpha\rVert_1(\hat{f}_+ - \hat{f}_-)$.
                \EndCase
            \EndSwitch
        \EndCase
      \EndSwitch
    \end{algorithmic}
\end{algorithm}

\begin{table}[t]
    \centering
    \caption{
        Total cost required to ensure $\varepsilon$ additive precision for each of the algorithms below. We highlight \emph{list-and-sum} with fixed budget via sampling as the \emph{naive} approach ($\naive$), \emph{all-at-once} via quantum amplitude estimation as \emph{query-optimal} ($\optimal$), and \emph{list-and-sum} with adaptive budget via quantum amplitude estimation as \emph{gate-optimal} ($\diamond$).
    }
    \label{tab:all_algorithms}
    \renewcommand{\arraystretch}{1.0}
    \begin{tabular}{@{}lccc@{}}
        \toprule
        \textbf{Strategy} & \textbf{Sampling} & \textbf{Quantum amplitude} & \textbf{Section reference} \\
        & & \textbf{estimation} & \\
        \midrule
        List-and-sum, fixed & $\calO\!\left(\tfrac{N\lVert\alpha\rVert_2^2}{\varepsilon^2}\right)^{\!\naive}$ [Cor.~\ref{cor:list-sum-fixed-sampling}] & $\tcalO\!\left(\tfrac{N\lVert\alpha\rVert_2}{\varepsilon}\right)$ [Cor.~\ref{cor:list-sum-fixed-qae}] & \ref{aa:list-and-sum} \\[2pt]
        List-and-sum, adaptive & $\calO\!\left(\tfrac{\lVert\alpha\rVert_1^2}{\varepsilon^2}\right)$ [Cor.~\ref{cor:list-sum-adapt-sampling}] & $\tcalO\!\left(\tfrac{\lVert\alpha\rVert_{\nicefrac{2}{3}}}{\varepsilon}\right)^{\diamond}$ [Cor.~\ref{cor:list-sum-adapt-qae}] & \ref{aa:list-and-sum} \\[2pt]
        All-at-once & $\calO\!\left(\tfrac{\lVert\alpha\rVert_1^2}{\varepsilon^2}\right)$ [Thm.~\ref{thm:all-at-once-sampling}] & $\calO\!\left(\tfrac{\lVert\alpha\rVert_1}{\varepsilon}\right)^{\!\optimal}$ [Thm.~\ref{thm:AAOQ}] & \ref{aa:all-at-once} \\
        \bottomrule
    \end{tabular}
\end{table}

In this appendix, we provide detailed analyses and proofs for each of the algorithmic combinations presented in Algorithm~\ref{alg:unified} and summarized in Theorem~\ref{thm:meta}. For each combination of choices -- how to approximate the sum (\emph{list-and-sum} with 
\emph{fixed} or \emph{adaptive} budget, or \emph{all-at-once}) and how to estimate expectation values (\emph{sampling} or \emph{quantum amplitude estimation}) -- we state the corresponding algorithm, derive the optimal parameter choices where applicable, and prove the resulting runtime complexity. The results are collected in Table~\ref{tab:all_algorithms}.

\subsection{List-and-sum}\label{aa:list-and-sum}

We first provide a template for all \emph{list-and-sum} theorems and algorithms, in Theorem~\ref{thm:list-sum-template} and Algorithm~\ref{alg:list-sum-template}.
We next specialize the template into the four possible choices.
We collect these results in Table~\ref{tab:list_and_sum}.

\begin{theorem}[List-and-sum algorithm template analysis]
    \label{thm:list-sum-template}
    Let $X_1,\ldots,X_N\in[0,1]$ and $X=\sum_{i=1}^N\alpha_iX_i$, let $r\in\{1,2\}$, and let $\varepsilon>0$.
    Let $\beta_1,\ldots,\beta_N\geq0$ be bias parameters and let $\gamma\geq0$ be a variance floor.
    Assume we have mutually independent $M_i$-query estimators $\hX_i$ of $X_i$ obeying
    \begin{align}\label{eq:template-hypotheses}
        \bigl\lvert\bbE_{\hX}[\hX_i]-X_i\bigr\rvert\leq\beta_i
        \qquad\text{and}\qquad
        \operatorname{Var}_{\hX}[\hX_i]\leq\frac{c}{M_i^{r}}+\gamma
    \end{align}
    for some constant $c>0$.
    Define the $M$-query estimator $\hX\coloneqq\sum_{i=1}^N\alpha_i\hX_i$, with $M=\sum_{i=1}^NM_i$ and $M\geq N$, and write $B\coloneqq\sum_{i=1}^N\lvert\alpha_i\rvert\beta_i$ for the accumulated bias.
    Assume the variance floor obeys
    \begin{align}\label{eq:template-floor}
        \gamma\lVert\alpha\rVert_2^2 \leq \frac{\varepsilon^2}{48}.
    \end{align}
    Then the following numbers of queries $M$ are sufficient to ensure $\lvert\hX-X\rvert<\varepsilon+B$ with probability at least $\nicefrac{2}{3}$:
    \begin{enumerate}[noitemsep]
        \item With fixed budget $M_i=M/N$ for all $i\in\{1,\ldots,N\}$:
            $M\in\calO(\varepsilon^{-\nicefrac{2}{r}}N\lVert\alpha\rVert_2^{\nicefrac{2}{r}})$.
        \item With optimal query allocation
            $M_i\in\calO(\varepsilon^{-\nicefrac{2}{r}}\lvert\alpha_i\rvert^{\nicefrac{2}{(r+1)}}\lVert\alpha\rVert_{\nicefrac{2}{(r+1)}}^{\nicefrac{2}{(r(r+1))}})$,
            which in turn yields
            $M\in\calO(\varepsilon^{-\nicefrac{2}{r}}\lVert\alpha\rVert_{\nicefrac{2}{(r+1)}}^{\nicefrac{2}{r}})$.
    \end{enumerate}
    For any $\delta\in(0,1)$, a median-of-means amplification -- running $\calO(\log(\delta^{-1}))$ independent copies of the estimator and returning their median -- reduces the failure probability to $\delta$ at a multiplicative overhead of $\calO(\log(\delta^{-1}))$ in the number of queries, without changing the $\varepsilon$-dependence.
\end{theorem}

\begin{proof}
    Write $\mu_i\coloneqq\bbE_{\hX}[\hX_i]$ and $Z\coloneqq\sum_{i=1}^N\alpha_i(\hX_i-\mu_i)$, and split the estimation error into a centered part and a deterministic bias,
    \begin{align}\label{eq:template-split}
        \lvert\hX-X\rvert
         \leq \lvert Z\rvert+\sum_{i=1}^N\lvert\alpha_i\rvert\lvert\mu_i-X_i\rvert
         \leq \lvert Z\rvert+B,
    \end{align}
    where the second inequality uses the first hypothesis in Eq.~\eqref{eq:template-hypotheses}.
    The bias contribution $B$ is deterministic, so it suffices to ensure $\lvert Z\rvert<\varepsilon$ with probability at least $\nicefrac{2}{3}$.
    By construction $\bbE_{\hX}[Z]=0$, and by independence of the estimators together with the second hypothesis in Eq.~\eqref{eq:template-hypotheses} and the floor condition Eq.~\eqref{eq:template-floor},
    \begin{align}\label{eq:template-variance}
        \operatorname{Var}_{\hX}[Z]=\sum_{i=1}^N\alpha_i^2\operatorname{Var}_{\hX}[\hX_i]
         \leq c\sum_{i=1}^N\frac{\alpha_i^2}{M_i^{r}}+\gamma\lVert\alpha\rVert_2^2
         \leq c\sum_{i=1}^N\frac{\alpha_i^2}{M_i^{r}}+\frac{\varepsilon^2}{48}.
    \end{align}
    Chebyshev's inequality gives $\bbP[\lvert Z\rvert\geq\varepsilon]\leq\operatorname{Var}_{\hX}[Z]/\varepsilon^2$, so it is enough to certify $\operatorname{Var}_{\hX}[Z]\leq\nicefrac{\varepsilon^2}{3}$.
    By Eq.~\eqref{eq:template-variance}, this holds as soon as the query budgets satisfy
    \begin{align}\label{eq:template-constraint}
        \sum_{i=1}^N\frac{\alpha_i^2}{M_i^{r}} \leq \kappa,
        \qquad \kappa\coloneqq\frac{\varepsilon^2}{4c}\in\Theta(\varepsilon^2),
    \end{align}
    since then $\operatorname{Var}_{\hX}[Z]\leq\nicefrac{\varepsilon^2}{4}+\nicefrac{\varepsilon^2}{48}=\nicefrac{13\varepsilon^2}{48}\leq\nicefrac{\varepsilon^2}{3}$, whence $\bbP[\lvert Z\rvert<\varepsilon]\geq\nicefrac{35}{48}>\nicefrac{2}{3}$.
    We stress that the variance floor $\gamma$ does not depend on the budgets $M_i$ and therefore plays no role in the allocation problem below; it enters solely through the standing hypothesis Eq.~\eqref{eq:template-floor}.
    
    We next give shot allocations realizing Eq.~\eqref{eq:template-constraint}.
    In the simpler case of \emph{fixed budget} $M_i=M/N$ for all $i$, we directly obtain
    \begin{align}
        \sum_{i=1}^N\frac{\alpha_i^2}{M_i^r}
        = \frac{N^r}{M^r}\sum_{i=1}^N\alpha_i^2
        = \frac{N^r\lVert\alpha\rVert_2^2}{M^r}
         \leq \kappa
        \qquad\Longleftrightarrow\qquad
        M  \geq  \frac{N\lVert\alpha\rVert_2^{\nicefrac{2}{r}}}{\kappa^{\nicefrac{1}{r}}},
    \end{align}
    so that $M\in\calO(\varepsilon^{-\nicefrac{2}{r}}N\lVert\alpha\rVert_2^{\nicefrac{2}{r}})$ suffices.
    To find the optimal query allocation (partly inspired by \citet{rubin2018application}), we consider the optimization problem
    \begin{align}
        \min_{M_i>0}   M =\sum_{i=1}^N M_i,
        \qquad
        \text{s.t. } \sum_{i=1}^N\frac{\alpha_i^2}{M_i^r}  \leq  \kappa,
    \end{align}
    in which integrality and the lower bounds $M_i\geq1$ are temporarily dropped; both are restored by the rounding step at the end.
    We may assume $\alpha_i\neq0$ for all $i$, as indices with $\alpha_i=0$ contribute to neither the objective nor the constraint and can be dropped.
    Each map $M_i\mapsto\alpha_i^2/M_i^r$ is convex and strictly decreasing on $M_i>0$, so the feasible region -- a sublevel set of a convex function -- is convex and the linear objective makes this a convex program; since the objective is increasing in each $M_i$ while the constraint function is decreasing, the constraint is active at any minimum, so it may be imposed as an equality, and the stationary point of the Lagrangian computed below is the global minimum.
    The Lagrangian of the problem is
    \begin{align}
        \calL=\sum_{i=1}^NM_i +\lambda\left(\sum_{i=1}^N\frac{\alpha_i^2}{M_i^r}- \kappa\right),
    \end{align}
    and the first order condition yields
    \begin{align}
        \pder[\calL]{M_i} &= 1 - r \lambda\frac{\alpha_i^2}{M_i^{r+1}} = 0, \\
        M_i^{r+1} &= r\lambda\alpha_i^2, \\
        M_i&=\left( r\lambda\alpha_i^2\right)^{\frac{1}{(r+1)}}.
    \end{align}
    Introducing this condition into the constraint, we obtain
    \begin{align}
        \sum_{i=1}^N\frac{\alpha_i^2}{M_i^r} &= \sum_{i=1}^N \frac{\alpha_i^2}{\left(r\lambda\alpha_i^2\right)^{\frac{r}{r+1}}} = \frac{1}{\left(r\lambda\right)^{\frac{r}{r+1}}}\sum_{i=1}^N\left(\alpha_i^2\right)^{1-\frac{r}{r+1}} = \frac{\sum_{i=1}^N \lvert\alpha_i\rvert^{\frac{2}{r+1}}}{\left(r\lambda\right)^{\frac{r}{r+1}}} = \kappa, \\
        \frac{\sum_{i=1}^N\lvert\alpha_i\rvert^{\frac{2}{r+1}}}{\kappa r^{\frac{r}{r+1}}} &=\lambda^{\frac{r}{r+1}},\\
        \lambda &= \frac{\left(\sum_{i=1}^N\lvert\alpha_i\rvert^{\frac{2}{r+1}}\right)^{\frac{r+1}{r}}}{r \kappa^{\frac{r+1}{r}}}.
    \end{align}
    We can use the value of the multiplier $\lambda$ to reach the final result given by
    \begin{align}
        M_i &=\left(r\lambda\alpha_i^2\right)^{\frac{1}{r+1}} = \left(r\frac{\left(\sum_{i=1}^N\lvert\alpha_i\rvert^{\frac{2}{r+1}}\right)^{\frac{r+1}{r}}}{r \kappa^{\frac{r+1}{r}}}\alpha_i^2\right)^{\frac{1}{r+1}} \\
        &= \frac{\lvert\alpha_i\rvert^{\frac{2}{r+1}}}{\sqrt[r]{\kappa}}\left(\sum_{i=1}^N\lvert\alpha_i\rvert^{\frac{2}{r+1}}\right)^{\frac{1}{r}} = \frac{\lvert\alpha_i\rvert^{\frac{2}{r+1}}}{\sqrt[r]{\kappa}}\lVert\alpha\rVert_{\frac{2}{r+1}}^{\frac{2}{r(r+1)}}.
        \nonumber
    \end{align}
    Spelling out the $\varepsilon$ dependence of $\kappa$, we obtain
    \begin{align}
        M_i &\in \calO\!\left(\varepsilon^{-\frac{2}{r}}\lvert\alpha_i\rvert^{\frac{2}{r+1}}\lVert\alpha\rVert_{\frac{2}{r+1}}^{\frac{2}{r(r+1)}}\right).
    \end{align}
    With this, we can reach the final result
    \begin{align}
        M&=\sum_{i=1}^N M_i = \frac{1}{\sqrt[r]{\kappa}}\left(\sum_{i=1}^N\lvert\alpha_i\rvert^{\frac{2}{r+1}}\right)\lVert\alpha\rVert_{\frac{2}{r+1}}^{\frac{2}{r(r+1)}} \\
        \nonumber
        &= \frac{1}{\sqrt[r]{\kappa}}\left(\sum_{i=1}^N\lvert\alpha_i\rvert^{\frac{2}{r+1}}\right)\left(\sum_{i=1}^N\lvert\alpha_i\rvert^{\frac{2}{r+1}}\right)^{\frac{1}{r}} \\
        &= \frac{1}{\sqrt[r]{\kappa}}\left(\sum_{i=1}^N\lvert\alpha_i\rvert^{\frac{2}{r+1}}\right)^{\frac{r+1}{r}} = \frac{\lVert\alpha\rVert_{\frac{2}{r+1}}^{\frac{2}{r}}}{\sqrt[r]{\kappa}},
         \nonumber
    \end{align}
    the full expression including $\varepsilon$ in turn becoming
    \begin{align}
        M &\in\calO\!\left(\varepsilon^{-\frac{2}{r}}\lVert\alpha\rVert_{\frac{2}{r+1}}^{\frac{2}{r}}\right).
    \end{align}
    Rounding each budget up to $\max\{1,\lceil M_i\rceil\}$ restores integrality and the constraint $M_i\geq1$ dropped above, only tightens Eq.~\eqref{eq:template-constraint}, and increases the total budget by at most $N$, which is absorbed because the continuous total budget before rounding is at least $N$.

    Finally, we address the amplification of the success probability.
    Let $K=2\lceil5\log(\delta^{-1})\rceil+1$, let $\hX^{(1)},\ldots,\hX^{(K)}$ be independent copies of the estimator, each with budget $M$, and let $\tX$ denote their median, i.e., the middle order statistic, $K$ being odd.
    Call a copy \emph{good} if it lands in the open interval $I\coloneqq\left(X-\varepsilon-B,X+\varepsilon+B\right)$, which by the above happens independently with probability at least $\nicefrac{35}{48}$ for each copy.
    We stress that $I$ is centred at the true value $X$ and is the same for all copies, so the bias does not interfere with the argument.
    If more than $\nicefrac{K}{2}$ copies are good, then $\tX\in I$: were $\tX$ to reach or exceed $\sup I$, the $\nicefrac{(K+1)}{2}$ largest copies would all reach or exceed $\sup I$ and hence be bad, leaving at most $\nicefrac{(K-1)}{2}<\nicefrac{K}{2}$ good copies -- a contradiction -- and symmetrically below $\inf I$.
    By Hoeffding's inequality applied to the number of good copies,
    \begin{align}
        \bbP\left[\#\{\text{good copies}\}\leq\frac{K}{2}\right]
         \leq \exp\!\left(-2K\left(\frac{35}{48}-\frac{1}{2}\right)^{\!2}\right)
         = \exp\!\left(-\frac{121K}{1152}\right)
         \leq \delta,
    \end{align}
    where the last inequality uses $K\geq10\log(\delta^{-1})\geq\nicefrac{1152}{121}\log(\delta^{-1})$.
    Hence $\lvert\tX-X\rvert<\varepsilon+B$ holds with probability at least $1-\delta$ at a total cost of $\calO(M\log(\delta^{-1}))$ queries.
\end{proof}

    \begin{table}[t]
    \centering
    \caption{Runtime scaling of \emph{list-and-sum} algorithms.}
    \label{tab:list_and_sum}
    \renewcommand{\arraystretch}{1.0}
    \begin{tabular}{@{}lcc@{}}
        \toprule
        \textbf{Strategy} & \textbf{Sampling} ($r=1$) & \textbf{Quantum amplitude estimation} ($r=2$) \\
        \midrule
        Fixed budget & $M\in \calO\!\left(\varepsilon^{-2}N\lVert\alpha\rVert_2^2\right)$ & $M\in \tcalO\!\left(\varepsilon^{-1}N\lVert\alpha\rVert_2\right)$ \\[2pt]
        \hline
        \multirow{2}{*}{Adaptive budget} & $M_i\in\calO\!\left(\varepsilon^{-2}\lvert\alpha_i\rvert\lVert\alpha\rVert_1\right)$ & $M_i\in\tcalO\!\left(\varepsilon^{-1}\lvert\alpha_i\rvert^{\nicefrac{2}{3}}\lVert\alpha\rVert_{\nicefrac{2}{3}}^{\nicefrac{1}{3}}\right)$ \\
        & $M\in\calO\!\left(\varepsilon^{-2}\lVert\alpha\rVert_1^2\right)$ & $M\in\tcalO\!\left(\varepsilon^{-1}\lVert\alpha\rVert_{\nicefrac{2}{3}}\right)$ \\
        \bottomrule
    \end{tabular}
\end{table}

    \begin{algorithm}[t]
        \caption{List-and-sum algorithm template.}
        \label{alg:list-sum-template}
        \begin{algorithmic}[1]
            \Require $x \in \calX$ \Comment New input.
            \Require $\varepsilon>0$ \Comment Required additive precision.
            \Require $b\geq0$ \Comment Bias parameter; $b=0$ for $r=1$, Eq.~\eqref{eq:bias-parameter} for $r=2$.
            \Require $M_1,\ldots,M_N$ with $M=\sum_{i=1}^NM_i\geq N$ \Comment Per-term query budgets.
            \Ensure $f_\alpha(x)=\sum_{i=1}^N\alpha_ik(x,x_i)$ to $\varepsilon + B$ additive precision, $B\coloneqq b\lVert\alpha\rVert_1$ \Comment Inference, Eq.~\eqref{eq:kernel_function_setup}.
            \For{$i\in\{1,\ldots,N\}$}:
                \State Reading oracle $O_R(\alpha,i)$: $\alpha_i$
                \State Estimate $k(x,x_i)$ using $M_i$ queries to single-input oracle: $\hk^{(M_i)}(x,x_i)$ 
                \Statex \hfill \Comment Bias $\leq b$, variance $\calO(1/{M_i^r}+b)$, with $r\in\{1,2\}$.
                \State Multiply $\hk^{(M_i)}(x,x_i)$ with the coefficient $\alpha_i$ \Comment Bias $\leq\lvert\alpha_i\rvert b$, variance $\calO\left(\alpha_i^2\left(\tfrac{1}{M_i^r}+b\right)\right)$.
            \EndFor \Comment Total number of queries is $M=\sum_{i=1}^NM_i$.
            \State\Return $\hf^{(M)}_\alpha(x)\coloneqq\sum_{i=1}^N\alpha_i\hk^{(M_i)}(x,x_i)$ 
            \Statex \hfill \Comment Total bias $B=b\lVert\alpha\rVert_1$, $\calO\left(\sum_{i=1}^N\tfrac{\alpha_i^2}{M_i^r}+b\lVert\alpha\rVert_2^2\right)$.
            \Statex \hfill \Comment Sufficient: $b\lVert\alpha\rVert_2^2\leq\varepsilon^2/48$ and $M\in\calO(\varepsilon^{-\nicefrac{2}{r}}\lVert\alpha\rVert_{\nicefrac{2}{(r+1)}}^{\nicefrac{2}{r}})$.
        \end{algorithmic}
    \end{algorithm}
    
    Below, we spell out the resulting costs for all four \emph{list-and-sum} variants as corollaries
    of Theorem~\ref{thm:list-sum-template}.
    The two sampling variants instantiate the template with an exactly unbiased estimator and no
    variance floor, so that both the accumulated bias $B$ and the floor condition
    Eq.~\eqref{eq:template-floor} are vacuous; the quantum-amplitude-estimation variants instead go
    through Lemma~\ref{l:qae-instantiation} and acquire the polylogarithmic factor recorded there.

    \begin{corollary}[Fixed budget, sampling]\label{cor:list-sum-fixed-sampling}
    Consider Algorithm~\ref{alg:list-sum-fixed-sampling} and the problem 
    set-up in Definition~\ref{def:problem_setup}. To estimate the quantum kernel model prediction in Eq.~(\ref{eq:kernel_function_setup}) to additive precision $\varepsilon$ with constant success probability, the runtime scaling of Algorithm~\ref{alg:list-sum-fixed-sampling} is dominated by the query complexity
        \begin{align}
            M&\in\calO\!\left(\frac{N\lVert\alpha\rVert_2^2}{\varepsilon^2}\right).
        \end{align}
    \end{corollary}
    \begin{proof}
        Apply Theorem~\ref{thm:list-sum-template} with $X_i=k(x,x_i)\in[0,1]$ and the estimators of Theorem~\ref{thm:sampling_variance_concentration}, which are exactly unbiased with $\operatorname{Var}\leq\nicefrac{1}{4M_i}$: the hypotheses of Eq.~\eqref{eq:template-hypotheses} thus hold with $r=1$, $\beta_i=0$, $\gamma=0$ and $c=\nicefrac14$. The floor condition Eq.~\eqref{eq:template-floor} is satisfied trivially and the accumulated bias vanishes, $B=0$, so the fixed-budget case of the theorem gives the claim.
    \end{proof}

    \begin{algorithm}[t] 
        \caption{List-and-sum with fixed budget, via sampling.} \label{alg:list-sum-fixed-sampling}
        \begin{algorithmic}[1] 
            \Require $x \in \calX$ \Comment New input.
            \Require $\varepsilon>0$ \Comment Required additive precision.
            \Require $M\geq N$ \Comment Total shot budget, split evenly.
            \Ensure $f_\alpha(x)=\sum_{i=1}^N\alpha_ik(x,x_i)$ to $\varepsilon$ additive precision.
            \Comment Inference, Eq.~\eqref{eq:kernel_function_setup}. 
            \For{$i\in\{1,\ldots,N\}$}: 
            \State Reading oracle $O_R(\alpha,i)$: $\alpha_i$
            \State Single-input oracle for $U(x)$ and $U^\dagger(x_i)$: estimate $k(x,x_i)$ using $M/N$ shots: $\hk^{(M/N)}(x,x_i)$. 
            \Statex \hfill \Comment Unbiased, variance $\calO(N/M)$. 
            \State Multiply $\hk^{(M/N)}(x,x_i)$ with the coefficient $\alpha_i$ 
            \Comment Variance $\calO\left(\tfrac{N\alpha_i^2}{M}\right)$. 
            \EndFor 
            \Comment Total number of shots is $M$. 
            \State\Return $\hf^{(M)}_\alpha(x)=\sum_{i=1}^N\alpha_i\hk^{(M/N)}(x,x_i)$ 
            \Statex \hfill \Comment Bias $B=0$, total variance $\calO\left(\tfrac{N\lVert\alpha\rVert_2^2}{M}\right)$
            \Statex \hfill Sufficient: $M\in\calO(\varepsilon^{-2}N\lVert\alpha\rVert_2^2)$. 
        \end{algorithmic} 
    \end{algorithm} 
    \begin{corollary}[Adaptive budget, sampling]\label{cor:list-sum-adapt-sampling}
        Consider Algorithm~\ref{alg:list-sum-adaptive-sampling} and the problem set-up in Definition~\ref{def:problem_setup}. To estimate the quantum kernel model prediction in Eq.~(\ref{eq:kernel_function_setup}) to additive precision $\varepsilon$ with constant success probability, the runtime scaling of Algorithm~\ref{alg:list-sum-adaptive-sampling} is dominated by the query complexity
        \begin{align}
            M &\in\calO\!\left(\frac{\lVert\alpha\rVert_1^2}{\varepsilon^2}\right).
        \end{align}
        The optimal allocation of shots is
        \begin{align}
            M_i &\in\calO\!\left(\frac{\lvert\alpha_i\rvert\lVert\alpha\rVert_1}{\varepsilon^2}\right).
        \end{align}
    \end{corollary}
    \begin{proof}
        As in Corollary~\ref{cor:list-sum-fixed-sampling}, the estimators of Theorem~\ref{thm:sampling_variance_concentration} satisfy Eq.~\eqref{eq:template-hypotheses} with $r=1$, $\beta_i=0$, $\gamma=0$ and $c=\nicefrac14$, so that Eq.~\eqref{eq:template-floor} holds trivially and $B=0$. The claim is the optimal-allocation case of Theorem~\ref{thm:list-sum-template}, where $\nicefrac{2}{(r+1)}=1$ and $\nicefrac{2}{(r(r+1))}=1$ at $r=1$.
    \end{proof}

    \begin{algorithm}[t]
        \caption{List-and-sum with adaptive budget, via sampling.}
        \label{alg:list-sum-adaptive-sampling}
        \begin{algorithmic}[1]
            \Require $x \in \calX$ \Comment New input.
            \Require $\varepsilon>0$ \Comment Required additive precision.
            \Require $M_1,\ldots,M_N$ with $M=\sum_{i=1}^NM_i\geq N$ \Comment Per-term shot budgets.
            \Ensure $f_\alpha(x)=\sum_{i=1}^N\alpha_ik(x,x_i)$ to $\varepsilon$ additive precision. \Comment Inference, Eq.~\eqref{eq:kernel_function_setup}.
            \For{$i\in\{1,\ldots,N\}$}:
                \State Reading oracle $O_R(\alpha,i)$: $\alpha_i$
                \State Single-input oracle for $U(x)$ and $U^\dagger(x_i)$: estimate $k(x,x_i)$ using $M_i$ shots: $\hk^{(M_i)}(x,x_i)$ 
                \Statex \hfill \Comment Unbiased, variance $\calO(1/M_i)$.
                \State Multiply $\hk^{(M_i)}(x,x_i)$ with the coefficient $\alpha_i$ \Comment Variance $\calO(\alpha_i^2/M_i)$.
            \EndFor \Comment Total number of shots is $M=\sum_{i=1}^NM_i$.
            \State\Return $\hf^{(M)}_\alpha(x)\coloneqq\sum_{i=1}^N\alpha_i\hk^{(M_i)}(x,x_i)$ 
            \Statex \hfill \Comment Bias $B=0$, total variance $\calO(\sum_{i=1}^N {\alpha_i^2}/{M_i})$, 
            \Statex \hfill Sufficient: $M\in\calO(\varepsilon^{-2}\lVert\alpha\rVert_1^2)$.
        \end{algorithmic}
    \end{algorithm}

    For the variants using quantum amplitude estimation, we first give the nearly unbiased estimator that results from instantiating \citet{Cornelissen2023} in our problem set-up.
    \begin{lemma}[Nearly unbiased kernel-value estimator]\label{l:unbiased-kernel-est}
        Let $x,x_i\in\calX$ and $b\in(0,1)$.
        In the access model of Definition~\ref{def:access_model}, for every $M\geq C_0\log\log(M)\log(\tfrac{M}{b})$ there is an estimator $\hat{k}_i$ of $k(x,x_i)$ that uses at most $M$ queries in expectation to the single-input oracle (counting $U(\cdot)$, $U^\dagger(\cdot)$, and controlled versions thereof), satisfies $\lvert\hat{k}_i\rvert\leq2\pi$, and obeys
        \begin{align}
            \bigl\lvert\bbE[\hat{k}_i]-k(x,x_i)\bigr\rvert \leq b,
            \qquad
            \mathrm{Var}[\hat{k}_i] \leq \frac{c\log\log^{2}(M)\log^{2}(\tfrac{M}{b})}{M^{2}}+b,
        \end{align}
        for universal constants $C_0,c>0$.
    \end{lemma}
    \begin{proof}
        Instantiate Theorem~\ref{thm:ndub-qae} with $\lvert\psi\rangle=U(x)\lvert0\rangle^{\otimes n}$ and $\Pi=U(x_i)\lvert0\rangle^{\otimes n}\!\langle0\rvert^{\otimes n}U^\dagger(x_i)$, so that $p=\lvert\langle\psi(x_i)\vert\psi(x)\rangle\rvert^{2}=k(x,x_i)$.
        The reflection $\bbI-2\lvert\psi\rangle\!\langle\psi\rvert=U(x)\bigl(\bbI-2\lvert0\rangle\!\langle0\rvert^{\otimes n}\bigr)U^\dagger(x)$ costs two single-input queries plus oracle-independent gates, and likewise for $\bbI-2\Pi$; controlled reflections decompose into controlled single-input queries, which Definition~\ref{def:access_model} grants, and preparing $\lvert\psi\rangle$ costs one further query.
        The query count of Theorem~\ref{thm:ndub-qae} is stated in expectation; we keep it so and defer the conversion to a worst-case bound to the point of use (Lemma~\ref{l:qae-instantiation}), where a single truncation over the whole algorithm suffices. Choosing $t=\Theta\bigl(M/(\log\log(M)\log(\tfrac{M}{b}))\bigr)$ and invoking Theorem~\ref{thm:ndub-qae} with bias parameter $b$ yields the claim, using $p\leq1$ in the variance bound; the bias and variance guarantees of Theorem~\ref{thm:ndub-qae} are unconditional, so no interaction with the state-restoration branch needs to be tracked.
    \end{proof}

    Next, we spell out how the poly-logarithmic factors affect the scaling of the complete estimator:

\begin{lemma}[Instantiating the template with quantum amplitude estimation]\label{l:qae-instantiation}
    Consider the problem set-up in Definition~\ref{def:problem_setup} and let $\varepsilon>0$.
    The case $\alpha=0$ is solved exactly by returning zero. Assume henceforth that $\alpha\neq0$. Set the uniform bias parameter
    \begin{align}\label{eq:bias-parameter}
        b \coloneqq \min\left\{
            \frac14,
            \frac{\varepsilon}{4\lVert\alpha\rVert_1},
            \frac{\varepsilon^2}{192\lVert\alpha\rVert_2^2}
        \right\},
    \end{align}
    and let $\tilde M_1,\ldots,\tilde M_N$ be the budgets provided by Theorem~\ref{thm:list-sum-template} applied with $r=2$, variance floor $\gamma=b$, precision $\varepsilon/2$, and either of its two allocation rules, writing $\tilde M\coloneqq\sum_{i=1}^N\tilde M_i$. Run the estimator of Lemma~\ref{l:unbiased-kernel-est} for each $i$ at bias parameter $b$ with the inflated budget $M_i\coloneqq\lceil\Lambda\tilde M_i\rceil$, where
    \begin{align}\label{eq:polylog-overhead}
        \Lambda \coloneqq C_1\log\log(\tilde M)
        \log\!\left(\frac{\tilde M}{b}\right),
    \end{align}
    for a suitable universal constant $C_1$. Then $\lvert\hf_\alpha(x)-f_\alpha(x)\rvert<\varepsilon$ with probability at least $\nicefrac{2}{3}$, at a total query cost $M=\sum_{i=1}^NM_i\in\calO(\Lambda\tilde M)$.
\end{lemma}
\begin{proof}
    The case $\alpha=0$ is immediate, so assume $\alpha\neq0$.
    Let $\tilde M_1,\ldots,\tilde M_N\geq1$ be the template budgets, and write $\tilde M=\sum_{i=1}^N\tilde M_i$.
    We may assume $\tilde M\geq16$ by increasing $\tilde M_1$ by a constant if necessary. This preserves the allocation constraint and ensures that the iterated logarithms below are positive.

    By Eq.~\eqref{eq:template-constraint}, applied at precision $\varepsilon/2$, the budgets satisfy
    \begin{align}
        \sum_{i=1}^N\frac{\alpha_i^2}{\tilde M_i^2}
        &\leq \frac{\varepsilon^2}{16c}.
    \end{align}
    Hence, by Cauchy--Schwarz,
    \begin{align}
        \lVert\alpha\rVert_1
        &\leq
        \left(
            \sum_{i=1}^N\frac{\alpha_i^2}{\tilde M_i^2}
        \right)^{1/2}
        \left(
            \sum_{i=1}^N\tilde M_i^2
        \right)^{1/2} \\
        &\leq \frac{\varepsilon\tilde M}{4\sqrt{c}}.
    \end{align}
    Since $\lVert\alpha\rVert_2\leq\lVert\alpha\rVert_1$, Eq.~\eqref{eq:bias-parameter} implies
    \begin{align}
        \frac{1}{b}
        &=
        \max\left\{
            4,\,
            \frac{4\lVert\alpha\rVert_1}{\varepsilon},\,
            \frac{192\lVert\alpha\rVert_2^2}{\varepsilon^2}
        \right\}
        \in\calO(\tilde M^2).
    \end{align}
    Consequently, for a fixed universal constant $C_1$,
    \begin{align}
        \Lambda
        &= C_1\log\log(\tilde M)\log(\tilde M/b)
        \in\calO\!\left(
            \log\log(\tilde M)\log(\tilde M)
        \right).
    \end{align}

    Set $M_i=\lceil\Lambda\tilde M_i\rceil$. Since $M_i\leq\Lambda\tilde M+1$ and $b^{-1}\in\calO(\tilde M^2)$, we can choose $C_1$ sufficiently large that $\Lambda\geq1$ and, for every $i$,
    \begin{align}
        \max\{1,C_0\}\,
        \log\log(M_i)\log(M_i/b)
        &\leq \Lambda.
    \end{align}
    Indeed, increasing $C_1$ increases $\Lambda$ linearly, whereas its contribution to the logarithmic factors on the left grows only polylogarithmically. Because $\tilde M_i\geq1$, this choice ensures
    \begin{align}
        M_i
        &\geq C_0\log\log(M_i)\log(M_i/b),
    \end{align}
    as required by Lemma~\ref{l:unbiased-kernel-est}.

    We first analyze the algorithm without truncation. Apply Lemma~\ref{l:unbiased-kernel-est} independently for each $i$, using bias parameter $b$ and budget $M_i$. The resulting estimators $\hat k_i$ satisfy
    \begin{align}
        \bigl\lvert
            \bbE[\hat k_i]-k(x,x_i)
        \bigr\rvert
        &\leq b
    \end{align}
    and
    \begin{align}
        \operatorname{Var}[\hat k_i]
        &\leq
        \frac{
            c\bigl(\log\log(M_i)\bigr)^2
            \bigl(\log(M_i/b)\bigr)^2
        }{M_i^2}
        +b \\
        &\leq \frac{c}{\tilde M_i^2}+b.
    \end{align}
    The second inequality follows from $M_i\geq\Lambda\tilde M_i$ and the choice of $C_1$.

    Define the untruncated prediction estimator
    \begin{align}
        \widehat F
        &\coloneqq \sum_{i=1}^N\alpha_i\hat k_i.
    \end{align}
    Its accumulated bias obeys
    \begin{align}
        \bigl\lvert
            \bbE[\widehat F]-f_\alpha(x)
        \bigr\rvert
        &\leq b\lVert\alpha\rVert_1
        \leq \frac{\varepsilon}{4}.
    \end{align}
    By independence, the allocation constraint, and Eq.~\eqref{eq:bias-parameter},
    \begin{align}
        \operatorname{Var}[\widehat F]
        &=
        \sum_{i=1}^N
        \alpha_i^2\operatorname{Var}[\hat k_i] \\
        &\leq
        c\sum_{i=1}^N\frac{\alpha_i^2}{\tilde M_i^2}
        +b\lVert\alpha\rVert_2^2 \\
        &\leq
        \frac{\varepsilon^2}{16}
        +\frac{\varepsilon^2}{192}
        =\frac{13\varepsilon^2}{192}.
    \end{align}
    Chebyshev's inequality therefore gives
    \begin{align}
        \bbP\!\left(
            \bigl\lvert
                \widehat F-\bbE[\widehat F]
            \bigr\rvert
            \geq \frac{\varepsilon}{2}
        \right)
        &\leq \frac{13}{48}.
    \end{align}
    Thus, with probability at least $35/48$,
    \begin{align}
        \bigl\lvert
            \widehat F-f_\alpha(x)
        \bigr\rvert
        &<
        \frac{\varepsilon}{2}
        +\frac{\varepsilon}{4}
        <\varepsilon.
    \end{align}

    It remains to bound the query cost. Let $Q$ be the total number of queries used by the untruncated algorithm. Lemma~\ref{l:unbiased-kernel-est} bounds each estimator's query count in expectation, so
    \begin{align}
        \bbE[Q]
        &\leq \sum_{i=1}^N M_i
        \leq \Lambda\tilde M+N
        \leq 2\Lambda\tilde M,
    \end{align}
    where we used $\tilde M\geq N$ and $\Lambda\geq1$.

    Set
    \begin{align}
        Q_{\max}
        &\coloneqq \lceil48\Lambda\tilde M\rceil.
    \end{align}
    Truncate the algorithm if it would require more than $Q_{\max}$ queries, returning zero on truncation. Otherwise, return $\widehat F$. By Markov's inequality,
    \begin{align}
        \bbP(Q>Q_{\max})
        &\leq \frac{\bbE[Q]}{Q_{\max}}
        \leq \frac{1}{24}.
    \end{align}
    Whenever truncation does not occur, the output agrees with the untruncated estimator. Therefore, by a union bound, the truncated algorithm succeeds with probability at least
    \begin{align}
        \frac{35}{48}-\frac{1}{24}
        &=\frac{11}{16}
        >\frac{2}{3}.
    \end{align}
    Its worst-case query cost is $Q_{\max}\in\calO(\Lambda\tilde M)$.
\end{proof}

    We can finally give the concise results that characterize the query-complexity scaling of both \emph{list-and-sum} strategies based on quantum amplitude estimation.
    
    \begin{corollary}[Fixed budget, quantum amplitude estimation]\label{cor:list-sum-fixed-qae}
        Consider Algorithm~\ref{alg:list-sum-fixed-qae} and the problem set-up in Definition~\ref{def:problem_setup}.
        To estimate the quantum kernel model prediction in Eq.~(\ref{eq:kernel_function_setup}) to additive precision $\varepsilon$ with constant success probability, the runtime scaling of Algorithm~\ref{alg:list-sum-fixed-qae} is dominated by the query complexity
            \begin{align}
        M &\in\tcalO\!\left(\frac{N\lVert\alpha\rVert_2}{\varepsilon}\right).
    \end{align}
    \end{corollary}
    \begin{proof}
        Theorem~\ref{thm:list-sum-template} with $r=2$ and the fixed-budget allocation gives $\tilde M\in\calO(\varepsilon^{-1}N\lVert\alpha\rVert_2)$, and Lemma~\ref{l:qae-instantiation} yields $M\in\calO(\Lambda\tilde M)$. The $\tcalO$ suppresses the polylogarithmic factor $\Lambda$.
    \end{proof}
    \begin{algorithm}[t]
        \caption{List-and-sum with fixed budget, via quantum amplitude estimation.}
        \label{alg:list-sum-fixed-qae}
        \begin{algorithmic}[1]
            \Require $x \in \calX$ \Comment New input.
            \Require $\varepsilon>0$ \Comment Required additive precision.
            \Require $b$ as in Eq.~\eqref{eq:bias-parameter} \Comment Bias parameter and variance floor.
            \Require $M\geq N$, with $M/N$ admissible for Lemma~\ref{l:unbiased-kernel-est}
            
            \Ensure $f_\alpha(x)=\sum_{i=1}^N\alpha_ik(x,x_i)$ to $\varepsilon$ additive precision. \Comment Inference, Eq.~\eqref{eq:kernel_function_setup}.
            \For{$i\in\{1,\ldots,N\}$}:
                \State Reading oracle $O_R(\alpha,i)$: $\alpha_i$
                \State Single-input oracle for $U(x)$ and $U^\dagger(x_i)$: estimate $k(x,x_i)$ using $M/N$ queries via Lemma~\ref{l:unbiased-kernel-est} at bias parameter $b$: $\tk^{(M/N)}(x,x_i)$ \Comment Bias $\leq b$, variance $\tcalO\left(\tfrac{N^2}{M^2}+b\right)$.
                \State Multiply $\tk^{(M/N)}(x,x_i)$ with the coefficient $\alpha_i$ \Comment Bias $\leq \lvert \alpha_i\rvert b$, variance $\tcalO\left(\alpha_i^2\left(\tfrac{N^2}{M^2}+b\right)\right)$.
            \EndFor \Comment Total number of queries is $M$.
            \State\Return $\tf^{(M)}_\alpha(x)=\sum_{i=1}^N\alpha_i\tk^{(M/N)}(x,x_i)$ 
            \Statex \hfill \Comment Bias $B=b\lVert\alpha\rVert_1\leq\varepsilon/4$, variance $\tcalO\left(\tfrac{N^2\lVert\alpha\rVert_2^2}{M^2}+b\lVert\alpha\rVert_2^2\right)$.
            \Statex \hfill \Comment Sufficient: $M\in\tcalO(\varepsilon^{-1}N\lVert\alpha\rVert_2)$.
        \end{algorithmic}
    \end{algorithm}

    \begin{corollary}[Adaptive budget, quantum amplitude estimation]\label{cor:list-sum-adapt-qae}
        Consider Algorithm~\ref{alg:list-sum-adaptive-qae} and the problem set-up in Definition~\ref{def:problem_setup}.
        To estimate the quantum kernel model prediction in Eq.~\eqref{eq:kernel_function_setup} to additive precision $\varepsilon$ with constant success probability, the runtime scaling of Algorithm~\ref{alg:list-sum-adaptive-qae} is dominated by the query complexity
        \begin{align}
            M &\in\tcalO\!\left(\frac{\lVert\alpha\rVert_{\nicefrac{2}{3}}}{\varepsilon}\right).
        \end{align}
        The optimal allocation of queries is
        \begin{align}
            M_i &\in\tcalO\!\left(\frac{\lvert\alpha_i\rvert^{\nicefrac{2}{3}}\lVert\alpha\rVert_{\nicefrac{2}{3}}^{\nicefrac{1}{3}}}{\varepsilon}\right).
        \end{align}
    \end{corollary}
    \begin{proof}
        Theorem~\ref{thm:list-sum-template} with $r=2$ and the optimal query allocation gives $\tilde M_i\in\calO(\varepsilon^{-1}\lvert\alpha_i\rvert^{\nicefrac{2}{3}}\lVert\alpha\rVert_{\nicefrac{2}{3}}^{\nicefrac{1}{3}})$ and hence $\tilde M\in\calO(\varepsilon^{-1}\lVert\alpha\rVert_{\nicefrac{2}{3}})$, and Lemma~\ref{l:qae-instantiation} inflates each budget to $M_i=\lceil\Lambda\tilde M_i\rceil$, for a total of $M\in\calO(\Lambda\tilde M)$. The $\tcalO$ suppresses the polylogarithmic factor $\Lambda$.
    \end{proof}
    \begin{algorithm}[t]
        \caption{List-and-sum with adaptive budget, via quantum amplitude estimation.}
        \label{alg:list-sum-adaptive-qae}
        \begin{algorithmic}[1]
            \Require $x \in \calX$ \Comment New input.
            \Require $\varepsilon>0$ \Comment Required additive precision.
            \Require $b$ as in Eq.~\eqref{eq:bias-parameter} \Comment Bias parameter and variance floor.
            \Require $M_1,\ldots,M_N$ admissible for Lemma~\ref{l:unbiased-kernel-est}, $M=\sum_{i=1}^NM_i\geq N$
            \Statex \hfill \Comment Per-term query budgets.
            \Ensure $f_\alpha(x)=\sum_{i=1}^N\alpha_ik(x,x_i)$ to $\varepsilon$ additive precision. \Comment Inference, Eq.~\eqref{eq:kernel_function_setup}.
            \For{$i\in\{1,\ldots,N\}$}:
                \State Reading oracle $O_R(\alpha,i)$: $\alpha_i$.
                \State Single-input oracle for $U(x)$ and $U^\dagger(x_i)$: estimate $k(x,x_i)$ using $M_i$ queries via Lemma~\ref{l:unbiased-kernel-est} at bias parameter $b$: $\tk^{(M_i)}(x,x_i)$ 
                \Statex \hfill \Comment Bias $\leq b$, variance $\tcalO\left(\tfrac{1}{M_i^2}+b\right)$.
                \State Multiply $\tk^{(M_i)}(x,x_i)$ with the coefficient $\alpha_i$ \Comment Bias $\leq\lvert\alpha_i\rvert b$, variance $\tcalO\left(\alpha_i^2\left(\tfrac{1}{M_i^2}+b\right)\right)$.
            \EndFor \Comment Total number of queries is $M=\sum_{i=1}^NM_i$.
            \State\Return $\tf^{(M)}_\alpha(x)=\sum_{i=1}^N\alpha_i\tk^{(M_i)}(x,x_i)$ \Comment 
            \Statex \hfill \Comment Bias $B=b\lVert\alpha\rVert_1\leq\varepsilon/4$, variance $\tcalO\left(\sum_{i=1}^N\tfrac{\alpha_i^2}{M_i^2}+b\lVert\alpha\rVert_2^2\right)$.
            \Statex \hfill \Comment Sufficient: $M\in\tcalO(\varepsilon^{-1}\lVert\alpha\rVert_{\nicefrac{2}{3}})$.
        \end{algorithmic}
    \end{algorithm}

\subsection{All-at-once}\label{aa:all-at-once}

    We next provide the algorithms and analysis for the \emph{all-at-once} approaches. Unlike for the case of list-and-sum, there is no single template from which all results follow, but rather we present each scenario individually. Several relevant definitions are in Section~\ref{ss:all-at-once} in the main text.

\subsubsection{Via sampling}\label{aaa:AOS}
    
We now analyze the query complexity of estimating the \emph{all-at-once} observable $\calM(\alpha,S)$ introduced in Theorem~\ref{thm:single-observable-existence} directly via sampling, using the circuit from Algorithm~\ref{alg:single-observable-sampling}. The resulting algorithm is specified in Algorithm~\ref{alg:AAOS} and the resulting runtime in Theorem~\ref{thm:all-at-once-sampling}.

    \begin{algorithm}[t]
        \caption{All-at-once via sampling.}
        \label{alg:AAOS}
        \begin{algorithmic}[1]
            \Require $x \in \calX$ \Comment New input.
            \Require $\varepsilon>0$ \Comment Required additive precision.
            \Ensure $f_\alpha(x)=\sum_{i=1}^N\alpha_ik(x,x_i)$ to $\varepsilon$ additive precision. \Comment Inference, Eq.~\eqref{eq:kernel_function_setup}.
            \State Single-input oracle $U(x)$: $\lvert\phi(x)\rangle$
            \State Estimate $\langle\phi(x)\rvert\tcalM(\alpha,S)\lvert\phi(x)\rangle$ via sampling (see Algorithm~\ref{alg:single-observable-sampling}), using $M$ shots: $\hf^{(M)}_\alpha(x)$ 
            \Statex \hfill \Comment Achieve precision $\calO({\lVert\tcalM(\alpha,S)\rVert_\infty}/{\sqrt{M}})$.
            \State\Return $\hf^{(M)}_\alpha(x)$ \Comment Total additive precision is in $\calO\!\left({\lVert\alpha\rVert_1}/{\sqrt{M}}\right)$.
        \end{algorithmic}
    \end{algorithm}

    \begin{theorem}[Query complexity of all-at-once via sampling]\label{thm:all-at-once-sampling}
        Consider Algorithm~\ref{alg:AAOS} and the problem set-up in Definition~\ref{def:problem_setup} for estimating the kernel model prediction in Eq.~\eqref{eq:kernel_function_setup} to additive precision $\varepsilon$ with constant success probability.
        The algorithm uses a total number of queries
        \begin{align}
            M&\in\calO\!\left(\frac{\lVert\alpha\rVert_1^2}{\varepsilon^2}\right).
        \end{align}
    \end{theorem}
    \begin{proof}
        Following Algorithms~\ref{alg:AAOS} and~\ref{alg:single-observable-sampling}, as well as Lemma~\ref{l:single-observable}, using $M$ shots produces an estimate $\hf^{(M)}_\alpha(x)$ such that, with probability at least $\nicefrac{2}{3}$,
        \begin{align}
            \left\lvert\hf^{(M)}_\alpha(x)-f_\alpha(x)\right\rvert
            &\in\calO\!\left(\frac{\lVert\alpha\rVert_1}{\sqrt{M}}\right).
        \end{align}
        To achieve $\varepsilon$ additive precision, it is thus sufficient to pick
        \begin{align}
            M\in\calO\!\left(\frac{\lVert\alpha\rVert_1^2}{\varepsilon^2}\right).
        \end{align}
        Since each shot constitutes three oracle queries (one execution of the circuit in Algorithm~\ref{alg:single-observable-sampling}), the total query complexity is $3M\in\calO\!\left({\lVert\alpha\rVert_1^2}/{\varepsilon^2}\right)$.
    \end{proof}

\subsubsection{Via quantum amplitude estimation}\label{aaa:QAE}

    We next provide a detailed description of the algorithm used to implement the amplitude-encoding unitary $V$ from Definition~\ref{def:amplitude_encoding_unitary}, with accompanying analysis in Lemma~\ref{l:single-observable-amplitude}.
    Below we analyze the query complexity of the \emph{all-at-once} algorithm using quantum amplitude estimation.
    The procedure is specified in Algorithm~\ref{alg:AAOQ} and the resulting runtime in Theorem~\ref{thm:AAOQ}.

    \begin{algorithm}[t]
        \caption{Amplitude-encoding unitary.}
        \label{alg:amplitude-encoding-observable}
        \begin{algorithmic}[1]
            \Require $x \in \calX$ \Comment New input.
            \Ensure Amplitude-encoding unitary $V$ \Comment From Definition~\ref{def:amplitude_encoding_unitary}.
            \State Initialize $\lvert0\rangle_\mathrm{data}\lvert0\rangle_\mathrm{idx}\lvert0\rangle_\mathrm{sign}\lvert0,0\rangle_\mathrm{QP}$. \Comment $n+\lceil\log_2(N)\rceil+3$ qubits.
            \State Single-input oracle $U(x)_\mathrm{data}$: $\lvert\tphi(x)\rangle=\lvert\phi(x)\rangle_\mathrm{data}\lvert0\rangle_\mathrm{idx}\lvert0\rangle_\mathrm{sign}$ \Comment $G$ gates.
            \State Quantum-state coefficient oracle $W(\alpha)_\mathrm{idx,sign}$: \Comment $N$ gates.
            \begin{align}\lvert\phi(x)\rangle_\mathrm{data}\lvert0\rangle_\mathrm{idx}\lvert0\rangle_\mathrm{sign}\mapsto \sum_{i=1}^N \sqrt{p_i}\lvert\phi(x)\rangle_\mathrm{data}\lvert i\rangle_\mathrm{idx}\lvert s_i\rangle_\mathrm{sign}\nonumber \end{align} 
            \State Training-set oracle $O^\dagger(S)$: \Comment $\calO(N(G + \log N))$ gates.
            \begin{align}\sum_{i=1}^N \sqrt{p_i}\lvert\phi(x)\rangle_\mathrm{data}\lvert i\rangle_\mathrm{idx}\lvert s_i\rangle_\mathrm{sign}\mapsto \sum_{i=1}^N \sqrt{p_i} U^\dagger(x_i)\lvert\phi(x)\rangle_\mathrm{data}\lvert i\rangle_\mathrm{idx}\lvert s_i\rangle_\mathrm{sign} \nonumber \end{align}
            \State Coherently flag $\lvert0\rangle_\mathrm{data}$ on the $\lvert0,0\rangle_\mathrm{QP}$ branch with a multi-control $X$: \Comment $\calO(n)$ elementary gates
            \begin{align}
                &\sum_{i=1}^N \sqrt{p_i} U^\dagger(x_i)\lvert\phi(x)\rangle_\mathrm{data}\lvert i\rangle_\mathrm{idx}\lvert s_i\rangle_\mathrm{sign}\lvert0,0\rangle_\mathrm{QP} \nonumber \\
                & \mapsto\sum_{i=1}^N \sqrt{p_i} \lvert0\rangle\!\langle0\rvert U^\dagger(x_i)\lvert\phi(x)\rangle_\mathrm{data}\lvert i\rangle_\mathrm{idx}\lvert s_i\rangle_\mathrm{sign}\lvert0,0\rangle_\mathrm{QP} + \cdots\lvert1,0\rangle_\mathrm{QP}\nonumber \\
                &= \sum_{i=1}^N \sqrt{p_i}\langle\phi(x_i)\vert\phi(x)\rangle\lvert0\rangle_\mathrm{data} \lvert i\rangle_\mathrm{idx}\lvert s_i\rangle_\mathrm{sign}\lvert0,0\rangle_\mathrm{QP} + \cdots\lvert1,0\rangle_\mathrm{QP}\nonumber
            \end{align}
            \State Separate the positive and negative terms in the sum with a $\CNOT$ \Comment $\calO(1)$ elementary gates.
            \begin{align}
                &\sum_{i=1}^N \sqrt{p_i}\langle\phi(x_i)\vert\phi(x)\rangle\lvert0\rangle_\mathrm{data} \lvert i\rangle_\mathrm{idx}\lvert s_i\rangle_\mathrm{sign}\lvert0,0\rangle_\mathrm{QP} + \cdots\lvert1,0\rangle_\mathrm{QP} \nonumber \\ & \mapsto \sum_{i=1}^N \sqrt{p_i}\langle\phi(x_i)\vert\phi(x)\rangle\lvert0\rangle_\mathrm{data} \lvert i\rangle_\mathrm{idx}\lvert s_i\rangle_\mathrm{sign}\lvert0,\frac{1-s_i}{2}\rangle_\mathrm{QP} \nonumber \\
                &\hphantom{\sum_{i=1}^N\sqrt{p_i}}+ \cdots\lvert1,0\rangle_\mathrm{QP}+ \cdots\lvert1,1\rangle_\mathrm{QP}\nonumber
            \end{align}
            \State\Return $\sum_{i=1}^N \sqrt{p_i}\langle\phi(x_i)\vert\phi(x)\rangle\lvert0\rangle_\mathrm{data} \lvert i\rangle_\mathrm{idx}\lvert s_i\rangle_\mathrm{sign}\lvert0,\frac{1-s_i}{2}\rangle_\mathrm{QP}+ \cdots\lvert1,0\rangle_\mathrm{QP}+ \cdots\lvert1,1\rangle_\mathrm{QP}$ 
            \Statex \hfill \Comment{Number of gates $\calO(N( G + \log N)+ n)$.}
        \end{algorithmic}
    \end{algorithm}

    \begin{customlemma}{\ref{l:single-observable-amplitude}}[Single-observable encoded in amplitude]
        Under the conditions of Theorem~\ref{thm:single-observable-existence}, there exists a quantum circuit using $n+\lceil\log_2(N)\rceil+3$ qubits which implements the amplitude-encoding unitary in Definition~\ref{def:amplitude_encoding_unitary}.
    \end{customlemma}
    \begin{proof}
        We refer to Fig.~\ref{fig:circuit_AAO_QAE} as visual proof.
        The quantum circuit follows the principles from Lemma~\ref{l:single-observable}, though replacing the observables $\lvert0\rangle\!\langle0\rvert_\text{data}$ and $Z_\text{sign}$ with controlled bitflips on auxiliary qubits.
        
        We rigorously prove the statement, by following the steps of Algorithm~\ref{alg:amplitude-encoding-observable}.
        Using the same notation as above for the different oracles, we now introduce two controlled operations as
        \begin{align}
            C_0X_{\text{data, QP,trash}}&=\lvert0\rangle\!\langle0\rvert_\text{data}\otimes \bbI_{\text{QP,trash}} + \left(\bbI-\lvert0\rangle\!\langle0\rvert\right)_\text{data}\otimes X_{\text{QP,trash}},\\
            \CNOT_{\text{sign, QP,}\pm}&=\lvert1\rangle\!\langle1\rvert_\text{sign}\otimes\bbI_{\text{QP,}\pm} + \lvert-1\rangle\!\langle-1\rvert_\text{sign}\otimes X_{\text{QP,}\pm}.
        \end{align}
        Recall we defined the sign register as the eigenbasis of the Pauli $Z$ operator $Z\lvert s_i\rangle=s_i\lvert s_i\rangle$.
        We claim that $V$ is of the form
        \begin{align}
            V&=\CNOT_{\text{sign, QP,}\pm} \cdot C_0X_{\text{data, QP,trash}} \cdot O^\dagger(S)_\text{data,idx,sign} \cdot W(\alpha)_\text{idx,sign} \cdot U(x)_\text{data},
        \end{align}
        as depicted in Fig.~\ref{fig:circuit_AAO_QAE}.
        We confirm this by applying a sequence of matrix-vector operations starting from the all-$0$ state, by performing the calculation
        \begin{align}
            V\lvert0\rangle &= \CNOT_{\text{sign, QP,}\pm} \cdot C_0X_{\text{data, QP,trash}} \cdot O^\dagger(S)_\text{data,idx,sign} \cdot W(\alpha)_\text{idx,sign} \cdot U(x)_\text{data} \lvert0\rangle_\text{data,idx,sign,QP} 
            \\
            \nonumber
            &= \CNOT_{\text{sign, QP,}\pm} \cdot C_0X_{\text{data, QP,trash}} \cdot O^\dagger(S)_\text{data,idx,sign} \left(\sum_{i=1}^N \sqrt{p_i} \lvert\phi(x)\rangle_\text{data}\lvert i\rangle_\text{idx}\lvert s_i\rangle_\text{sign}\lvert0,0\rangle_\text{QP}\right) \\
             \nonumber
            &= \CNOT_{\text{sign, QP,}\pm} \cdot C_0X_{\text{data, QP,trash}} \left(\sum_{i=1}^N \sqrt{p_i} \left(U^\dagger(x_i)\lvert\phi(x)\rangle_\text{data}\right)\lvert i\rangle_\text{idx}\lvert s_i\rangle_\text{sign}\lvert0,0\rangle_\text{QP}\right) \\
             \nonumber
            &= \CNOT_{\text{sign, QP,}\pm} \left(\lvert0\rangle\!\langle0\rvert_\text{data}\otimes \bbI_{\text{QP,trash}} + \left(\bbI-\lvert0\rangle\!\langle0\rvert\right)_\text{data}\otimes X_{\text{QP,trash}}\right) \nonumber \\
            & \quad \quad \times \left(\sum_{i=1}^N \sqrt{p_i} \left(U^\dagger(x_i)\lvert\phi(x)\rangle_\text{data}\right)\lvert i\rangle_\text{idx}\lvert s_i\rangle_\text{sign}\lvert0,0\rangle_\text{QP}\right) \nonumber \\
             \nonumber
            &= \CNOT_{\text{sign, QP,}\pm} \left(\lvert0\rangle\!\langle0\rvert_\text{data}\otimes \bbI_{\text{QP,trash}}\left(\sum_{i=1}^N \sqrt{p_i} \left(U^\dagger(x_i)\lvert\phi(x)\rangle_\text{data}\right)\lvert i\rangle_\text{idx}\lvert s_i\rangle_\text{sign}\lvert0,0\rangle_\text{QP}\right)\right. \\
             \nonumber
            &
            \quad \quad \left. + \left(\bbI-\lvert0\rangle\!\langle0\rvert\right)_\text{data}\otimes X_{\text{QP,trash}}\left(\sum_{i=1}^N \sqrt{p_i} \left(U^\dagger(x_i)\lvert\phi(x)\rangle_\text{data}\right)\lvert i\rangle_\text{idx}\lvert s_i\rangle_\text{sign}\lvert0,0\rangle_\text{QP}\right)\right) \\
             \nonumber
            &= \CNOT_{\text{sign, QP,}\pm}\left(\left(\sum_{i=1}^N \sqrt{p_i}\underbrace{\langle0\rvert U^\dagger(x_i)}_{\langle\phi(x_i)\rvert}\lvert\phi(x)\rangle \lvert0\rangle_\text{data}\lvert i\rangle_\text{idx}\lvert s_i\rangle_\text{sign}\right)\lvert0,0\rangle_\text{QP} + \cdots \lvert1,0\rangle_\text{QP}\right) \\
             \nonumber
            &= \left(\lvert1\rangle\!\langle1\rvert_\text{sign}\otimes\bbI_{\text{QP,}\pm} + \lvert-1\rangle\!\langle-1\rvert_\text{sign}\otimes X_{\text{QP,}\pm}\right) \nonumber \\
            & \quad \quad \times \left(\left(\sum_{i=1}^N \sqrt{p_i}\underbrace{\langle0\rvert U^\dagger(x_i)}_{\langle\phi(x_i)\rvert}\lvert\phi(x)\rangle \lvert0\rangle_\text{data}\lvert i\rangle_\text{idx}\lvert s_i\rangle_\text{sign}\right)\lvert0,0\rangle_\text{QP} + \cdots \lvert1,0\rangle_\text{QP}\right) \nonumber \\
             \nonumber
            &= \sum_{i=1}^N \sqrt{p_i}\langle\phi(x_i)\vert\phi(x)\rangle \lvert0\rangle_\text{data}\lvert i\rangle_\text{idx}\lvert s_i\rangle_\text{sign}\lvert0,\frac{1-s_i}{2}\rangle_\text{QP} + \cdots \lvert1,0\rangle_\text{QP} + \cdots\lvert1,1\rangle_\text{QP}.
             \nonumber
        \end{align}
        According to our convention for the definition of the sign register, we have that $\frac{1-s_i}{2}$ is $0$ if $s_i=1$, and $1$ in the case $s_i=-1$.
        For us to confirm that the last line corresponds to the r.h.s. of Eq.~\eqref{eq:amplitude-encoding-observable}, we must only compute its $\lvert0,0\rangle_\text{QP}$ and $\lvert0,1\rangle_\text{QP}$ components.
        We denote the projectors as $P_+ = \bbI_\text{data,idx,sign}\otimes\lvert0,0\rangle\!\langle 0,0\rvert_\text{QP}$, and $P_- = \bbI_\text{data,idx,sign}\otimes\lvert0,1\rangle\!\langle 0,1\rvert_\text{QP}$.
        Then, we find
        \begin{align}
            P_+ &\left(\sum_{i=1}^N \sqrt{p_i}\langle\phi(x_i)\vert\phi(x)\rangle \lvert0\rangle_\text{data}\lvert i\rangle_\text{idx}\lvert s_i\rangle_\text{sign}\lvert0,\frac{1-s_i}{2}\rangle_\text{QP} + \cdots \lvert1,0\rangle_\text{QP} + \cdots\lvert1,1\rangle_\text{QP}\right) \\
            \nonumber
            &= \sum_{i=1}^N\delta_{s_i,1}\sqrt{p_i}\langle\phi(x_i)\vert\phi(x)\rangle\lvert0\rangle_\text{data}\lvert i\rangle_\text{idx}\lvert s_i\rangle_\text{sign}\lvert0,0\rangle_\text{QP} = \lvert\tpsi_+\rangle_\text{data,idx,sign},\\
            P_- &\left(\sum_{i=1}^N \sqrt{p_i}\langle\phi(x_i)\vert\phi(x)\rangle \lvert0\rangle_\text{data}\lvert i\rangle_\text{idx}\lvert s_i\rangle_\text{sign}\lvert0,\frac{1-s_i}{2}\rangle_\text{QP} + \cdots \lvert1,0\rangle_\text{QP} + \cdots\lvert1,1\rangle_\text{QP}\right) \\
            &= \sum_{i=1}^N\delta_{s_i,-1}\sqrt{p_i}\langle\phi(x_i)\vert\phi(x)\rangle\lvert0\rangle_\text{data}\lvert i\rangle_\text{idx}\lvert s_i\rangle_\text{sign}\lvert0,1\rangle_\text{QP} = \lvert\tpsi_-\rangle_\text{data,idx,sign},
            \nonumber
        \end{align}
        (note that $\lvert\tpsi_\pm\rangle$ are unnormalized)
        we also compute the measurement probability for each component
        \begin{align}
            &\langle\tpsi_+\rvert P_+\lvert\tpsi_+\rangle\\
            &= \sum_{i,j=1}^N \delta_{s_i,1}\delta_{s_j,1}\sqrt{p_ip_j}\langle\phi(x_i)\vert\phi(x)\rangle\langle\phi(x)\vert\phi(x_j)\rangle\langle0\vert0\rangle_\text{data}\underbrace{\langle j\vert i\rangle_\text{idx}}_{\delta_{i,j}}\langle s_j\vert s_i\rangle_\text{sign}\langle0,0\vert0,0\rangle_\text{QP} \nonumber\\
            &= \sum_{i=1}^N \delta_{s_i,1} p_i \underbrace{\lvert\langle\phi(x_i)\vert\phi(x)\rangle\rvert^2}_{k(x,x_i)}.
            \nonumber
        \end{align}
        We notice that the last line corresponds exactly to the definition of $f_+$ from Definition~\ref{def:short-hand}.
        The same is true for $f_-$:
        \begin{align}
            &\langle\tpsi_-\rvert P_-\lvert\tpsi_-\rangle\\ &= \sum_{i,j=1}^N \delta_{s_i,-1}\delta_{s_j,-1}\sqrt{p_ip_j}\langle\phi(x_i)\vert\phi(x)\rangle\langle\phi(x)\vert\phi(x_j)\rangle\langle0\vert0\rangle_\text{data}\underbrace{\langle j\vert i\rangle_\text{idx}}_{\delta_{i,j}}\langle s_j\vert s_i\rangle_\text{sign}\langle01\vert01\rangle_\text{QP} \nonumber \\
            &= \sum_{i=1}^N \delta_{s_i,-1} p_i \underbrace{\lvert\langle\phi(x_i)\vert\phi(x)\rangle\rvert^2}_{k(x,x_i)}=f_-.
            \nonumber
        \end{align}
        This completes the proof.
    \end{proof}

    \begin{remark}
        Under the assumption that multi-controlled $X$ gates with $n$ controls require $\calO(n)$ elementary gates, Algorithm~\ref{alg:amplitude-encoding-observable} uses $\calO(N(G + \log N) + n)$ gates and $2$ auxiliary qubits.
    \end{remark}

    \begin{algorithm}[t]
        \caption{All-at-once via quantum amplitude estimation.}
        \label{alg:AAOQ}
        \begin{algorithmic}[1]
            \Require $x \in \calX$ \Comment New input.
            \Require $\varepsilon>0$ \Comment Required additive precision.
            \Ensure $f_\alpha(x)=\sum_{i=1}^N\alpha_ik(x,x_i)$ to $\varepsilon$ additive precision \Comment Inference, Eq.~\eqref{eq:kernel_function_setup}.
            \State Initialize $\lvert0\rangle_\mathrm{data}\lvert0\rangle_\mathrm{idx}\lvert0\rangle_\mathrm{sign}\lvert0,0\rangle_\mathrm{QP}$. \Comment $n+\lceil\log_2(N)\rceil+3$ qubits.
            \State $\ta^{(M)}_+\gets$ $M$-query-quantum amplitude estimation on $V$ from Algorithm~\ref{alg:amplitude-encoding-observable} with $\lvert\psi_\text{good}\rangle=\lvert\psi_+\rangle$ 
            \Statex \hfill \Comment Estimate $\sqrt{f_+}$ 
            to precision $\calO(1/M)$.
            \State $\ta^{(M)}_-\gets$ $M$-query-quantum amplitude estimation on 
            $V$ from Algorithm~\ref{alg:amplitude-encoding-observable} with $\lvert\psi_\text{good}\rangle=\lvert\psi_-\rangle$ 
            \Statex \hfill \Comment Estimate $\sqrt{f_-}$ to precision $\calO(1/M)$.
            \State Square estimates $\tf_\pm^{(M)}=\left(\ta_\pm^{(M)}\right)^2$. \Comment Estimate $f_\pm$ to precision $\calO(1/M)$.
            \State Call the norm oracle $O_N(\alpha)$ to obtain $\lVert\alpha\rVert_1$.
            \State\Return $\tf^{(2M)}_\alpha(x) = \lVert\alpha\rVert_1\left(\tf^{(M)}_+-\tf^{(M)}_-\right)$  
            \Statex \hfill \Comment Combined additive precision in $\calO\left(\tfrac{\lVert\alpha\rVert_1}{M}\right)$. The total number of queries is $2M$.
        \end{algorithmic}
    \end{algorithm}

    Our main contribution is the following result, which underpins the complexity of inference in quantum kernels via quantum amplitude estimation.

    \begin{theorem}[Query complexity of all-at-once via quantum amplitude estimation]\label{thm:AAOQ}
        Consider Algorithm~\ref{alg:AAOQ}.
        To estimate the quantum kernel model prediction in Eq.~\eqref{eq:kernel_function_setup} to additive precision $\varepsilon$ with constant success probability, it suffices to choose $M=\Theta\!\left(\lVert\alpha\rVert_1/\varepsilon\right)$.
        The algorithm uses $2M$ applications of $V$ and $V^\dagger$ in total.
    \end{theorem}
    \begin{proof}
        Algorithm~\ref{alg:AAOQ} evaluates Eq.~\eqref{eq:kernel_function_setup} by using quantum amplitude estimation twice as a subroutine, once for the positive terms in the sum $f_+$, and once for the negative terms $f_-$.
        Recall we defined these terms as $f_\alpha(x)=\lVert\alpha\rVert_1\left(f_+-f_-\right)$, so that $f_\pm\in[0,1]$.
        Each of the two quantum amplitude estimation subroutines uses $M$ applications of $V$ and $V^\dagger$, for a total query count of $2M$.
        By Theorem~\ref{thm:qae-runtime}, $M$ queries yield additive error $\calO(1/M)$ on each of the amplitudes $\sqrt{f_\pm}$, so that
        \begin{align}
            \lvert \ta^{(M)}_\pm-\sqrt{f_\pm}\rvert &\in\calO\!\left(\frac{1}{M}\right).
        \end{align}
        Here we choose the failure probability $\delta=1/6$ in Theorem~\ref{thm:qae-runtime} for each subroutine, with the constant confidence overhead included in the budget $M$ of each subroutine. By the union bound, both amplitude-error bounds hold simultaneously with probability at least $2/3$. The remainder of the argument is on this joint success event.
        We recover estimates of $f_\pm$ by squaring: $\hf^{(M)}_\pm \coloneqq \left(\ta^{(M)}_\pm\right)^2$.
        Since $\sqrt{f_\pm}\in[0,1]$ (as $f_\pm$ is a convex combination of kernel values in $[0,1]$) and $\ta_\pm^{(M)} \in [0,1]$, we have that the factor $\lvert \ta^{(M)}_\pm + \sqrt{f_\pm}\rvert$ is upper bounded by $2$.
        Therefore,
        \begin{align}
            \lvert \hf^{(M)}_\pm - f_\pm\rvert
            = \lvert \ta^{(M)}_\pm - \sqrt{f_\pm}\rvert\cdot\underbrace{\lvert \ta^{(M)}_\pm + \sqrt{f_\pm}\rvert}_{\leq2}
            \leq 2\lvert \ta^{(M)}_\pm - \sqrt{f_\pm}\rvert
            \in\calO\!\left(\frac{1}{M}\right).
        \end{align}
        Taking the difference and applying the triangle inequality, we get
        \begin{align}
            \left\lvert \left(\hf^{(M)}_+ - \hf^{(M)}_-\right) - \left(f_+ - f_-\right)\right\rvert
            \leq \lvert \hf^{(M)}_+ - f_+ \rvert + \lvert \hf^{(M)}_- - f_- \rvert
        \in \calO\!\left(\frac{1}{M}\right).
        \end{align}
        Multiplying by $\lVert\alpha\rVert_1$ to recover an estimate $\hf^{(2M)}_\alpha(x) = \lVert\alpha\rVert_1\left(\hf^{(M)}_+ - \hf^{(M)}_-\right)$ for our target function $f_\alpha$, we obtain
        \begin{align}
            \lvert \hf^{(2M)}_\alpha(x)-f_\alpha(x)\rvert \in \calO\!\left(\frac{\lVert\alpha\rVert_1}{M}\right).
        \end{align}
        To achieve a final precision $\varepsilon$, we, therefore, set
        \begin{align}
            M &\in\Theta\!\left(\frac{\lVert\alpha\rVert_1}{\varepsilon}\right),
        \end{align}
        which gives a total of $2M \in \Theta\!\left(\lVert\alpha\rVert_1/\varepsilon\right)$ applications of $V$ and $V^\dagger$.
    \end{proof}
    
    \begin{remark}
        Algorithm~\ref{alg:AAOQ} does not require reading access to $\alpha$ as the positive and negative parts are separated implicitly inside the algorithm.
    \end{remark}

\section{Matching lower bound: statements and proofs}\label{app:lower_bounds}
    This appendix contains the formal statement and proof of the query lower bound announced in Section~\ref{ss:lower_bounds}, together with the remarks delineating its scope.
    The bound holds for any algorithm equipped with the access model specified in Definition~\ref{def:access_model}.
    The mechanism is to place the hardness inside the feature-map oracle, so that the only way to learn $f_\alpha(x)$ is to query $U(\cdot)$, and to reduce to the problem of approximately counting the ones of a Boolean function rather than to amplitude estimation.
    \begin{theorem}[Query lower bound]\label{thm:lower-bound-full}
        Let $\varepsilon,A>0$ and let $n\in\bbN$ be such that
        \begin{align}\label{eq:lb-regime}
            64 \leq \frac{A}{\varepsilon} \leq 2^{n-1}.
        \end{align}
        Then there exists a family of oracle instances satisfying Definitions~\ref{def:problem_setup} and~\ref{def:access_model}, with $\lVert\alpha\rVert_1=A$, on which any quantum algorithm that estimates $f_\alpha(x)$ to additive precision $\varepsilon$ with constant success probability must make
        \begin{align}
            \Omega\!\left(\frac{\lVert\alpha\rVert_1}{\varepsilon}\right)
        \end{align}
        queries to the single-input oracle.
        Here, each application of $U(x')$, $U^\dagger(x')$, or a controlled version thereof, at any input $x'$, counts as one query; for $N=1$ this includes the training-set oracle $O^{(\dagger)}(S)=U^{(\dagger)}(x_1)$, where the notation $O^{(\dagger)}$ covers the oracle and its inverse; all remaining oracles of Definition~\ref{def:access_model} may be called free of charge.
    \end{theorem}
    \begin{proof}
        Throughout, write $D \coloneqq 2^n$, so that Eq.~\eqref{eq:lb-regime} reads
        $64  \leq  \frac{A}{\varepsilon}  \leq  \frac{D}{2}$; in particular $D\geq128$.
        Without loss of generality, the success probability is at least $2/3$.
        
        \emph{Step 1: The hard family.}
        Set $N=1$ and $\alpha_1=A$, so that $\lVert\alpha\rVert_1=A$ and $f_\alpha(x)=A k(x,x_1)$, and fix any query point $x\neq x_1$. For $g\colon\{0,1\}^n\to\{0,1\}$, let $O_g\lvert z\rangle=(-1)^{g(z)}\lvert z\rangle$ denote the phase oracle of $g$, and define the family
        \begin{align}
            U_g(x_1)=O_gH^{\otimes n}, \qquad U_g(x')=H^{\otimes n} \quad\text{for all } x'\neq x_1,
        \end{align}
        in particular at the query point $x$. Each $U_g$ is a valid input-dependent unitary quantum feature map in the sense of Definition~\ref{def:problem_setup}: writing $g^{-1}(1)=\{z_1,\ldots,z_t\}$, we have $O_g=\prod_{\ell=1}^{t}\left(\bbI-2\lvert z_\ell\rangle\!\langle z_\ell\rvert\right)$, and each such reflection is exactly realizable with $\calO(n)$ gates from a fixed finite gate set containing the Toffoli gate, using one auxiliary qubit~\citep{nielsen2000quantum}; padding with identity gates, the gate count can be taken as $G\in\calO(2^n n)$, uniformly over the family. The members of the family differ only at the single training input $x_1$, and since Definition~\ref{def:access_model} grants only black-box access to the map $x'\mapsto U(x')$, these are legitimate oracle instances for every $g$. Uniformity of $G$ across the family is essential, since $G$ is visible to the algorithm; padding makes all instances indistinguishable in gate count.
        
        \emph{Step 2: Kernel values.}
        Write $t\coloneqq\lvert g^{-1}(1)\rvert$ and $w\coloneqq t/D$. Then 
        \begin{align}
        \langle\psi(x)\vert\psi_g(x_1)\rangle=\langle0\rvert H^{\otimes n}O_gH^{\otimes n}\lvert0\rangle=\frac{1}{D}\sum_{z}(-1)^{g(z)}=1-2w,
        \end{align}
        and hence
        \begin{align}
            k(x,x_1)=(1-2w)^2, \qquad f_\alpha(x)=A (1-2w)^2.
        \end{align}
        We restrict the family to those $g$ with $w\in[{1}/{8},{3}/{8}]$. On this window, the map $\varphi(w)\coloneqq(1-2w)^2$ is strictly decreasing and maps $[{1}/{8},{3}/{8}]$ bijectively onto $[{1}/{16},{9}/{16}]$, and its inverse $\varphi^{-1}(k)=(1-\sqrt{k})/2$ satisfies $\lvert(\varphi^{-1})'(k)\rvert=1/(4\sqrt{k})\leq1$ for $k\geq {1}/{16}$, i.e., $\varphi^{-1}$ is $1$-Lipschitz on the window. In particular, all kernel values of the restricted family lie in $[{1}/{16},{9}/{16}]$, bounded away from $0$ and $1$.
        
        \emph{Step 3: From estimating $f_\alpha(x)$ to distinguishing two Hamming weights.}
        Suppose an algorithm outputs $\hat{f}$ with $\lvert\hat{f}-f_\alpha(x)\rvert\leq\varepsilon$ with probability at least $2/3$ on every instance of the restricted family. Post-process $\hat{f}$ as follows: clip $\hat{k}\coloneqq\min\{\max\{{\hat{f}}/{A},{1}/{16}\},{9}/{16}\}$, invert $\hat{w}\coloneqq\varphi^{-1}(\hat{k})$, and round to $\hat{t}$, the nearest integer to $D\hat{w}$. Whenever $\lvert\hat{f}-f_\alpha(x)\rvert\leq\varepsilon$: clipping cannot increase the distance to $k(x,x_1)\in[{1}/{16},{9}/{16}]$, so $\lvert\hat{k}-k(x,x_1)\rvert\leq \varepsilon/A$; the $1$-Lipschitz property gives $\lvert\hat{w}-w\rvert\leq \varepsilon/A$; and rounding gives $\lvert\hat{t}-t\rvert\leq\frac{\varepsilon D}{A}+\frac{1}{2}$. Hence, with probability at least $2/3$, the algorithm computes $t$ to additive accuracy
        \begin{align}
            \Delta\coloneqq\left\lceil\frac{\varepsilon D}{A}+\frac{1}{2}\right\rceil,
            \qquad
            1 \leq \Delta \leq 2 \frac{\varepsilon D}{A},
        \end{align}
        where the upper bound uses $\varepsilon D/A\geq2$, guaranteed by the right end of Eq.~\eqref{eq:lb-regime}. Now set $t_0\coloneqq\lceil D/4 \rceil$ and $t_1\coloneqq t_0+2\Delta+1$. Both weights lie in the window: $t_0\geq D/8$ is clear, and using $\Delta\leq 2\varepsilon D/A\leq D/32$ from the left end of Eq.~\eqref{eq:lb-regime},
        \begin{align}
            t_1 \leq \left(\frac{D}{4}+1\right)+\left(\frac{D}{16}+1\right) \leq \frac{3D}{8}
            \qquad\text{for } D\geq32.
        \end{align}
        A $\Delta$-accurate estimate distinguishes the two: given $\hat{t}$, declare weight $t_0$ if $\hat{t}\leq t_0+\Delta$ and weight $t_1$ otherwise; since $t_1-t_0=2\Delta+1$, this decision is correct with probability at least $2/3$ on every instance of either weight. The estimation algorithm therefore yields a bounded-error algorithm that, promised $\lvert g^{-1}(1)\rvert\in\{t_0,t_1\}$, decides which case holds, at the same query cost.
        
        \emph{Step 4: The counting lower bound.}
        For integers $0\leq\ell\neq\ell'\leq D$, let $f_{\ell,\ell'}$ denote the partial Boolean function on $\{0,1\}^D$ that evaluates to $1$ on inputs of Hamming weight $\ell$, to $0$ on inputs of Hamming weight $\ell'$, and is undefined otherwise. Identifying a $D$-bit input string with the truth table of a function $g\colon\{0,1\}^n\to\{0,1\}$, \cite{Nayak1999} prove that any quantum algorithm which computes $f_{\ell,\ell'}$ with bounded error, given access to the input via the standard bit oracle
        \begin{align}
            B_g\colon\lvert z\rangle\lvert b\rangle\mapsto\lvert z\rangle\lvert b\oplus g(z)\rangle,
        \end{align}
        must make
        \begin{align}\label{eq:nayak}
            \Omega\!\left(\sqrt{\frac{D}{\Delta'}}+\frac{\sqrt{m(D-m)}}{\Delta'}\right)
        \end{align}
        queries, where $\Delta'=\lvert\ell-\ell'\rvert$ and $m\in\{\ell,\ell'\}$ is the weight farther from $D/2$~\cite[Corollary~1.2]{Nayak1999}. The distinguishing task of Step~3 is exactly the computation of $f_{t_1,t_0}$, with $\Delta'=t_1-t_0=2\Delta+1\leq5\Delta$ and, since $t_0<t_1\leq 3D/8 <D/2$, with $m=t_0=\lceil D/4\rceil$. There, $\sqrt{t_0(D-t_0)}\geq\sqrt{\frac{D}{4}\cdot\frac{D}{2}}\geq\frac{D}{3}$; since Eq.~\eqref{eq:nayak} is a sum of non-negative terms, it is bounded below by its second term alone, which is at least ${D}/{(3\Delta')}\geq {D}/{(15\Delta)}$. Hence the distinguishing task, and by Step~3 the estimation task, requires
        \begin{align}
            \Omega\!\left(\frac{D}{\Delta}\right)=\Omega\!\left(\frac{A}{\varepsilon}\right)
        \end{align}
        queries to $B_g$, where the last equality uses $\Delta\leq {2\varepsilon D}/{A}$.\\
        \emph{Step 5: Back to the single-input oracle.}
        It remains to check that the single-input oracle is the only carrier of information about $g$, and that it carries it at unit rate. The classical oracles $O_R$, $O_N$, $O_C$ and the coefficient quantum-state oracle $W(\alpha)$ act on the fixed coefficient vector $\alpha=(A)$ and are therefore $g$-independent, as is $U_g(x')=H^{\otimes n}$ for every $x'\neq x_1$; for $N=1$ the training-set oracle is $O^\dagger(S)=U^\dagger_g(x_1)$, i.e.\ a single-input query by the convention of the theorem. The remaining calls are $U_g(x_1)=O_gH^{\otimes n}$, its inverse $U^\dagger_g(x_1)=H^{\otimes n}O_g$ (using $O_g^\dagger=O_g$), and controlled versions of either; we show that each can be simulated with exactly one query to $B_g$, together with $\calO(n)$ $g$-independent gates. For the uncontrolled calls this is standard phase kickback: on an auxiliary qubit prepared in $\lvert-\rangle$, one query to $B_g$ acts as $B_g\lvert z\rangle\lvert-\rangle=(-1)^{g(z)}\lvert z\rangle\lvert-\rangle$ and thereby implements $O_g$ on the data register; the preparation and uncomputation of the auxiliary qubit are $g$-independent. For the controlled calls, write $c\text{-}W$ for the version of a unitary $W$ controlled on a qubit $c$, and decompose $c\text{-}U_g(x_1)=(c\text{-}O_g)(c\text{-}H^{\otimes n})$ and $c\text{-}U^\dagger_g(x_1)=(c\text{-}H^{\otimes n})(c\text{-}O_g)$, where $c\text{-}H^{\otimes n}$ is $g$-independent and costs $\calO(n)$ gates. The controlled phase oracle $c\text{-}O_g$ is realized with one query to $B_g$ by the following gadget on an auxiliary qubit $a$ initialized in $\lvert0\rangle$: apply $\CNOT_{c\to a}$ followed by a Hadamard gate on $a$, so that $a$ carries $\lvert+\rangle$ on the control-$\lvert0\rangle$ branch and $\lvert-\rangle$ on the control-$\lvert1\rangle$ branch; query $B_g$ with target $a$, which, by
        \begin{align}
            B_g\lvert z\rangle\lvert+\rangle=\lvert z\rangle\lvert+\rangle,
            \qquad
            B_g\lvert z\rangle\lvert-\rangle=(-1)^{g(z)}\lvert z\rangle\lvert-\rangle,
        \end{align}
        acts as the identity on the first branch and imprints the phase on the second; then undo the Hadamard gate and the $\CNOT_{c\to a}$. The auxiliary qubit returns to $\lvert0\rangle$ coherently on both branches, so the net action is
        \begin{align}
            \lvert c\rangle\lvert z\rangle\lvert0\rangle_a \mapsto (-1)^{c g(z)}\lvert c\rangle\lvert z\rangle\lvert0\rangle_a,
        \end{align}
        i.e.\ exactly $c\text{-}O_g$, at the cost of one query and $\calO(1)$ $g$-independent gates; multiply controlled versions are handled identically, replacing $\CNOT_{c\to a}$ by the corresponding multi-controlled bit flip. Hence an algorithm making $q$ queries to the single-input oracle induces a bounded-error algorithm for the distinguishing task of Step~3 making $q$ queries to $B_g$, and $q\in\Omega({A}/{\varepsilon})=\Omega({\lVert\alpha\rVert_1}/{\varepsilon})$ follows from Step~4.
    \end{proof}

    \begin{remark}
        Neither end of the regime in Eq.~\eqref{eq:lb-regime} weakens the optimality conclusion. 
        Some cutoff at the lower end is necessary: for $A\leq\varepsilon$ the constant estimate $0$ is already $\varepsilon$-accurate, since $\lvert f_\alpha(x)\rvert\leq\lVert\alpha\rVert_1$, and for $A/\varepsilon\leq64$ Theorem~\ref{thm:AAOQ} solves the task with $\calO(1)$ queries, so no super-constant bound can hold there.
        The upper end, in turn, is an artifact of the hard family used in the proof, whose attainable values of $f_\alpha(x)$ are $\Theta(A/2^n)$-spaced. Since the theorem asserts the existence of hard instances and the number of feature qubits $n$ is a parameter of the construction, hard instances exist for every ratio $A/\varepsilon\geq64$ upon choosing $n$ with $2^{n-1}\geq A/\varepsilon$. Whether the full-access-model bound persists for fixed $n$ beyond the upper cutoff is a question we do not pursue.
    \end{remark}
    
    \begin{corollary}[Optimality of Algorithm~\ref{alg:AAOQ}]
    \label{cor:AAOQ-optimal}
        Consider any algorithm that accesses the problem instance only through applications of the amplitude-encoding unitary $V$ from Definition~\ref{def:amplitude_encoding_unitary}, of its inverse $V^\dagger$, and of controlled versions thereof, as is the case for Algorithm~\ref{alg:AAOQ} and for any quantum-amplitude-estimation routine run on $V$. In the regime of Eq.~\eqref{eq:lb-regime}, if such an algorithm estimates $f_\alpha(x)$ to additive precision $\varepsilon$ with constant success probability, it must apply $V$, $V^\dagger$, or controlled versions thereof $\Omega\left({\lVert\alpha\rVert_1}/{\varepsilon}\right)$ times. Together with the matching upper bound of Theorem~\ref{thm:AAOQ}, Algorithm~\ref{alg:AAOQ} is query-optimal up to constant factors.
    \end{corollary}
    \begin{proof}
        Instantiate the hard family of Theorem~\ref{thm:lower-bound-full}. By construction (Algorithm~\ref{alg:amplitude-encoding-observable}), each application of $V$ makes one call to the single-input oracle at the query point $x$ and one call to the training-set oracle $O^\dagger(S)$, which for $N=1$ is a single call to $U^\dagger(x_1)$; all further operations inside $V$ are calls to $W(\alpha)$ or fixed $g$-independent gates. Each application of $V^\dagger$ likewise makes one call to $U^\dagger(x)$ and one to $O(S)=U(x_1)$, and controlled applications decompose into controlled versions of the same oracle calls, each of which Theorem~\ref{thm:lower-bound-full} counts as a single query. A total of $q$ applications therefore induces at most $2q$ queries to the single-input oracle, and Theorem~\ref{thm:lower-bound-full} yields $2q\in\Omega({\lVert\alpha\rVert_1}/{\varepsilon})$, hence $q\in\Omega({\lVert\alpha\rVert_1}/{\varepsilon})$.
    \end{proof}
    Three comments delimit the scope of Theorem~\ref{thm:lower-bound-full}. First, the bound is unconditional within the access model of Definition~\ref{def:access_model}: it constrains every algorithm equipped with those oracles, with no requirement that the computation be routed through the amplitude-encoding unitary $V$. This is possible because the hardness is confined to the feature-map oracle at the single training input $x_1$, while the classical oracles $O_R$, $O_N$, $O_C$ and the coefficient quantum-state oracle $W(\alpha)$ act on the fixed vector $\alpha=(A)$ and are therefore instance-independent, as established in Step~5. It says nothing, however, about procedures that exploit a known structure of the feature-map unitaries: the family of Step~1 is hard only insofar as the map $x'\mapsto U(x')$ is accessible as a black box. Second, the hard family has $N=1$, where $\lVert\alpha\rVert_1=\lVert\alpha\rVert_2=\lVert\alpha\rVert_{\nicefrac{2}{3}}=\lVert\alpha\rVert_\infty$. What the theorem certifies is therefore the optimality of the rate of Theorem~\ref{thm:AAOQ} in the two parameters $\lVert\alpha\rVert_1$ and $\varepsilon$; it does not rule out algorithms whose cost is governed by a finer feature of the coefficient vector, and we leave the tight rate for general $N$ open. Third, quantum amplitude estimation is optimal for each individual kernel term: applying Theorem~\ref{thm:lower-bound-full} with $N=1$ and $A=1$ shows that, for any target precision $\varepsilon_i$ with $\Omega(2^{-n})\leq\varepsilon_i\leq\calO(1)$, estimating $k(x,x_i)$ to additive error $\varepsilon_i$ requires $\Omega(\varepsilon_i^{-1})$ queries to the single-input oracle. Per-term estimation in \emph{list-and-sum} strategies is therefore already optimal, and any improvement must arise from changing how the sum is approximated.

\section{Sample-and-average algorithms}\label{a:sample-average}
    We introduce one final algorithmic choice along the axis of how to approximate the sum in Eq.~\eqref{eq:kernel_function_setup}, which we refer to as \emph{sample-and-average}.
    Having identified that we can express the sum as the expectation value of a quantum observable, we take a step back and notice we could have expressed it as the expectation value of a \emph{classical} observable.
    Following the notation in Definition~\ref{def:short-hand}, we could re-write inference as
    \begin{align}
        f_\alpha(x) &=\sum_{i=1}^N \alpha_i k(x,x_i) = \lVert\alpha\rVert_1\sum_{i=1}^Np_is_ik(x,x_i) = \lVert\alpha\rVert_1\bbE_{i\sim p}[s_i k(x,x_i)].
    \end{align}
    Intuitively, we could now decide to approximate the expectation value on the right-most part of the equation via its finite-sample mean.
    In doing this, we recall that we defined $k(x,x')=\lvert\langle\phi(x)|\phi(x')\rangle\rvert^2$ as a (quantum) expectation value itself, and hence this formula involves two sources of randomness: an \emph{inner source of randomness} associated to estimating $k(x,x_i)$ for a given $i$, and an \emph{outer source of randomness} in selecting $i$ from the distribution $p$.
    In the following, we give an algorithmic template as Algorithm~\ref{alg:sample-average-template}, where the \emph{inner} source of randomness (evaluating the kernel, either via sampling or via quantum amplitude estimation), is wrapped by the \emph{outer} source of randomness (evaluating the sum, via sampling).
    The template is given in terms of a free parameter $r\in\{1,2\}$ which can be further specialized to $r=1$ for sampling and $r=2$ for quantum amplitude estimation.
    We analyze the sample complexity of Algorithm~\ref{alg:sample-average-template} generically in Theorem~\ref{thm:sample-average-template}, and further provide Corollaries~\ref{cor:sample-average-sampling} and~\ref{cor:sample-average-qae} for the specific runtimes using sampling and quantum amplitude estimation, respectively.

    For the following result, we assume we have an \emph{unbiased} estimator for quantum amplitude estimation, as for instance the one given by \citet{ambainis2026tight}.

 \begin{theorem}[Sample-and-average with fixed budget algorithm template analysis]\label{thm:sample-average-template}
    Let $X_1,\ldots,X_N\in[0,1]$ be fixed values, let $a\in\bbR$ be a normalization factor, let $s=(s_1,\ldots,s_N)\in\{\pm1\}^N$ be a vector of signs, and let $p=(p_1,\ldots,p_N)$ be an $N$-outcome probability distribution.
    Define $F\coloneqq a \bbE_{i\sim p}[s_iX_i]$.
    Assume we have $T$-query independent estimators $\hX^{(T)}_i$ with $\bbE_{\hX}[\hX^{(T)}_i]=X_i$ and $\operatorname{Var}_{\hX}[\hX^{(T)}_i]\leq c/T^r$, for some constant $c>0$, and for $r\in\{1,2\}$.
    For $L\in\bbN$, consider i.i.d. samples $i_1,\ldots,i_L\sim p$, and define the intermediate $T$-query estimator $\hY^{(T)}_\ell$ corresponding to the $\ell^\text{th}$ sample: $\hY^{(T)}_\ell \coloneqq s_{i_\ell}\hX^{(T)}_{i_\ell}$.
    Finally, define the $M$-query estimator $\hF^{(M)}=a \sum_{\ell=1}^L \hY^{(T)}_\ell/L$, where $M=LT$.
    Then, for any $\varepsilon>0$, to ensure $\lvert\hF^{(M)}-F\rvert<\varepsilon$ with constant probability, it suffices to choose $M\in\calO(a^2 \varepsilon^{-2})$ total queries, with $T\in\calO(1)$ inner queries per sample.
    For any $\delta\in(0,1)$, by a median-of-means amplification (repeating the estimator $\calO(\log(\delta^{-1}))$ times and taking the median), the failure probability can be reduced to $\delta$ without changing the $\varepsilon$-dependence of the query complexity.
\end{theorem}

\begin{proof}
    We prove the statement directly by first deriving the variance of the estimator $\hF^{(M)}$ and then optimizing over the possible values of $T$ and $L$, with $M=LT$.
    We define the intermediate quantity $\hY^{(M)}=\sum_{\ell=1}^L\hY_\ell^{(T)}$.
    From the independence of the estimators, it follows that
    \begin{align}
        \operatorname{Var}\!\left(\hF^{(M)}\right)
        = \frac{a^2}{L^2}\operatorname{Var}\!\left(\hY^{(M)}\right)
        = \frac{a^2}{L^2} \sum_{\ell=1}^L \operatorname{Var}\!\left(\hY_\ell^{(T)}\right)
        = \frac{a^2}{L^2}\cdot L\cdot\operatorname{Var}\!\left(\hY_\ell^{(T)}\right)
        = \frac{a^2}{L}\operatorname{Var}\!\left(\hY_\ell^{(T)}\right).
    \end{align}
    By Chebyshev's inequality, this implies that $\lvert\hF^{(M)}-F\rvert<\varepsilon$ with constant probability whenever $\operatorname{Var}(\hF^{(M)})\in\calO(\varepsilon^2)$.
    We next invoke the law of total 
    variance, to obtain
    \begin{align}
    \operatorname{Var}\left(\hY^{(T)}_\ell\right)
        &= \bbE_{i\sim p}\left[\operatorname{Var}_{\hX}\left(\left.\hY^{(T)}_i\right\vert i\right)\right]
        + \operatorname{Var}_{i\sim p}\left(\bbE_{\hX}\left[\left.\hY^{(T)}_i\right\vert i\right]\right).
    \end{align}
    We can relate each term to the estimators for the initial quantities as
    \begin{align}
        \bbE_{i\sim p}\left[\operatorname{Var}_{\hX}\left(\left.\hY^{(T)}_i\right\vert i\right)\right]
        &= \bbE_{i\sim p}\left[\operatorname{Var}_{\hX}\left(\hX^{(T)}_i\right)\right]
        \leq \bbE_{i\sim p}\left[\frac{c}{T^r}\right]
        = \frac{c}{T^r}, \\
        \operatorname{Var}_{i\sim p}\left(\bbE_{\hX}\left[\left.\hY^{(T)}_i\right\vert i\right]\right)
        &= \operatorname{Var}_{i\sim p}\left(s_i X_i\right)
        \leq 1.
    \end{align}
    Plugging these back in the previous formulas, we obtain
    \begin{align}
    \operatorname{Var}\left(\hF^{(M)}\right)
        \leq \frac{a^2}{L}\left(1+\frac{c}{T^r}\right).
    \end{align}
    This formula holds in general, and we can exploit it to bound the required number of queries.
    By enforcing $M=LT$, we would like to minimize $\tfrac{a^2}{L}\left(1+\tfrac{c}{T^r}\right)$.
    We start by substituting $L=M/T$, to get
    \begin{align}
        \frac{a^2T}{M}\left(1+\frac{c}{T^r}\right)
        = \frac{a^2}{M}\left(T+\frac{c}{T^{r-1}}\right).
    \end{align}
    To find the minimum with respect to $T$, we take the derivative
    \begin{align}
        \der{T}\left(\frac{a^2}{M}\left(T+\frac{c}{T^{r-1}}\right)\right)
        &= \begin{cases}
            \frac{a^2}{M} &\text{if } r=1, \\
            \frac{a^2}{M}\left(1-\frac{c}{T^2}\right) &\text{if } r=2.
        \end{cases}
    \end{align}
    In the case $r=1$, it follows that there are no local extrema, and the first derivative is always positive, hence the minimum is achieved at the lower limit of the domain, $T=1$.
    In the case $r=2$, the derivative nullifies when $T=\sqrt{c}$, which is a constant.
    Therefore, in both cases $T\in\calO(1)$ achieves the optimal balance, for a total number of queries $M\in\calO(a^2 \varepsilon^{-2})$.
\end{proof}

    \begin{algorithm}[H]
        \caption{Sample-and-average with fixed budget, template.}
        \label{alg:sample-average-template}
        \begin{algorithmic}[1]
            \Require $x \in \calX$ \Comment New input.
            \Require $\varepsilon>0$ \Comment Required additive precision.
            \Ensure $f_\alpha(x)=\sum_{i=1}^N\alpha_ik(x,x_i)$ to $\varepsilon$ additive precision. \Comment Inference, Eq.~\eqref{eq:kernel_function_setup}.
            \For{$\ell\in\{1,\ldots,L\}$}:
                \State Sampling oracle $O_C(\alpha)$: $i_\ell\sim p$
                \State Estimate $k(x,x_{i_\ell})$ using $T$ queries to the single-input oracle: $\hk^{(T)}(x,x_{i_\ell})$
                \Statex \hfill \Comment Variance $\calO(\frac{1}{T^r})$, with $r\in\{1,2\}$.
                \State Multiply $\hk^{(T)}(x,x_{i_\ell})$ with the sign $s_{i_\ell}$: $\hY_\ell^{(T)}=s_{i_\ell}\hk^{(T)}(x,x_{i_\ell})$. \Comment Variance remains $\calO(\frac{1}{T^r})$.
            \EndFor \Comment Total number of queries is $M=LT$.
            
            \State Sum $\hY^{(M)}=\sum_{\ell=1}^L \hY_\ell^{(T)}$ \Comment Variance $\calO(1+\frac{c}{T^r})$.
            \State Norm oracle $O_N(\alpha)$: $\lVert\alpha\rVert_1$
            \State Multiply $\hY^{(M)}$ by $\lVert\alpha\rVert_1/L$: obtain $\hf^{(M)}_\alpha(x)=\frac{\lVert\alpha\rVert_1}{L}\hY^{(M)}$.
            \State\Return $\hf^{(M)}_\alpha(x)$ 
            \Statex \hfill
            \Comment Total variance $\calO(\frac{\lVert\alpha\rVert_1^2}{L}(1+\frac{c}{T^r}))$, sufficient $T\in\calO(1)$ and $M\in\calO(\varepsilon^{-2}\lVert\alpha\rVert_1^2)$.
        \end{algorithmic}
    \end{algorithm}

    \begin{corollary}[Sampling]\label{cor:sample-average-sampling}
        Consider Algorithm~\ref{alg:sample-average-template} with $r=1$ and the problem set-up in Definition~\ref{def:problem_setup}.
        To estimate the quantum kernel model prediction in Eq.~\eqref{eq:kernel_function_setup} to additive precision $\varepsilon$ with constant success probability, the runtime scaling is dominated by the query complexity
        \begin{align}
            M&\in\calO\!\left(\frac{\lVert\alpha\rVert_1^2}{\varepsilon^2}\right).
        \end{align}
        The optimal allocation of queries is $T=1$, $L=M$.
    \end{corollary}
    \begin{proof}
        Immediate from Theorem~\ref{thm:sample-average-template}.
    \end{proof}

    \begin{corollary}[Quantum amplitude estimation]\label{cor:sample-average-qae}
        Consider Algorithm~\ref{alg:sample-average-template} with $r=2$ and the problem set-up in Definition~\ref{def:problem_setup}.
        Assuming that quantum amplitude estimation provides an unbiased estimator: to estimate the quantum kernel model prediction in Eq.~\eqref{eq:kernel_function_setup} to additive precision $\varepsilon$ with constant success probability, the runtime scaling is dominated by the query complexity
        \begin{align}
            M&\in\calO\!\left(\frac{\lVert\alpha\rVert_1^2}{\varepsilon^2}\right).
        \end{align}
        The optimal allocation of queries is $T\in\calO(1)$, $L\in\calO(M)$.
    \end{corollary}
    \begin{proof}
        Immediate from Theorem~\ref{thm:sample-average-template}.
    \end{proof}
    Corollaries~\ref{cor:sample-average-qae} and~\ref{cor:sample-average-sampling} therefore establish the same query-complexity upper bound for sampling and quantum amplitude estimation.

    \begin{remark}[Sample-and-average recovers previous algorithms.]
        The proof of Theorem~\ref{thm:sample-average-template} confirms that the optimal balance between the inner and outer numbers of samples, $T$ and $L$ respectively, is struck by $T\in\calO(1)$ (and even $T=1$ for $r=1$).
        Explicitly, this is independent of whether the inner estimator is based on sampling or quantum amplitude estimation $r\in\{1,2\}$.
        The resulting estimator achieves the same query complexity scaling as both \emph{all-at-once} and \emph{list-and-sum} with adaptive budget, via sampling.
        
        On the one hand, we can interpret Algorithm~\ref{alg:sample-average-template} as the \emph{all-at-once} observable where the linear combination over indices is instantiated as a classical expectation value, instead of a quantum superposition.
        On the other hand, we can also interpret it as the \emph{list-and-sum} algorithm with adaptive 
        budget, where the number of classical samples $M_i$ for the $i^\text{th}$ entry is automatically selected by the probability distribution $i\sim p$.
    \end{remark}



\end{document}